\documentclass[11pt]{article}
\usepackage[utf8]{inputenc}
\usepackage[T1]{fontenc}

\usepackage[english,activeacute]{babel}
\usepackage{natbib}
\usepackage{comment}
\usepackage{float}
\usepackage[hidelinks]{hyperref}
\usepackage{mathrsfs}
\usepackage{enumitem}
\usepackage[font={small,it}]{caption}
\usepackage{amsmath,amsfonts,amsthm,amssymb}
\usepackage{bm,rotating,multirow,dsfont,graphicx}
\usepackage[usenames, dvipsnames]{color}
\usepackage{url}
\usepackage{multicol}
\usepackage{multirow}
\usepackage[T1]{fontenc}
\usepackage{flafter}
\usepackage{appendix}
\usepackage{subfigure}
\usepackage{xcolor}
\usepackage{soul}
\usepackage{setspace}
\usepackage{booktabs}
\usepackage{tabularx}
\usepackage{setspace}

\usepackage{algorithm}
\usepackage{algpseudocode}

\usepackage{tikz}
\usetikzlibrary{arrows.meta}
\usetikzlibrary{positioning}
\usetikzlibrary{calc}

\makeatletter
\def\hlinewd#1{%
	\noalign{\ifnum0=`}\fi\hrule \@height #1 %
	\futurelet\reserved@a\@xhline}
\makeatother
\def\spacingset#1{\renewcommand{\baselinestretch}{#1}\small\normalsize}
\spacingset{1}

\begin{document}

\title{\textbf{Variational Inference for Fully Bayesian Hierarchical Linear Models}}

\date{}

\author{
    Cristian Parra-Aldana\qquad\qquad Juan Sosa\footnote{Corresponding author: jcsosam@unal.edu.co.} \\
    \hspace{1cm}\\
    Universidad Nacional de Colombia, Colombia 
}

\maketitle

\begin{abstract}
Bayesian hierarchical linear models provide a natural framework to analyze nested and clustered data. Classical estimation with Markov chain Monte Carlo produces well calibrated posterior distributions but becomes computationally expensive in high dimensional or large sample settings. Variational Inference and Stochastic Variational Inference offer faster optimization based alternatives, but their accuracy in hierarchical structures is uncertain when group separation is weak. This paper compares these two paradigms across three model classes, the Linear Regression Model, the Hierarchical Linear Regression Model, and a Clustered Hierarchical Linear Regression Model. Through simulation studies and an application to real data, the results show that variational methods recover global regression effects and clustering structure with a fraction of the computing time, but distort posterior dependence and yield unstable values of information criteria such as WAIC and DIC. The findings clarify when variational methods can serve as practical surrogates for Markov chain Monte Carlo and when their limitations make full Bayesian sampling necessary, and they provide guidance for extending the same variational framework to generalized linear models and other members of the exponential family.

\end{abstract}


\spacingset{1.1} 

\newpage

\section{Introduction}

Statistical modeling often begins with the analysis of the relationship between a response variable $\bm{y}$ and one or more predictors $\bm{x}$. In the classical linear regression setting, the focus is on the conditional distribution $p(\bm{y}\mid\bm{x},\bm{\Theta})$, where $\bm{\Theta} = (\bm{\beta}, \sigma^2)$ is the parameter vector, with $\bm{\beta}$ the regression coefficients and $\sigma^2$ the error variance. Under exchangeability assumptions (e.g., \citealt{bernardo1994bayesian}), the model can be written as $\bm{y} = \bm{X}\bm{\beta} + \bm{\epsilon}$, with $\bm{\epsilon} \sim \mathsf{N}_n(\bm{0}, \sigma^2 \bm{I}_n)$, where $\bm{X}$ is the design matrix \citep{GELMAN14}. This formulation provides a powerful and interpretable framework, but it assumes conditional independence across observations and a common error variance, which is not realistic in many applied settings.

The hierarchical linear regression model (HLRM) extends this framework by allowing data to be organized into groups or clusters, such as students within schools, patients within hospitals, or repeated measurements within individuals. For group $j$ and observation $i$, the model is $y_{ij} = \bm{x}_{ij}^\top \bm{\beta}_j + \epsilon_{ij}$, with $\epsilon_{ij} \sim \mathsf{N}(0,\sigma_j^2)$, where the group specific regression parameters $\bm{\beta}_j$ are drawn from a population distribution. This nested specification captures both within group and between group variability and avoids the extremes of complete pooling and separate estimation for each group \citep{HOFF,SOSA}. Bayesian hierarchical modeling formalizes this perspective by treating regression coefficients and variance components as random quantities with prior distributions that propagate uncertainty across levels.

Estimation in Bayesian HLRMs often relies on Markov chain Monte Carlo (MCMC) \citep{albert2009bayesian}. MCMC is asymptotically exact and produces well calibrated posterior distributions, but its computational cost can be prohibitive in high dimensional or large scale problems \citep{GELMAN14}. These limitations have motivated the development of scalable alternatives based on variational inference (VI), for example \citep{NIPS2000_77369e37,beal2003variational,Wainwright2007-ul}. VI is a deterministic optimization based approach that approximates intractable posterior distributions with a tractable family. It has its roots in variational calculus \citep{gelfand2000calculus}, and modern formulations use exponential family representations \citep{Diaconis1979-lb} and conjugate structures to enable efficient Bayesian computation.

The core idea of VI is to posit a family of candidate densities $q(\bm{\Theta})$ and to select the member that minimizes the Kullback–Leibler (KL) divergence to the posterior. This leads to the Evidence Lower Bound (ELBO) as the optimization objective. Early work shows the usefulness of VI in Bayesian networks and latent variable models \citep{Blei2006-kd,Ormerod2010-bc}, and comprehensive reviews emphasize both its computational advantages and its statistical limitations \cite{Blei2016VariationalIA,zhang2018advancesvariationalinference}. A key requirement in classical VI algorithms is that the model belongs to the natural exponential family, which enables closed form coordinate updates and efficient implementations \cite{Diaconis1979-lb,NIPS2000_77369e37}. This tractability often reduces approximation quality, since VI tends to underestimate posterior variances and can distort dependence among parameters.

Modern advances seek to address these shortcomings. Structured VI relaxes independence assumptions to improve statistical fidelity \cite{wang22g}, while amortized inference with neural networks enables scalable inference in deep generative models, most notably within the variational autoencoder (VAE) framework \cite{VAE}. Black box and stochastic variants, such as stochastic variational inference (SVI), extend applicability to massive datasets by combining stochastic gradients with Robbins--Monro updates \cite{HOFFMAN,RANGAN}. These developments highlight the trade off between scalability and accuracy and position VI and SVI as attractive but imperfect surrogates for MCMC.

This paper builds on this body of work by examining the performance of VI and SVI in hierarchical linear regression. We derive the algorithms under conjugate priors, implement coordinate ascent and stochastic variants, and benchmark them against MCMC in controlled simulations and real data applications. Our contributions are threefold. First, we provide a critical comparison of sampling based and optimization based inference across two models of increasing complexity, the Linear Regression Model (LRM) and the Clustered Hierarchical Linear Regression Model (CHLRM). Second, we present a detailed computational implementation that makes variational methods accessible for hierarchical regression analysis. Third, we offer empirical evidence that clarifies when VI and SVI can serve as practical substitutes for MCMC and when their limitations in posterior approximation become decisive. In doing so, the paper contributes to the methodological literature by linking statistical and computational perspectives on variational inference and plays a pedagogical role by clarifying its theoretical foundations and practical performance.

The structure of this paper is as follows. Section 2 introduces the theoretical background and reviews the foundations of Bayesian inference, the role of MCMC methods, and the main ideas of VI, including the KL divergence, the ELBO, and coordinate ascent and stochastic variational approaches. Section 3 presents the methodological framework and describes the LRM and the CHLRM, together with their prior specifications, posterior inference, and illustrative applications. Section 4 discusses the main findings and highlights the methodological and computational trade offs between MCMC, VI, and SVI, and closes with implications for future research.

\section{Theoretical Background}

This section presents the theoretical foundations of Bayesian statistics, simulation based inference and variational inference, and provides the basis for the analyses and discussions in the following sections.

\subsection{Bayesian Inference}

A statistical model is a collection of probability distributions indexed by an unknown parameter vector $\bm{\theta} = (\theta_1,\ldots,\theta_p)\in\bm{\Theta}$, which must be estimated from the data $\bm{y} = (y_1,\ldots,y_n)\in\bm{\mathcal{Y}}$. In the Bayesian framework, $\bm{\theta}$ is treated as a random quantity with a probability distribution that is updated using the information provided by the observations, yielding an updated distribution that summarizes uncertainty about the parameter based on the data.

Bayesian inference represents uncertainty about $\bm{y}$ and $\bm{\theta}$ with probability distributions on the sample space $\bm{\mathcal{Y}}$ and the parameter space $\bm{\Theta}$. The prior distribution $p(\bm{\theta})$ summarizes beliefs about the parameter external to the data, and the sampling model $p(\bm{y}\mid\bm{\theta})$ describes the probabilistic behavior of the data for each $\bm{\theta}$. 
Using Bayes's theorem, we update our knowledge about $\bm{\theta}$ with the information in the dataset $\bm{y}$ as
\begin{equation*}
p(\bm{\theta}\mid\bm{y})
= \frac{p(\bm{y}\mid\bm{\theta})\,p(\bm{\theta})}{\int_{\bm{\Theta}} p(\bm{y}\mid\bm{\theta})\,p(\bm{\theta})\,\textsf{d}\bm{\theta}}
\propto p(\bm{y}\mid\bm{\theta})\,p(\bm{\theta}),
\end{equation*}
because the denominator does not depend on $\bm{\theta}$.
This framework is particularly appealing because it yields the posterior predictive distribution
\begin{equation*} 
\label{bayes_pred}
p(\tilde{y}\mid\bm{y}) = \int_{\bm{\Theta}} p(\tilde{y},\bm{\theta}\mid\bm{y}) \,\textsf{d}\bm{\theta} 
= \int_{\bm{\Theta}} p(\tilde{y}\mid\bm{\theta})\,p(\bm{\theta}\mid\bm{y})\,\textsf{d}\bm{\theta}.
\end{equation*}

\subsection{Markov Chain Monte Carlo Methods}

In practice point and interval estimates of $\bm{\theta}$, or of any function of it, are often obtained using MCMC methods. These methods draw values from the posterior distribution $p(\bm{\theta}\mid\bm{y})$ by constructing a Markov chain whose stationary distribution is $p(\bm{\theta}\mid\bm{y})$ (e.g., \citealt{GAMERMAN}, \citealt{HOFF}, and \citealt{GELMAN14}). 
Suppose we are able to obtain $B$ draws of $\bm{\theta}$ from the posterior distribution $p(\bm{\theta}\mid\bm{y})$, that is $\bm{\theta}^{(b)} \sim p(\bm{\theta}\mid\bm{y})$, for $b=1,\ldots,B$. If these samples come from an ergodic Markov chain whose stationary distribution is $p(\bm{\theta}\mid\bm{y})$, then by the ergodic theorem the empirical distribution of $\bm{\theta}^{(1)},\ldots,\bm{\theta}^{(B)}$ approximates the true posterior distribution as $B \rightarrow \infty$. In this case, the posterior mean of $\bm{\theta}$ can be approximated as
\begin{equation*} 
\label{MCMC_g}
\mathsf{E}\bigl(\bm{\theta}\mid\bm{y}\bigr)
= \int_{\bm{\Theta}} \bm{\theta}\,p(\bm{\theta}\mid\bm{y})\,\textsf{d}\bm{\theta}
\approx \frac{1}{B}\sum_{b=1}^{B} \bm{\theta}^{(b)}.
\end{equation*}
This methodology also allows prediction of new values and imputation of missing values through the posterior predictive distribution $p(\tilde{y}\mid\bm{y})$, by generating draws $\tilde{y}^{(b)} \sim p(\tilde{y}\mid\bm{\theta}^{(b)})$ for each posterior sample $\bm{\theta}^{(b)}$.

\subsubsection{Gibbs Sampling}

Gibbs sampling (e.g., \citealt{gelfand2000gibbs}) is a very popular MCMC algorithm for generating samples from a joint posterior distribution when the full conditional distributions (fcd's) of each parameter are available in closed form or can be sampled from. Specifically, let $\bm{\theta} = \left( \theta_1, \ldots, \theta_p \right)$ denote the full parameter vector, and let $\bm{\theta}_{-j} = \left( \theta_1, \ldots, \theta_{j-1}, \theta_{j+1}, \ldots, \theta_p \right)$ be the vector that excludes the $j$th component. The full conditional distribution of $\theta_j$ is $p\left(\theta_{j} \mid \bm{\theta}_{-j}, \bm{y}\right)$, from which the Gibbs sampler iteratively draws, cycling through $j = 1,\ldots,p$.

Given an initial value $\bm{\theta}^{(0)}$, the Gibbs sampler produces a dependent sequence $\bm{\theta}^{(1)}, \ldots, \bm{\theta}^{(B)}$, where each $\bm{\theta}^{(b)}$ is obtained by successively sampling each coordinate from its corresponding fcd, conditioned on the most recent values of the other coordinates:
\begin{align*}
\theta_1^{(b)} &\sim p(\theta_1 \mid \theta_2^{(b-1)}, \theta_3^{(b-1)}, \ldots, \theta_p^{(b-1)}, \bm{y}), \\
\theta_2^{(b)} &\sim p(\theta_2 \mid \theta_1^{(b)}, \theta_3^{(b-1)}, \ldots, \theta_p^{(b-1)}, \bm{y}), \\
&\ \vdots \\
\theta_p^{(b)} &\sim p(\theta_p \mid \theta_1^{(b)}, \theta_2^{(b)}, \ldots, \theta_{p-1}^{(b)}, \bm{y}).
\end{align*}
This iterative procedure defines an ergodic Markov chain with stationary distribution $p(\bm{\theta}\mid\bm{y})$, which in turn allows posterior inference (see Algorithm \ref{alg_gibbs_sampling}).

\begin{algorithm}[!htb]
\footnotesize
\caption{Gibbs Sampling.}\label{alg_gibbs_sampling}
\begin{algorithmic}
\State \textbf{Input:} Initial values $\theta_j^{(0)}$ for $j = 1,\dots,p$, data $\bm{y}$, number of iterations $B$
\For{$b = 1,\dots,B$}
  \For{$j = 1,\dots,p$}
    \State Sample
    $$
    \theta_j^{(b)} \sim p(\theta_j \mid \theta_1^{(b)}, \ldots, \theta_{j-1}^{(b)}, \theta_{j+1}^{(b-1)}, \ldots, \theta_p^{(b-1)}, \bm{y})
    $$
  \EndFor
\EndFor
\State \textbf{Output:} Posterior draws $\bm{\theta}^{(1)},\ldots,\bm{\theta}^{(B)}$
\end{algorithmic}
\end{algorithm}

\subsubsection{Metropolis–Hastings}

When the fcd of a parameter does not have a closed form or is difficult to sample from directly, the Metropolis--Hastings (MH; e.g., \citealt{chib1995understanding}) algorithm provides a general strategy to obtain draws from the target distribution. Under this sampling scheme, a candidate value is generated from a proposal density and then accepted or rejected according to an acceptance probability that corrects for the discrepancy between the proposal and the target posterior (see Algorithm \ref{alg_mh}).

\begin{algorithm}[!htb]
\footnotesize
\caption{Metropolis–Hastings}\label{alg_mh}
\begin{algorithmic}
\State \textbf{Input:} Initial values $\theta_j^{(0)}$ for $j = 1,\dots,p$, data $\bm{y}$, proposal density $q(\cdot\mid\cdot)$, number of iterations $B$
\For{$b = 1,\dots,B$}
  \For{$j = 1,\dots,p$}
    \State Generate a candidate value
    \[
      \theta_j^{*(b)} \sim q\big(\theta_j^{*}\mid\theta_j^{(b-1)}\big)
    \]
    \State Compute the acceptance ratio
    \[
      r_j = 
      \frac{
        p\big(\theta_j^{*(b)} \mid \bm{\theta}_{-j}^{(b)},\bm{y}\big)\,
        q\big(\theta_j^{(b-1)}\mid\theta_j^{*(b)}\big)
      }{
        p\big(\theta_j^{(b-1)} \mid \bm{\theta}_{-j}^{(b)},\bm{y}\big)\,
        q\big(\theta_j^{*(b)}\mid\theta_j^{(b-1)}\big)
      }
    \]
    \State Draw $u \sim \mathsf{U}(0,1)$
    \If{$u < \min\{1,r_j\}$}
      \State Accept $\theta_j^{(b)} = \theta_j^{*(b)}$
    \Else
      \State Reject $\theta_j^{(b)} = \theta_j^{(b-1)}$
    \EndIf
  \EndFor
\EndFor
\State \textbf{Output:} Posterior draws $\bm{\theta}^{(1)},\ldots,\bm{\theta}^{(B)}$
\end{algorithmic}
\end{algorithm}

\subsubsection{Hamiltonian Monte Carlo}

Hamiltonian Monte Carlo (HMC; e.g., \citealt{neal2011mcmc}) is an MCMC method designed to reduce random walk behavior and slow convergence that often occur in classical algorithms such as Gibbs sampling and Metropolis--Hastings, especially when the posterior distribution has complex geometry \citep{GELMAN14}. By introducing an auxiliary momentum variable $\bm{\phi}$, HMC simulates Hamiltonian dynamics to generate proposals that explore the posterior distribution more efficiently. In this framework, samples are drawn from the joint distribution $p(\bm{\theta},\bm{\phi}\mid\bm{y}) = p(\bm{\theta}\mid\bm{y}) \, p(\bm{\phi})$, where $\bm{\phi}\sim \mathsf{N}_p(\bm{0},\bm{M})$ and $\bm{M}$ is a positive definite mass matrix with the same dimension as $\bm{\theta}$ (See Algorithm \ref{alg_hmc}). 

\begin{algorithm}[!htb]
\footnotesize
\caption{Hamiltonian Monte Carlo}\label{alg_hmc}
\begin{algorithmic}
\State \textbf{Input} Initial values $\bm{\theta}^{(0)}$, data $\bm{y}$, tunning parameters $\epsilon$, $L$, mass matrix $\bm{M}$, number of iterations $B$
\For{$b = 1,\dots,B$}
  \State Sample momentum $\bm{\phi} \sim \mathsf{N}_p(\bm{0},\bm{M})$
  \State Set $\bm{\theta} \leftarrow \bm{\theta}^{(b-1)}$ and $\bm{\phi} \leftarrow \bm{\phi}$
  \For{$l = 1,\dots,L$}
    \State Update momentum $\bm{\phi}$
    $$
        \bm{\phi} \leftarrow \bm{\phi}-\frac{\epsilon}{2}\frac{\partial}{\partial\bm{\theta}}\log p(\bm{\theta}\mid \bm{y})
        $$
    \State Update position $\bm{\theta}$
    $$
        \bm{\theta}\leftarrow \bm{\theta}+\epsilon\bm{M}\bm{\phi}
        $$
  \EndFor
  \State Compute acceptance ratio 
  $$
  r=\frac{p(\bm{\theta}^{*}\mid \bm{y})\,p(\bm{\phi}^{*})}{p(\bm{\theta}^{(b-1)}\mid\bm{y})\,p(\bm{\phi}^{(b-1)})}
  $$ 
  \State Draw $u \sim \mathsf{U}(0,1)$
  \If{$u < \min\{1,r\}$}
    \State Accept $\bm{\theta}^{(b)} = \bm{\theta}^*$
  \Else
    \State Reject $\bm{\theta}^{(b)} = \bm{\theta}^{(b-1)}$
  \EndIf
\EndFor
\State \textbf{Output} Posterior draws $\bm{\theta}^{(1)},\ldots,\bm{\theta}^{(B)}$
\end{algorithmic}
\end{algorithm}

\subsection{Kullback-Leibler Divergence}

The Kullback--Leibler (KL) divergence \citep{KL} is a measure from information theory that quantifies the difference between a probability distribution $p(\bm{y})$ and a reference distribution $q(\bm{y})$. It is defined as
$$
\mathsf{KL}\bigl(p(\bm{y})\mid\mid q(\bm{y})\bigr)
= \mathsf{E}_{p(\bm{y})}\left[\log\left(\frac{p(\bm{y})}{q(\bm{y})}\right)\right]
= \int_{\bm{\mathcal{Y}}} p(\bm{y}) \log\left(\frac{p(\bm{y})}{q(\bm{y})}\right)\,\textsf{d}\bm{y},
$$
for a continuous random vector. Under this definition, it is straightforward to show that
$\mathsf{KL}\left(p(\bm{y})\mid\mid q(\bm{y})\right)\geq0$ (non negativity),
$\mathsf{KL}\left(p(\bm{y})\mid\mid q(\bm{y})\right)=0$ if and only if $p(\bm{y})=q(\bm{y})$ (identity of indiscernibles), and
$\mathsf{KL}\left(p(\bm{y})\mid\mid q(\bm{y})\right)\neq \mathsf{KL}\left(q(\bm{y})\mid\mid p(\bm{y})\right)$ in general (asymmetry). Therefore, the KL divergence is not a distance or metric in the strict sense.

\subsection{Evidence Lower Bound}

Following \cite{Ormerod2010-bc}, consider a Bayesian model with parameter
$\bm{\theta}\in\bm{\Theta}$ and observed data $\bm{y}\in\bm{\mathcal{Y}}$. The posterior
distribution of $\bm{\theta}$ is $p(\bm{\theta}\mid\bm{y}) = p(\bm{y},\bm{\theta}) / p(\bm{y})$,
where $p(\bm{y})$ is the marginal likelihood or model evidence. Introducing an
arbitrary probability density function $q(\bm{\theta})$ on $\bm{\Theta}$, the marginal
likelihood can be written as
\begin{equation}
\label{lnPy}
\begin{split}
\log p(\bm{y})
&= \int q(\bm{\theta}) \log p(\bm{y}) \,\textsf{d}\bm{\theta} \\
&= \int q(\bm{\theta}) \log \left(\frac{p(\bm{y},\bm{\theta})}{p(\bm{\theta}\mid\bm{y})}\right) \,\textsf{d}\bm{\theta} \\
&= \int q(\bm{\theta}) \log \left(\frac{p(\bm{y},\bm{\theta})}{q(\bm{\theta})}\right) \,\textsf{d}\bm{\theta}
  + \int q(\bm{\theta}) \log \left(\frac{q(\bm{\theta})}{p(\bm{\theta}\mid\bm{y})}\right) \,\textsf{d}\bm{\theta} \\
&= \mathcal{L}(q) + \mathsf{KL}\bigl(q(\bm{\theta})\mid\mid p(\bm{\theta}\mid\bm{y})\bigr),
\end{split}
\end{equation}
where
\begin{equation*}
\label{ELBO}
\begin{split}
\mathcal{L}(q)
&= \mathsf{E}_{q(\bm{\theta})}\left[ \log \left(\frac{p(\bm{y},\bm{\theta})}{q(\bm{\theta})}\right)\right] \\
&= \mathsf{E}_{q(\bm{\theta})}\left[\log p(\bm{y},\bm{\theta}) \right]
  - \mathsf{E}_{q(\bm{\theta})}\left[\log q(\bm{\theta}) \right]. 
\end{split}
\end{equation*}
Since $\mathsf{KL}\bigl(q(\bm{\theta})\mid\mid p(\bm{\theta}\mid\bm{y})\bigr)\ge 0$, it follows that
$\mathcal{L}(q) \le \log p(\bm{y})$, so $\mathcal{L}(q)$ is known as the Evidence Lower Bound (ELBO) and provides a lower
bound for the log marginal likelihood.

\subsection{Variational Inference}

According to \cite{ZETER}, the variational approach has a different focus from MCMC methods. Instead of drawing samples from a distribution, it aims to find a probability distribution $q^*(\bm{\theta})\in\bm{\mathcal{Q}}$ that is as close as possible to the target distribution $p(\bm{\theta}\mid\bm{y})$. This leads to an optimization problem that requires specifying a cost function and a similarity measure to quantify the discrepancy between these two probability distributions.

Typically, $q^{*}(\bm{\theta})$ is chosen such that
\begin{equation*}
    \label{argmin}
    q^{*}(\bm{\theta})
    = \underset{q(\bm{\theta})\in\bm{\mathcal{Q}}}{\arg \min}\,
      \mathsf{D}\!\left(q(\bm{\theta}),p(\bm{\theta}\mid\bm{y})\right),
\end{equation*}
where $\mathsf{D}(\cdot,\cdot)$ denotes a discrepancy measure between probability distributions. In variational inference (VI), a common choice is the KL divergence.
Since $\log p(\bm{y})$ does not depend on $q(\bm{\theta})$, minimizing $\mathsf{KL}\left(q(\bm{\theta})\mid\mid p(\bm{\theta}\mid\bm{y})\right)$ is equivalent to maximizing the evidence lower bound $\mathcal{L}(q)$, because the equality given in Eq. \eqref{lnPy} and the KL divergence is non negative, so $\mathcal{L}(q)$ is always a lower bound on $\log p(\bm{y})$.

\subsubsection{Mean Field Approach}

The mean field approach assumes that the elements of $\bm{\theta}$ are independent and factorizes the variational distribution $q(\bm{\theta})$ as
\begin{equation}\label{MFA}
q(\bm{\theta}) = \prod_{j=1}^{p} q\left(\theta_j\right).    
\end{equation}
This assumption allows a tractable solution for each $\theta_j$ and facilitates efficient optimization. The approach is effective in capturing marginal effects, but it tends to underestimate the variance and higher order moments because it does not account for correlations among the parameters.

\subsubsection {Coordinate Ascent Variational Inference}

To optimize the ELBO, the Coordinate Ascent Variational Inference (CAVI) algorithm iteratively optimizes each factor of the variational distribution in Eq. \eqref{MFA}, holding the other components fixed. This procedure resembles Gibbs sampling and uses the chain rule of probability to write the joint distribution $p(\bm{y},\bm{\theta})$ as
$$
p(\bm{y},\bm{\theta}) = p(\bm{y})\prod_{j=1}^{p}p(\theta_j\mid \bm{\theta}_{-j},\bm{y}).
$$

Consider $\mathcal{L}(q)$ as a function of $q(\theta_j)$ for some $j\in\lbrace1,\dots,p\rbrace$, while keeping the other $p-1$ factors fixed. With some algebra it can be shown that Eq. \eqref{ELBO} can be rewritten as
\begin{equation}
\label{CAVI}
    \mathcal{L}(q)
    = -\mathsf{KL}\left(q(\theta_j)\mid\mid
       \exp(\mathsf{E}_{- j}\left[\log p(\bm{y},\bm{\theta})\right]\right)) + \text{constant},
\end{equation}
where $\mathsf{E}_{{j}}[\cdot]$ and $\mathsf{E}_{{-j}}[\cdot]$ denote expectations with respect to $q(\theta_{j})$ and $\prod_{i\neq j}q(\theta_i)$, respectively.
Since Eq. \eqref{CAVI} involves the KL divergence between $q(\theta_j)$ and $\exp(\mathsf{E}_{- j}[\log p(\bm{y},\bm{\theta})])$, this divergence is minimized when it is equal to zero. In that case,
\begin{equation}
\label{MVFB_opt}
   q(\theta_j)\propto \exp\left(\mathsf{E}_{- j}\left[\log p(\bm{y},\bm{\theta})\right]\right).   
\end{equation}
This expression corresponds to the optimal choice of $q(\theta_j)$ under the mean field assumption \cite{BISHOP}. Each variational factor depends on expectations over the remaining $p-1$ latent variables, which leads to a system of coupled updates. These dependencies define the iterative structure of the CAVI algorithm (see Algorithm \ref{alg_cavi}).

\begin{algorithm}[!htb]
\footnotesize
\caption{Coordinate Ascent Variational Inference}\label{alg_cavi}
\begin{algorithmic}
\Statex \textbf{Input} Initial values $\theta_j^{(0)}$ for $j = 1,...,p$, data $\bm{y}$
\While{ELBO has not converged}
    \For{$j = 1,...,p$}
        \State Set 
        $$
        q\left(\theta_j\right)\propto\exp{\left(\mathsf{E}_{{-j}}\left[\log{p(\bm{y},\bm{\theta})}\right]\right)}
        $$
    \EndFor
    \State Compute ELBO 
    $$
    \mathcal{L}(q)=\mathsf{E}_{q(\bm{\theta})}\left[\log (p(\bm{y},\bm{\theta})) \right]-\mathsf{E}_{q(\bm{\theta})}\left[\log (q(\bm{\theta}) )\right]
    $$
\EndWhile
\State \textbf{Output} Variational distribution $q(\bm{\theta})$
\end{algorithmic}
\label{CAVI_alg}
\end{algorithm}

Using this result, the goal is to maximize the $\mathsf{ELBO}$. To ensure tractability and efficiency, it is common to choose $q(\bm{\theta}) \in \bm{\mathcal{Q}}$, where $\bm{\mathcal{Q}}$ is a family of tractable probability distributions \citep{HOFFMAN}. Noting that $p(\bm{y},\bm{\theta}) \propto p(\theta_j,\bm{\theta}_{-j},\bm{y}) \propto p(\theta_j\mid \bm{\theta}_{-j},\bm{y})$, and that the denominator of $p(\theta_j\mid \bm{\theta}_{-j},\bm{y})$ does not depend on $\theta_j$, the optimal solution in Eq. \eqref{MVFB_opt} becomes
\begin{equation*}
\label{MVFB_opt_post}
q\left(\theta_j\right)\propto \exp\left(\mathsf{E}_{{- j}}\left[\log p(\theta_j\mid \bm{\theta}_{-j},\bm{y})\right]\right).
\end{equation*}

If $p(\theta_j\mid \bm{\theta}_{-j},\bm{y})$ belongs to the natural exponential family (NEF; e.g., \citealt{Wainwright2007-ul}, see Appendix \ref{Appendix_A}), then
\begin{equation*}
\label{fullconditionalfed}
    p(\theta_j\mid \bm{\theta}_{-j},\bm{y})
    = h(\theta_j)\exp\bigl\{\bm{\eta}_j(\bm{\theta}_{-j},\bm{y})^\top\bm{t}(\theta_{j})
      - a\big(\bm{\eta}_j(\bm{\theta}_{-j},\bm{y})\big)\bigr\}.
\end{equation*}
Substituting this expression into \eqref{MVFB_opt} gives
\[
   \mathsf{E}_{{-j}}\big[ \log p(\theta_j\mid \bm{\theta}_{-j},\bm{y})\big]
   = \log h(\theta_j)
     + \mathsf{E}_{{-j}}\left[\bm{\eta}_j(\bm{\theta}_{-j},\bm{y})^\top\bm{t}(\theta_{j})
     - a\big(\bm{\eta}_j(\bm{\theta}_{-j},\bm{y})\big)\right].
\]
Terms that do not depend on $\theta_j$ can be absorbed into the normalizing constant, so the optimal variational factor is
\begin{equation*}
   q\left(\theta_j\right)
   \propto h(\theta_j)\exp\bigl\{\mathsf{E}_{{-j}}\big[\bm{\eta}_j(\bm{\theta}_{-j},\bm{y})\big]^\top\bm{t}(\theta_{j})\bigr\},
\end{equation*}
which implies that each $q(\theta_j)=q(\theta_j\mid\nu_j)$ also belongs to the NEF, with natural parameter
\begin{equation}
\label{NMVFB_opt}
\nu_{j}=\mathsf{E}_{{-j}}\big[\bm{\eta}_j(\bm{\theta}_{-j},\bm{y})\big].
\end{equation}
This result shows the direct relationship between the variational distribution and its full conditional distribution and provides a simple mechanism to update the natural parameters when the full conditional is known.

\subsubsection{Stochastic Variational Inference}\label{SVI}

One of the most important classes of exponential family models are the conditionally conjugate models \citep{Blei2016VariationalIA}, in which the parameter vector $\bm{\theta}$ is partitioned into global and local components, $\bm{\theta} = (\bm{\theta_g},\bm{\theta_l})$. The vector $\bm{\theta_g}$ represents global latent variables associated with the overall model structure, and $\bm{\theta_l}$ contains local latent variables, one for each data point \citep{REGUEIRO}. In this framework, the joint distribution factorizes as
\begin{equation}
    p(\bm{y},\bm{\theta})
    = p(\bm{\theta_g},\bm{\theta_l},\bm{y})
    = p(\bm{\theta_l},\bm{y}\mid \bm{\theta_g})\,p(\bm{\theta_g})
    = p(\bm{\theta_g})\prod_{i=1}^{n} p(\theta_{l_i},y_i\mid\bm{\theta_g}),
\end{equation}
where both $p(\bm{\theta_l},\bm{y}\mid \bm{\theta_g})$ and $p(\bm{\theta_g})$ belong to conjugate exponential family distributions.

Specifically, the conditional likelihood factors as
\[
p(\theta_{l_i},y_i\mid\bm{\theta}_g)
\propto \exp\bigl\{\bm{\theta}_g^\top\bm{t}(\theta_{l_i},y_i)-a_l(\bm{\theta}_g)\bigr\},
\]
with prior distribution for the global parameter given by $p(\bm{\theta}_g) \propto \exp\bigl\{\bm{\lambda}^\top\bm{t}(\bm{\theta}_g)-a_g(\bm{\lambda})\bigr\}$, where $\bm{\lambda}=(\bm{\lambda}_{1},\lambda_{2})^\top$ is the natural parameter vector. By conjugacy, the full conditional distribution of $\bm{\theta}_g$ belongs to the same exponential family as the prior,
\[
p(\bm{\theta}_g \mid \bm{\theta_l},\bm{y} )
\propto \exp\bigl\{\bm{\eta}_g(\bm{\theta_l},\bm{y})^\top\bm{t}(\bm{\theta}_g)\bigr\},
\]
where $\bm{t}(\bm{\theta}_g) = \bigl(\bm{\theta}_g - a_l(\bm{\theta}_g)\bigr)^\top$ and $\bm{\eta}_g(\bm{\theta_l},\bm{y}) = \bm{\lambda}+\sum_{i=1}^{n}\bigl(\bm{t}(\theta_{l_i},y_i),1\bigr)^\top$ is the corresponding natural parameter.

The variational distribution factorizes as $q(\bm{\theta})=q(\bm{\theta}_l)\,q(\bm{\theta}_g)$. Applying the mean field optimality result in Eq. \eqref{NMVFB_opt}, the variational distribution for $\bm{\theta}_g$ is given by
\[
q(\bm{\theta}_{g}\mid\bm{h})
\propto \exp\big(\mathsf{E}_{l}[\bm{\eta}_g(\bm{\theta}_l,\bm{y})]^\top\bm{\theta}_{g}\big),
\]
so the global variational parameters are
\begin{equation}
\label{global_updates}
\bm{h}
= \bm{\lambda}+\sum_{i=1}^{n}\bigl(\mathsf{E}_{l}[\bm{t}(\theta_{l_i},y_i)],1\bigr)^\top.
\end{equation}

Assume now that each local full conditional $p(\theta_{l_i}\mid y_i,\bm{\theta}_g)$ also belongs to the NEF, so that
\[
p(\theta_{l_i}\mid y_i,\bm{\theta}_g)
\propto \exp\left\{\bm{\eta}_l(\bm{\theta}_g,y_i)^\top\theta_{l_i}
      - a\bigl(\bm{\eta}_l(\bm{\theta}_g,y_i)\bigr)\right\}.
\]
The corresponding local variational distribution is
\[
q(\theta_{l_i}\mid\bm{\varphi_i})
\propto \exp\bigl(\mathsf{E}_{g}[\bm{\eta}_l(\bm{\theta}_g,y_i)]^\top\theta_{l_i}\bigr),
\]
with local variational parameters
\begin{equation}
\label{local_updates}
\bm{\varphi_i}=\mathsf{E}_{g}\left[\bm{\eta}_l(\bm{\theta_g},y_i)\right].
\end{equation}
The global parameters in Eq. \eqref{global_updates} depend on expectations with respect to the local factors in Eq. \eqref{local_updates}, and the local parameters depend on expectations with respect to the global factor. This mutual dependence produces a system of coupled updates that defines the iterative structure of the CAVI algorithm for conditionally conjugate models (see Algorithm \ref{CAVI_FEDN_alg}).

\begin{algorithm}[!htb]
\footnotesize
\caption{CAVI for conditionally conjugate models}
\begin{algorithmic}
\Statex \textbf{Input} Initial values $\bm{h}^{(0)},\bm{\varphi}^{(0)}$ , data $\bm{y}$ 
\While{ELBO has not converged}
    \For{$i = 1,...,n$}
        \State Optimize local variational parameters 
        $$
        \bm{\varphi_i}=\mathsf{E}_{g}\left[\bm{\eta}_l(\bm{\theta_g},y_i)\right]
        $$
    \EndFor
    \State Update global variational parameters 
    $$
    \bm{h}=\bm{\lambda}+\sum_{i=1}^{n}(\mathsf{E}_{l}\left[\bm{t}\left(\theta_{l_i},y_i\right)\right],1)^\top
    $$
    \State Calculate ELBO 
    $$
    \mathcal{L}(q)=\mathsf{E}_{q(\bm{\theta})}\left[\log (p(\bm{y},\bm{\theta})) \right]-\mathsf{E}_{q(\bm{\theta})}\left[\log (q(\bm{\theta}) )\right]
    $$
\EndWhile
\Statex \textbf{Output} Optimal variational parameters $\bm{h}$ and $\bm{\varphi}$
\end{algorithmic}
\label{CAVI_FEDN_alg}
\end{algorithm}

Despite its flexibility, the CAVI algorithm has an important limitation because each iteration requires a full pass through all data points, which restricts its scalability for large datasets. To address this issue, \cite{HOFFMAN} provides an approach called Stochastic Variational Inference (SVI), which optimizes the ELBO using stochastic gradient methods based on an unbiased estimator of its natural gradient. Instead of computing expectations over the full dataset at every step, SVI uses mini batches or single observations to estimate the gradient and combines classical CAVI style updates with the stochastic approximation rule of \cite{Robbins&Monro:1951}. Under mild conditions this approach converges and greatly improves computational efficiency, which enables variational inference for massive datasets (see Algorithm \ref{alg_svi}).

\textit{Natural Gradients of ELBO}

In this context, the natural gradient of the ELBO in Eq. \eqref{ELBO} with respect to the global variational parameters $\bm{h}$ is
\[
\widetilde{\nabla}_{\bm{h}} \mathcal{L}(q)
= \mathsf{E}_{l}\left[\bm{\eta}_g(\bm{\theta}_{l}, \bm{y})\right] - \bm{h},
\]
and, similarly, for each local variational parameter $\bm{\varphi}_i$ the natural gradient is
\[
\widetilde{\nabla}_{\bm{\varphi}_i} \mathcal{L}(q)
= \mathsf{E}_{g}\left[\bm{\eta}_l(\bm{\theta}_g, y_i)\right] - \bm{\varphi}_i.
\]
These quantities give the directions of steepest ascent in the Riemannian geometry induced by the exponential family and can be more efficient than standard Euclidean gradients in variational inference optimization \citep{AMARI}

\textit{Noisy Gradient Estimator}

In practice, computing the exact gradient of the ELBO over the entire dataset is often computationally prohibitive. To address this, SVI uses a mini batch strategy. A random subsample $\mathcal{S} \subset \{1,\dots,\mathcal{N}\}$ of size $|\mathcal{S}|$ is selected, where $\mathcal{N}$ is the total number of sampling units (for example observations or groups), and the indices in $\mathcal{S}$ are sampled uniformly at random. 

To keep the gradient estimator unbiased with respect to the full dataset, a scaling factor $\mathcal{C} = \mathcal{N} / |\mathcal{S}|$ is introduced to rescale the contribution of the mini batch. Using this factor, an unbiased estimator of the natural gradient with respect to $\bm{h}$, often called the noisy gradient, is
\[
\widehat{\widetilde{\nabla}}_{\bm{h}} \mathcal{L}(q)
= \bm{\lambda}
  + \mathcal{C} \, \sum_{i \in \mathcal{S}} \bigl(\mathsf{E}_{l}[\bm{t}(\theta_{l_i},y_i)],1\bigr)^\top
  - \bm{h},
\]
so that
\[
\mathsf{E}_{\mathcal{S}}\bigl[\widehat{\widetilde{\nabla}}_{\bm{h}} \mathcal{L}(q)\bigr]
= \widetilde{\nabla}_{\bm{h}} \mathcal{L}(q).
\]
This stochastic approximation enables scalable updates of the global variational parameters without processing the full dataset at every iteration, which makes SVI well suited for large scale Bayesian inference.

\textit{Robbins-Monro Update Rule}

Let $\bm{h}^{(t)}$ denote the value of the global variational parameter at iteration $t$ and let $\rho_t \in (0,1)$ be the learning rate. The update rule for $\bm{h}$ follows the stochastic approximation scheme
\begin{equation}
\label{Robbins-Monro}
\begin{split}
\bm{h}^{(t)}
=&\, \bm{h}^{(t-1)} + \rho_t \,\widehat{\widetilde{\nabla}}_{\bm{h}}\mathcal{L}\bigl(q(\bm\theta^{(t)})\bigr)\\
=&\, \bm{h}^{(t-1)} + \rho_t \Big(\bm{\lambda}
     + \mathcal{C} \, \sum_{i \in \mathcal{S}}\bigl(\mathsf{E}_{l}[\bm{t}(\theta_{l_i}^{(t)},y_i)],1\bigr)^\top
     - \bm{h}^{(t-1)}\Big)\\
=&\, (1-\rho_t)\,\bm{h}^{(t-1)}
   + \rho_t \Big(\bm{\lambda}
     + \mathcal{C}\,\sum_{i \in \mathcal{S}}\bigl(\mathsf{E}_{l}[\bm{t}(\theta_{l_i}^{(t)},y_i)],1\bigr)^\top\Big).
\end{split}
\end{equation}

Convergence of this update is guaranteed under the Robbins and Monro conditions \citep{Robbins&Monro:1951}, which require that $\sum_{t=1}^{\infty} \rho_t = \infty$ and $\sum_{t=1}^{\infty} \rho_t^2 < \infty$. A commonly used learning rate schedule \citep{HOFFMAN}, is
\[
\rho_t = (t + \tau)^{-\chi},
\]
where $\tau \ge 0$ is a delay parameter that downweights early iterations to improve stability, and $\chi \in (0.5,1]$ is the forgetting rate that controls how quickly past updates are discounted.

\begin{algorithm}[!htb]
\footnotesize
\caption{Stochastic Variational Inference}\label{alg_svi}
\begin{algorithmic}
\Statex \textbf{Input} Initial values $\bm{h}^{(0)},\bm{\varphi}^{(0)}$, data $\bm{y}$, learning rate parameters $\tau$ and $\chi$, minibatch size $|\mathcal{S}|$, number of iterations $T$
\For{$t = 1,\dots,T$}
    \State Sample a minibatch $\mathcal{S} \subset \{1,\dots,\mathcal{N}\}$
    \For{$i \in \mathcal{S}$}
        \State Optimize local variational parameter
        $$
        \bm{\varphi}_{i}^{(t)} = \mathsf{E}_{g} \left[ \bm{\eta}_l(\bm{\theta}_g^{(t-1)}, y_i) \right]
        $$
    \EndFor
    \State Compute $\rho_t = (t+\tau)^{-\chi}$
    \State Compute the intermediate global parameters
    $$
    \widehat{\bm{h}}^{(t)} = \bm{\lambda} +\mathcal{C}\,\sum_{i \in \mathcal{S}}\left(\mathsf{E}_{l} \left[ \bm{t}(\theta_{l_i}^{(t)}, y_i) \right],1\right)^\top
    $$
    \State Update global variational parameter
    $$
    \bm{h}^{(t)} = (1 - \rho_t)\bm{h}^{(t-1)} + \rho_t \widehat{\bm{h}}^{(t)}
    $$
\EndFor
\Statex \textbf{Output} Optimal variational parameters $\bm{h}$ and $\bm{\varphi}$
\end{algorithmic}
\label{alg:SVI_minibatch}
\end{algorithm}

\section{Illustration: Hierarchical Linear Models}

Building on the theoretical foundations of Bayesian inference and variational methods presented in the previous section, this section formalizes the Linear Regression Model (LRM), the Hierarchical Linear Regression Model (HLRM), and the Clustering Linear Regression Model (CHLRM) and specifies their prior and posterior structures. These models provide the basis for implementing and comparing Markov chain Monte Carlo (MCMC), Variational Inference (VI), and Stochastic Variational Inference (SVI) and establish the link between theory and application that guides the empirical analysis in this section.

We compare models and estimation algorithms on synthetic and real datasets using posterior predictive checks and information criteria that trade off fit and complexity. Posterior predictive $p$ values, which are obtained by comparing discrepancy statistics on replicated and observed data, assess model adequacy \citep{GELMAN14}. As relative criteria we use the Deviance Information Criterion (DIC; \citealt{spiegelhalter2002bayesian} and \citealt{spiegelhalter2014deviance}), which penalizes complexity through an effective number of parameters but can be unstable under multimodality or weak identifiability, and the Watanabe--Akaike Information Criterion (WAIC; \citealt{watanabe2010asymptotic} and \citealt{watanabe2013widely}), a fully Bayesian measure closely related to cross validation and usually more robust in hierarchical settings. Together, posterior predictive $p$ values, DIC and WAIC provide complementary evidence to assess model fit and predictive performance across competing Bayesian estimation methods.

\subsection{Linear Regression Model}

The classical linear regression model is described as the workhorse of statistics and supervised machine learning \citep{ML_MURPHY}. This model explains variation in a continuous response variable y using one or more predictors $\bm{x} = (x_{1},\dots,x_{p})^\top$ by specifying the mean of the response as a linear combination of the predictors \citep{Christensen2010-ps}. We consider the Linear Regression Model (LRM) given by
\begin{equation*}
    \label{LRM_i}
    y_i \mid \bm{x}_i,\bm{\beta},\sigma^2
    \overset{\text{ind}}{\sim}
    \mathsf{N}\big(\bm{x}_i^\top\bm{\beta},\sigma^2\big),
    \quad i = 1,\dots,n,
\end{equation*}
with $\bm{x}_i = (x_{i1},\dots,x_{ip})^\top$, $\bm{\beta} = (\beta_1,\dots,\beta_p)^\top$ is the regression coefficient vector, and $\sigma^2$ is the error variance. This model can be written as $\bm{y} \mid \bm{X},\bm{\beta},\sigma^2 \sim \mathsf{N}_n(\bm{X}\,\bm{\beta},\sigma^2\,\bm{I})$, where $\bm{y} = (y_1,\dots,y_n)^\top$ is the response vector, and $\bm{X} = [\bm{x}_1,\dots,\bm{x}_n]^\top$ is the $n\times p$ design matrix.

\subsubsection{Prior Distributions}

\begin{figure}[!htb]
    \centering
    \includegraphics[scale=0.45]{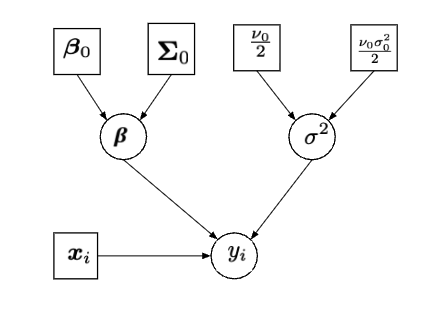}
    \caption{Directed acyclic graph (DAG) representation of the Linear Regression Model (LRM).}
    \label{fig:DAG_LRM}
\end{figure}

From a Bayesian perspective, we define the parameter vector as $\bm{\Theta} = (\bm{\beta},\sigma^2)$ and specify the prior distributions
\begin{equation*}
    \bm{\beta} \sim \mathsf{N}_p(\bm{\beta}_0, \bm{\Sigma}_0)
    \quad \text{and} \quad
    \sigma^2 \sim \mathsf{IG}\left(\nu_0 / 2, \nu_0 \sigma_0^2 / 2\right),
\end{equation*}
where $\bm{\beta}_0$, $\bm{\Sigma}_0$, $\nu_0$ and $\sigma_0^2$ are hyperparameters of the model (see Figure \ref{fig:DAG_LRM}).

Prior elicitation plays a key role in hierarchical linear regression because it controls how much external information is incorporated into the posterior. According to \cite{SOSA}, two strategies are common depending on the amount of prior knowledge. When little or no prior information is available, it is convenient to use weakly informative priors such as the unit information prior of \cite{KassWasserman1995}, which contributes information roughly equivalent to a single observation so that the prior does not dominate the likelihood while keeping the posterior well behaved. In regression models, \cite{HOFF} suggests centering the prior mean at the ordinary least squares (OLS) estimator by setting
\[
\bm{\beta}_0  = \hat{\bm{\beta}}_{\text{OLS}}, \quad
\bm{\Sigma}_0 = n \, \sigma^2_{0} \big(\bm{X}^\top \bm{X}\big)^{-1}, \quad
\nu_0 = 1, \quad 
\sigma_0^2 = \hat{\sigma}^2_{\text{OLS}},
\]
where
\[
\hat{\bm{\beta}}_{\text{OLS}} = \big(\bm{X}^\top \bm{X}\big)^{-1}\bm{X}^\top \bm{y}
\quad\text{and}\quad
\hat{\sigma}^2_{\text{OLS}}  = \frac{1}{n - p} \sum_{i=1}^n\big( y_i - \bm{x}_i^\top\hat{\bm{\beta}}_{\text{OLS}}\big)^2.
\]
With this choice, the prior on $\bm{\beta}$ carries information roughly equivalent to one data point. A second strategy is to use a scale invariant prior of the form $\bm{\beta}_0 = \bm{0}$ and $\bm{\Sigma}_0 = g \,\sigma_0^2 (\bm{X}^\top \bm{X})^{-1}$, with $g = n$, which yields Zellner's $g$ prior \cite{Zellner1986}. Both specifications give concrete and interpretable ways to encode weak or moderate prior information that are consistent with the model structure and straightforward to implement in practice.

\subsubsection{MCMC}

Under the LRM, the posterior distribution is
\begin{equation*}
    \label{POST_LRM}
    \begin{split}
    p(\bm{\Theta} \mid \bm{y}, \bm{X}) 
    &\propto
    \mathsf{N}_n\bigl(\bm{y}\mid\bm{X}\bm{\beta},\sigma^2\bm{I}\bigr)
    \times \mathsf{N}_p(\bm{\beta}\mid\bm{\beta}_0, \bm{\Sigma}_0)
    \times \mathsf{IG}\bigl(\sigma^2\mid\nu_0/ 2, \nu_0 \sigma_0^2 / 2\bigr).
    \end{split}
\end{equation*}
Thus, we develop a Gibbs sampler using the fcd's of each parameter in the model (see Algorithm \ref{alg_mcmc_lrm}). Let $p(\cdot \mid \text{rest})$ denote the distribution of a parameter given all remaining parameters. From the posterior the fcd's are:
{\footnotesize
\begin{itemize}
  \item $\bm{\beta}\mid\text{rest} \sim \mathsf{N}_p(\bm{m},\bm{V})$, with
  \begin{equation}
    \label{LRM_MCMC_Beta}
    \bm{m} = \bm{V} \left( \bm{\Sigma}_0^{-1} \bm{\beta}_0 + \sigma^{-2} \bm{X}^\top \bm{y} \right), \qquad
    \bm{V} = \left( \bm{\Sigma}_0^{-1} + \sigma^{-2} \bm{X}^\top \bm{X} \right)^{-1}.
  \end{equation}

  \item $\sigma^2\mid\text{rest}\sim\textsf{IG}(a,b)$, with:
  \begin{equation}
    \label{LRM_MCMC_sigma2}
    a = \frac{n + \nu_0}{2}, \qquad
    b = \frac{\nu_0 \sigma_0^2 + (\bm{y} - \bm{X} \bm{\beta})^\top (\bm{y} - \bm{X} \bm{\beta})}{2}.
  \end{equation}
\end{itemize}}\hspace{-3pt}

\begin{algorithm}[!htb]
\footnotesize
\caption{MCMC for Linear Regression Model}\label{alg_mcmc_lrm}
\begin{algorithmic}
\State \textbf{Input} Initial values $\bm{\beta}^{(0)},(\sigma^2)^{(0)}$, hyperparameters $\bm{\beta}_0,\bm{\Sigma}_0,\nu_0,\sigma_0^2$, data $\bm{y},\bm{X}$
\For{$b = 1, \dots, B$}
    \State Sample $\bm{\beta}^{(b)} \sim p\bigl(\bm{\beta} \mid (\sigma^2)^{(b-1)},\bm{y},\bm{X}\bigr)$ with Eq.  \eqref{LRM_MCMC_Beta}
    \State Sample $(\sigma^2)^{(b)} \sim p\bigl(\sigma^{2} \mid \bm{\beta}^{(b)},\bm{y},\bm{X}\bigr)$ with Eq.  \eqref{LRM_MCMC_sigma2}
\EndFor
\State \textbf{Output} Posterior draws $\big\{\bm{\beta}^{(b)}, (\sigma^2)^{(b)}\big\}_{b=1}^{B}$
\end{algorithmic}
\end{algorithm}

\subsubsection{VI}

We use VI deriving updates for each parameter in $\bm{\Theta} = (\bm{\beta}, \sigma^2)$ (see Algorithm~\ref{alg_cavi_lrm}). Under the mean--field assumption \eqref{MFA}, the variational distribution factorizes as
$q(\bm{\Theta}) = q(\bm{\beta})\,q(\sigma^2)$, and each factor is optimized in turn while the others are kept fixed, as described below:
{\footnotesize
\begin{itemize}
\item $q(\bm{\beta}) = \mathsf{N}_p\bigl(\bm{\mu}_{\bm{\beta}},\bm{\Sigma}_{\bm{\beta}}\bigr)$, with
\begin{equation}
\label{LRM_CAVI_Mu_Beta_sigma2}
\bm{\mu}_{\bm{\beta}}
=
\bm{\Sigma}_{\bm{\beta}}
\left(\bm{\Sigma}_0^{-1} \bm{\beta}_0 + \mathsf{E}_{\sigma^2}\Bigl[ \tfrac{1}{\sigma^2}\Bigr]\bm{X}^\top \bm{y}\right),\qquad
\bm{\Sigma}_{\bm{\beta}}
=
\left(\bm{\Sigma}_0^{-1} + \mathsf{E}_{\sigma^2}\Bigl[ \tfrac{1}{\sigma^2} \Bigr]\bm{X}^\top \bm{X} \right)^{-1}.
\end{equation}

\item $q(\sigma^2) = \mathsf{IG}\bigl( a, b \bigr)$, with
\begin{equation}
\label{LRM_CAVI_a_b_sigma2}
a = \frac{n + \nu_0}{2}, \qquad
b = \frac{\nu_0 \sigma_0^2
+ \mathsf{E}_{\bm{\beta}}\bigl[(\bm{y} - \bm{X} \bm{\beta})^{\top}
(\bm{y} - \bm{X} \bm{\beta})\bigr]}{2}.
\end{equation}
\end{itemize}}

The ELBO is
{\footnotesize
\begin{equation}
\label{LRM_CAVI_ELBO}
\begin{split}
\mathcal{L}\bigl(q(\bm{\Theta})\bigr)
&=
\mathsf{E}_{\bm{\Theta}}\bigl[\log p(\bm{y} \mid \bm{X}, \bm{\beta}, \sigma^2)\bigr] \\
&\quad\quad+
\left(\mathsf{E}_{\bm{\Theta}}\bigl[\log  p(\bm{\beta})\bigr]
      -\mathsf{E}_{\bm{\beta}}\bigl[\log q(\bm{\beta})\bigr]\right) 
+
\left(\mathsf{E}_{\bm{\Theta}}\bigl[\log p(\sigma^2)\bigr]
      -\mathsf{E}_{\sigma^2}\bigl[\log q(\sigma^2)\bigr]\right),
\end{split}
\end{equation}
}\hspace{-3pt}
with
{\footnotesize
\begin{align*}
\mathsf{E}_{\bm{\Theta}}\bigl[\log p(\bm{y} \mid \bm{X}, \bm{\beta}, \sigma^2)\bigr]
&=
-\frac{n}{2}\log(2\pi)
-\frac{n}{2}\bigl(\log b -\psi(a)\bigr) \\
&\quad\quad
-\frac{1}{2}\left(\frac{a}{b}\right)
\big((\bm{y} - \bm{X}  \bm\mu_{\bm{\beta}})^{\top}
(\bm{y} - \bm{X} \bm\mu_{\bm{\beta}})
+ \operatorname{tr}(\bm{X}^{\top}\bm{X}\bm{\Sigma}_{\bm{\beta}})\big),\\
\mathsf{E}_{\bm{\Theta}}\bigl[\log  p(\bm{\beta})\bigr]
&=
-\frac{p}{2}\log(2\pi)
-\frac{1}{2}\log|\bm\Sigma_{0}| \\
&\quad\quad 
-\frac{1}{2}
\big((\bm\mu_{\bm\beta}-\bm\beta_{0})^{\top}
\bm\Sigma_{0}^{-1}(\bm\mu_{\bm\beta}-\bm\beta_{0})
+ \operatorname{tr}(\bm\Sigma_0^{-1}\bm\Sigma_{\bm{\beta}})\big),\\
\mathsf{E}_{\bm{\beta}}\bigl[\log q(\bm{\beta})\bigr]
&=
-\frac{p}{2}\log(2\pi e)
-\frac{1}{2}\log|\bm{\Sigma}_{\bm{\beta}}|,\\
\mathsf{E}_{\bm{\Theta}}\bigl[\log p(\sigma^2)\bigr]
&=
\frac{\nu_{0}}{2}\log \left(\frac{\nu_0\sigma_{0}^2}{2}\right)
- \log\Gamma\left(\frac{\nu_{0}}{2}\right) \\
&\quad\quad 
- \left(\frac{\nu_{0}}{2}+1\right)\bigl(\log b -\psi(a)\bigr)
- \frac{\nu_0\sigma_{0}^2}{2}\left(\frac{a}{b}\right),\\
\mathsf{E}_{\sigma^2}\bigl[\log q(\sigma^2)\bigr]
&=
- \log b - \log\Gamma(a) + (a+1)\psi(a) - a,
\end{align*}
}\hspace{-3pt}
where $\Gamma(\cdot)$ and $\psi(\cdot)$ denote the gamma and digamma functions, respectively. 

\begin{algorithm}[!htb]
\footnotesize
\caption{VI for Linear Regression Model}\label{alg_cavi_lrm}
\begin{algorithmic}
\State \textbf{Input} Initial values $\bm{\mu}_{\bm{\beta}}^{(0)}, \bm{\Sigma}_{\bm{\beta}}^{(0)}, a^{(0)}, b^{(0)}$, data $\bm{y},\bm{X}$
\While{ELBO has not converged}
\State Update $q(\bm{\beta})$ setting $\bm{\mu}_{\bm{\beta}}$ and $\bm{\Sigma}_{\bm{\beta}}$ with Eq. \eqref{LRM_CAVI_Mu_Beta_sigma2}
\State Update $q(\sigma^2)$ setting $a$ and $b$ with  Eq. \eqref{LRM_CAVI_a_b_sigma2}
\State Compute ELBO $\mathcal{L}(q(\bm{\Theta}))$ with Eq. \eqref{LRM_CAVI_ELBO}
\EndWhile
\State \textbf{Output} Optimal variational parameters $\bm{\mu}_{\bm{\beta}}, \bm{\Sigma}_{\bm{\beta}}, a, b$
\end{algorithmic}
\end{algorithm}

\subsubsection{Illustration: Iris Data}

To illustrate the implementation of MCMC and VI methods, we use Fisher's Iris dataset \citep{rizzo2019statistical}, a standard benchmark in statistics and machine learning. This dataset contains $150$ observations of iris flowers, each with four morphological measurements (sepal length, sepal width, petal length and petal width) and a species label. Here we focus on modeling petal length as a function of sepal length.

\begin{figure}[!htb]
    \centering
    \includegraphics[scale=0.5]{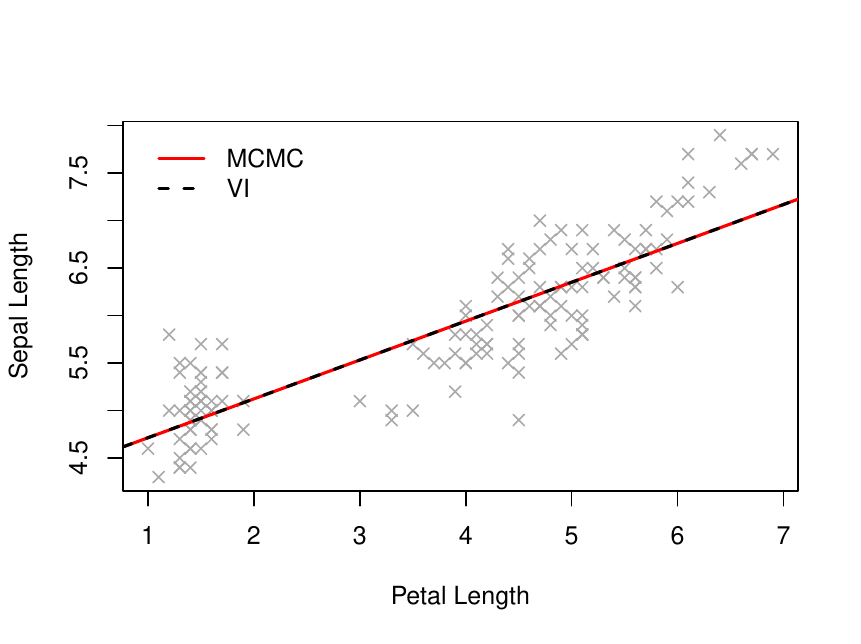}
    \includegraphics[scale=0.5]{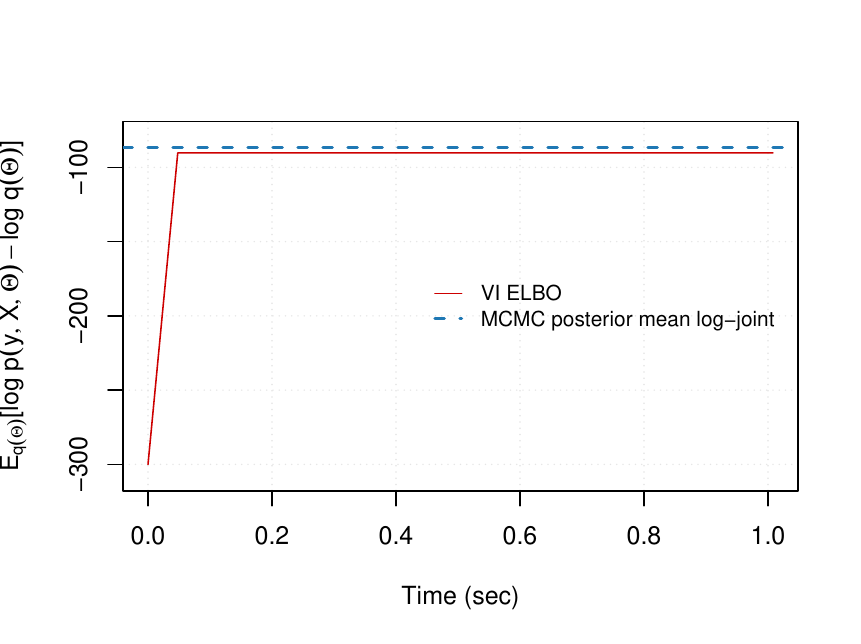}
    \caption{Left: Scatterplot of sepal length against petal length for the Iris dataset, with regression lines estimated using MCMC and VI. Right: Evolution of the ELBO and the posterior mean log–joint probability estimated via MCMC for the LRM using the Iris data.}
    \label{fig:lrm_scatter}
\end{figure}

Left panel of Figure \ref{fig:lrm_scatter} shows that MCMC and VI produce very similar fitted regression lines, and Table \ref{tab:fit_lr} reports almost identical values for WAIC, DIC, $\text{R}^2$ and MSE, indicating that VI approximates the posterior distribution well in this simple regression setting. The main difference is computational cost, since MCMC takes about 21 seconds while VI converges in about 1 second. Right panel of Figure \ref{fig:lrm_scatter} displays the evolution of the ELBO over time, with a solid red curve for the VI optimization path and a dashed blue line for the posterior mean log joint estimated via MCMC. The ELBO under VI quickly reaches a stable plateau that closely matches the MCMC benchmark, highlighting the fast convergence and efficiency of VI compared with MCMC.

\begin{table}[!htb]
\centering
\footnotesize
\caption{Goodness of fit and runtime for the LRM with 1000 posterior samples based, Iris data}
\label{tab:fit_lr}
\begin{tabular}{lrr}
\toprule
Metric & MCMC & VI \\
\midrule
WAIC       & 160.061 & 160.259 \\
DIC        & 160.028 & 160.215 \\
$\text{R}^2$      & 0.760   & 0.760   \\
MSE        & 0.164   & 0.164   \\
Time (sec) & 21.184  & 1.008   \\
\bottomrule
\end{tabular}
\end{table}

\begin{figure}[!htb]
    \centering
    \includegraphics[scale=0.5]{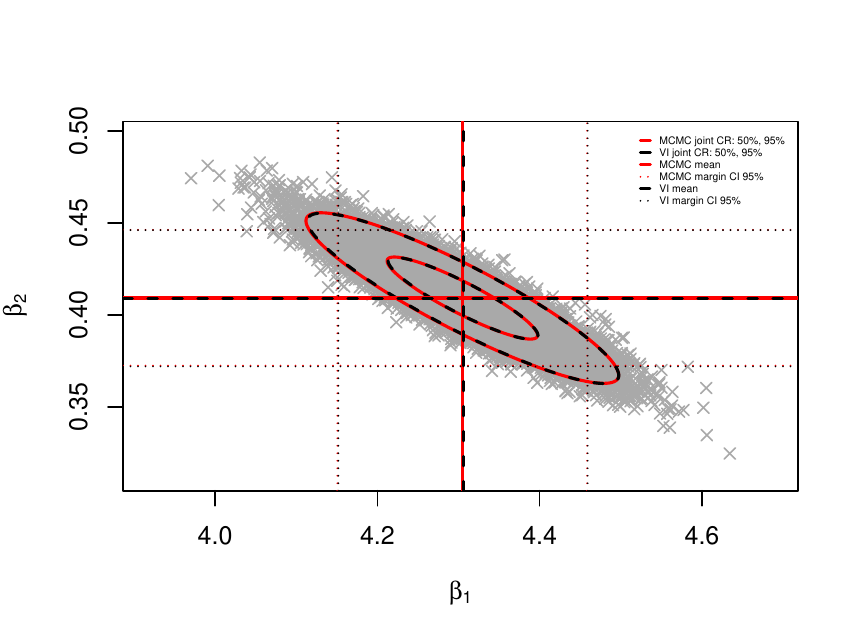}
    \includegraphics[scale=0.5]{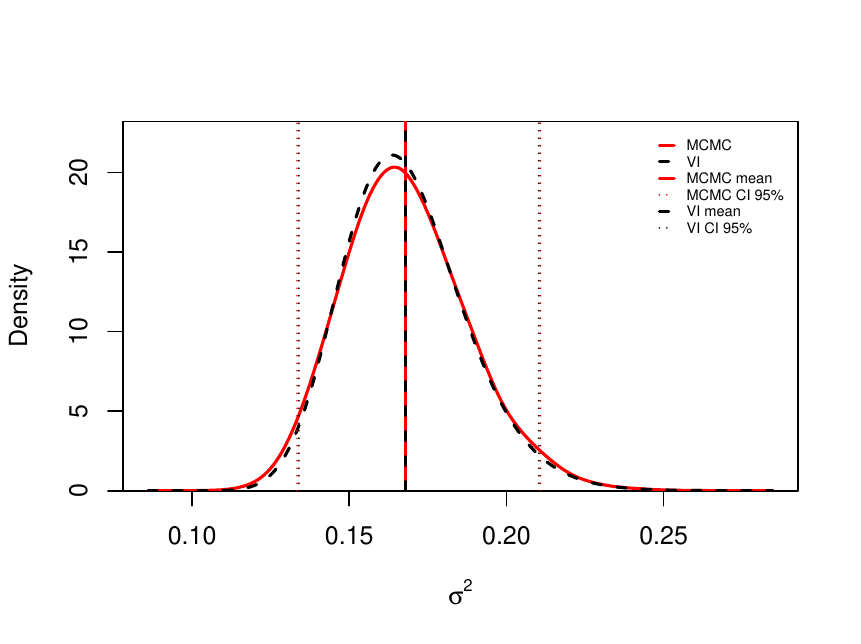}
    \caption{Left: Joint posterior density estimates for $\bm{\beta} = (\beta_1, \beta_2)$ in the LRM using the Iris data, with Mahalanobis credible regions from MCMC and VI and marginal credible intervals. Right: Posterior density estimates for the error variance $\sigma^2$ in the LRM using the Iris data.}
    \label{fig:lrm_betas_joint}
\end{figure}

Left panel of Figure \ref{fig:lrm_betas_joint} shows the joint posterior density of $\bm{\beta} = (\beta_1, \beta_2)$ in the Iris dataset, with gray points for MCMC draws and solid and dashed contours for the $50\%$ and $95\%$ Mahalanobis credible regions under MCMC and VI. The strong overlap of these regions and the almost identical posterior means and marginal credible intervals indicate that VI closely matches the MCMC posterior at much lower computational cost. Right panel of Figure \ref{fig:lrm_betas_joint} shows the posterior density of $\sigma^2$ under both methods, with nearly identical curves, posterior means around $0.168$ and very similar $95\%$ credible intervals, confirming that VI provides an accurate approximation to the MCMC posterior in this example.

\subsubsection{Illustration: Simulation}

Finally, table \ref{tab:method_quality} summarizes the performance of MCMC and VI across six simulation scenarios defined by sample sizes $n \in \{50,100,1000\}$ and numbers of predictors $p \in \{2,3\}$, with true values $\bm{\beta} = (25,10,-30)^\top$ and $\sigma^2 = 100^2$ for $p = 3$, and $\bm{\beta} = (2,-12)^\top$ and $\sigma^2 = 4^2$ for $p = 2$. For each case we report the proportion of parameters whose $95\%$ posterior credible interval includes zero, the coverage of the true parameter values and the runtime. Both methods display very high coverage of the true parameters in all scenarios, and the inclusion of zero in the credible intervals behaves consistently with the magnitude of the true effects. The main difference is computational cost, since MCMC takes between $11$ and $15$ seconds in all settings, whereas VI finishes in well under one second, confirming that VI attains accuracy comparable to MCMC while providing substantial gains in efficiency, especially for larger $n$.

\begin{table}[!htb]
\centering
\footnotesize
\caption{Posterior zero coverage in the LRM under MCMC and VI within the $95\%$ CI, and runtime across simulation scenarios.}
\label{tab:method_quality}
\begin{tabular}{ccccccccc}
\toprule
 & & & \multicolumn{3}{c}{MCMC} & \multicolumn{3}{c}{VI} \\
\cmidrule(lr){4-6}\cmidrule(lr){7-9}
$n$ & $p$ & \# Param. & $0 \in \text{CI}$ & Truth $\in$ CI & Time (s) & $0 \in \text{CI}$ & Truth $\in$ CI & Time (s) \\
\midrule
50   & 3 & 4 & 3/4 & 4/4 & 13.73 & 3/4 & 4/4 & 0.59 \\
100  & 3 & 4 & 3/4 & 4/4 & 13.57 & 3/4 & 4/4 & 0.59 \\
1000 & 3 & 4 & 3/4 & 4/4 & 14.67 & 3/4 & 4/4 & 0.69 \\
50   & 2 & 3 & 1/3 & 3/3 & 11.69 & 1/3 & 2/3 & 0.00 \\
100  & 2 & 3 & 0/3 & 3/3 & 11.80 & 0/3 & 3/3 & 0.00 \\
1000 & 2 & 3 & 0/3 & 3/3 & 12.70 & 0/3 & 3/3 & 0.00 \\
\bottomrule
\end{tabular}
\end{table}

\subsection{Clustering Hierarchical Linear Regression Model}

Hierarchical linear models extend classical linear regression by explicitly representing grouping structures such as persons within cities, students within schools, or repeated measurements on the same individual \citep{GELMAN}. Observations are nested within groups and regression coefficients, including intercepts and slopes, are allowed to vary by group, which yields a more realistic description of individual responses by combining within-group information with systematic differences between groups. As described in \cite{HOFF}, the hierarchical linear model generalizes the normal hierarchical model to compare multiple groups while accounting for both within- and between-group heterogeneity. For groups $j = 1,\ldots,m$ with $n_j$ observations each, the Hierarchical Linear Regression Model (HLRM) is given by
\begin{equation*}
\label{HRLM}
y_{ij} \mid \bm{x}_{ij}, \bm{\beta}_j, \sigma_j^2 \overset{\text{ind}}{\sim}
\mathsf{N}\bigl(\bm{x}_{ij}^{\top} \bm{\beta}_j, \sigma_j^2\bigr),
\quad i = 1,\ldots,n_j,\quad j = 1,\ldots,m,
\end{equation*}
where $\bm{x}_{ij} = (x_{ij1},\dots,x_{ijp})^\top$ is the predictor vector for observation $i$ in group $j$, $\bm{\beta}_j=(\beta_{j1},\dots,\beta_{jp})^\top$ is the group-specific regression coefficient vector, and $\sigma^2_j$ is the group-specific error variance. This model can be written  as $\bm{y}_j \mid \bm{X}_j,\bm{\beta}_j, \sigma_j^2
\overset{\text{ind}}{\sim} \mathsf{N}_{n_j}(\mathbf{X}_j \, \bm{\beta}_j, \sigma_j^2 \, \bm{I})$, where $\bm{y}_j = (y_{1j}, \ldots, y_{n_j j})^\top$ is the response vector in group $j$, and $\mathbf{X}_j = [\bm{x}_{1j}, \ldots, \bm{x}_{n_j j}]^{\top}$ is the $n_j \times p$ design matrix.

The HLRM given above can be extended to uncover latent group structure by modeling the response through a finite Gaussian mixture \citep{SOSA}. The goal is to use the regression framework as a clustering device so that hierarchical units, for example cities or schools, are grouped into latent clusters that share regression coefficients and error variances. Under this setting, the Clustering Linear Regression Model (CLRM) can be specified as follows:
\begin{equation}
\label{CHRLM}
y_{ij} \mid \bm{x}_{ij}, \{\bm{\beta}_k\}, \{\sigma_k^2 \}
\overset{\text{ind}}{\sim}
\sum_{k=1}^{K}\omega_k\,\mathsf{N}\bigl(y_{ij}\mid\bm{x}_{ij}^{\top} \bm{\beta}_k, \sigma_k^2\bigr),
\quad i=1, \ldots, n_j,\quad j=1, \ldots, m,
\end{equation}
where $K$ is a fixed number of clusters, $\bm{\beta}_1,\dots,\bm{\beta}_K$ and $\sigma_1^2,\dots,\sigma_K^2$ are cluster specific regression parameters and variances, and $\omega_1,\dots,\omega_K$ are mixture weights with $\sum_{k=1}^{K}\omega_k=1$ and $0<\omega_k<1$. An equivalent formulation introduces a latent cluster indicator $\gamma_j\in\{1,\ldots,K\}$ such that $\Pr(\gamma_j = k\mid\omega_k)=\omega_k$ for $j=1,\dots,m$ and $k=1,\dots,K$, where $\gamma_j = k$ means that group $j$ belongs to cluster $k$. Conditional on $\gamma_j$, the model \eqref{CHRLM} can be written as
\begin{equation}
\label{CHRLM2}
y_{ij} \mid \bm{x}_{ij}, \gamma_j,\bm{\beta}_{\gamma_j},\sigma_{\gamma_j}^2
\overset{\text{ind}}{\sim}
\mathsf{N}\bigl(y_{ij}\mid\bm{x}_{ij}^{\top} \,\bm{\beta}_{\gamma_j}, \sigma_{\gamma_j}^2\bigr),
\quad i=1, \ldots, n_j,\quad j=1, \ldots, m.
\end{equation}

\subsubsection{Prior Distribution}

\begin{figure}[!htb]
    \centering
    \includegraphics[scale=0.45]{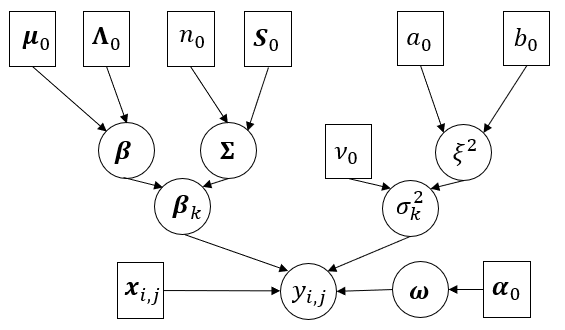}
    \caption{DAG representation of the Clustering Linear Regression Model (CLRM).}
    \label{fig:DAG_CHLRM}
\end{figure}

To adopt a Bayesian formulation for parameter estimation in the CLRM, we follow the prior specification provided in \cite{SOSA} for the parameter vector
$\bm{\Theta} = (\bm{\gamma}, \bm{\omega}, \bm{\zeta}, \bm{\sigma}^2, \bm{\beta}, \bm{\Sigma}, \xi^2)$, where 
$\bm{\gamma} = (\gamma_1,\ldots,\gamma_m)^\top$ denotes the cluster assignments for the $m$ hierarchical units, 
$\bm{\omega} = (\omega_1,\ldots,\omega_K)^\top$ are the mixture weights, 
$\bm{\zeta} = (\bm{\beta}_1,\ldots,\bm{\beta}_K)^\top$ collects the cluster specific regression coefficients, and 
$\bm{\sigma}^2 = (\sigma_1^2,\ldots,\sigma_K^2)^\top$ are the cluster specific variances. 
Conditionally on these parameters, the model in \eqref{CHRLM2} induces the following hierarchical prior structure  
\begin{equation}
  \label{hyper_CHRLM_1}
  \gamma_j \mid \bm{\omega} \sim \mathsf{Cat}(\bm{\omega}), 
  \quad
  \bm{\beta}_k \mid \bm{\beta}, \bm{\Sigma} \overset{\mathsf{iid}}{\sim} \mathsf{N}_p(\bm{\beta}, \bm{\Sigma}), 
  \quad
  \sigma_k^2 \mid \xi^2 \overset{\mathsf{iid}}{\sim} \mathsf{IG}\!\left(\frac{\nu_0}{2}, \frac{\nu_0 \xi^2}{2}\right),
\end{equation}
for $j = 1,\ldots,m$ and $k = 1,\ldots,K$, with
\begin{equation}
  \label{hyper_CHRLM_2}
  \bm{\omega} \sim \mathsf{Dir}(\bm{\alpha}_0), 
  \quad
  \bm{\beta} \sim \mathsf{N}_p(\bm{\mu}_0, \bm{\Lambda}_0), 
  \quad
  \bm{\Sigma} \sim \mathsf{IW}(n_0, \mathbf{S}_0^{-1}), 
  \quad
  \xi^2 \sim \mathsf{G}(a_0, b_0),
\end{equation}
where $\bm{\mu}_0, \bm{\Lambda}_0, n_0, \mathbf{S}_0, \nu_0, a_0, b_0$ and $\bm{\alpha}_0$ are hyperparameters of the model (see Figure~\ref{fig:DAG_CHLRM}).

To maintain consistency with the unit–information prior strategy used in the HLRM \citep{SOSA}, we adopt a weak but coherent prior specification for the CLRM. We set $g = \sum_{j=1}^{m} n_j$ as the total number of observations and fix $\nu_0 = 1$. The global prior mean is taken as $\bm{\mu}_0 = \hat{\bm{\beta}}_{\text{OLS}}$, and the scale matrix as $\bm{\Lambda}_0 = g\,\hat{\sigma}^2_{\text{OLS}}(\bm{X}^\top\bm{X})^{-1}$, where $\hat{\bm{\beta}}_{\text{OLS}}$ and $\hat{\sigma}^2_{\text{OLS}}$ are the OLS estimates obtained by fitting a single global regression to all stacked data. For the covariance matrix $\bm{\Sigma}$ of the cluster-level effects, we choose $n_0 = p + 2$ and $\bm{S}_0 = \bm{\Lambda}_0$, which induces a diffuse prior centered at $\bm{\mu}_0$ and aligned with the global regression structure. Each cluster-specific variance $\sigma_k^2$ receives an Inverse Gamma prior with shape $\nu_0 = 1$ and scale driven by $\xi^2$, while $\xi^2$ itself follows a Gamma prior with $a_0 = 1$ and $b_0 = 1/\hat{\sigma}^2_{\text{OLS}}$, so that $\mathsf{E}(\xi^2) = \hat{\sigma}^2_{\text{OLS}}$ and the prior on $\xi^2$ remains weakly informative. Finally, the mixture weights $\bm{\omega}$ follow a Dirichlet prior $\mathsf{Dir}(\bm{\alpha}_0)$ with $\bm{\alpha}_0 = (1/K,\ldots,1/K)^\top$, yielding a diffuse specification that does not favor any particular cluster a priori.

\subsubsection{MCMC}

Under the CHLRM, the posterior distribution is
\begin{equation*}\label{POST_CHLRM}
\begin{split}
p(\bm{\Theta} \mid \bm{y}, \bm{X}) 
&\propto
\prod_{j=1}^{m}\prod_{i=1}^{n_j}
\mathsf{N}\bigl(y_{ij}\mid \bm{x}_{ij}^{\top}\bm{\beta}_{\gamma_j},\sigma_{\gamma_j}^2\bigr)
\times
\prod_{j=1}^{m}\mathsf{Cat}(\gamma_j\mid\bm{\omega})
\times
\mathsf{Dir}(\bm{\omega}\mid\bm{\alpha}_0)\\
&\quad\quad\times
\prod_{k=1}^{K}\mathsf{N}_p(\bm{\beta}_k\mid\bm{\beta}, \bm{\Sigma})
\times
\mathsf{N}_p\bigl(\bm{\beta}\mid\bm{\mu}_0, \bm{\Lambda}_0\bigr)
\times
\mathsf{IW}\bigl(\bm{\Sigma}\mid n_0, \mathbf{S}_0^{-1}\bigr)\\
&\quad\quad\quad\times
\prod_{k=1}^{K}\mathsf{IG}\bigl(\sigma_k^2 \mid\nu_0 / 2, \nu_0 \xi^2 / 2\bigr)
\times
\mathsf{G}\bigl(\xi^2 \mid a_0, b_0\bigr).
\end{split}
\end{equation*}
Thus, we develop a Gibbs sampler using the fcd's of each parameter in the model (see Algorithm~\ref{alg_mcmc_chlrm}). From the posterior the fcd's are:
{\footnotesize
\begin{itemize}
  \item $\gamma_j \mid \text{rest} \sim \mathsf{Cat}\bigl(\pi_{j1},\dots,\pi_{jK}\bigr)$ for $j=1,\dots,m$, with
  \begin{equation}
    \label{CHLRM_MCMC_gammaj}
    \pi_{jk}
    \propto
      \omega_k \,
      \prod_{i=1}^{n_j}
      (\sigma_k^2)^{-1/2}
      \exp\left\{
        -\dfrac{1}{2\sigma_k^2}
        \bigl(y_{ij} - \bm{x}_{ij}^{\top} \bm{\beta}_k\bigr)^2
      \right\}, \quad k=1,\dots,K.
  \end{equation}

  \item $\bm{\omega} \mid \text{rest} \sim \mathsf{Dir}(\alpha_{01} + n_1,\dots,\alpha_{0K} + n_K)$, with
  \begin{equation}
    \label{CHLRM_MCMC_Omega}
    n_k = \#\{ j : \gamma_j = k\}, \quad k=1,\dots,K.
  \end{equation}

  \item $\bm{\beta}_k \mid \text{rest} \sim \mathsf{N}_p(\bm{m}_k,\bm{V}_k)$ for $k=1,\dots,K$, with
  \begin{equation}
    \label{CHLRM_MCMC_Betak}
    \bm{m}_k
    =
    \bm{V}_k\left(
      \bm{\Sigma}^{-1}\bm{\beta}
      + \sigma_k^{-2}\bm{X}_{k}^{\top}\bm{y}_{k}
    \right),
    \qquad
    \bm{V}_k
    =
    \left(
      \bm{\Sigma}^{-1}
      + \sigma_k^{-2}\bm{X}_{k}^{\top}\bm{X}_{k}
    \right)^{-1},
  \end{equation}
  where
  $\bm{X}_{k} = [\bm{X}_j^{\top} : \gamma_j = k]^{\top}$ and
  $\bm{y}_{k} = [\bm{y}_j^{\top} : \gamma_j = k]^{\top}$.
  If cluster $k$ is empty $(n_k = 0)$, then
  $\bm{\beta}_k \mid \text{rest} \sim \mathsf{N}_p(\bm{\beta}, \bm{\Sigma})$.

  \item $\sigma_k^2 \mid \text{rest} \sim \mathsf{IG}(a_k,b_k)$ for $k=1,\dots,K$, with
  \begin{equation}
    \label{CHLRM_MCMC_sigma2k}
    a_k
    =
    \frac{\nu_0 + \sum_{j:\gamma_j = k} n_j}{2},
    \qquad
    b_k
    =
    \frac{
      \nu_0 \xi^2
      + (\bm{y}_{k} - \bm{X}_{k} \bm{\beta}_k)^{\top}
        (\bm{y}_{k} - \bm{X}_{k} \bm{\beta}_k)
    }{2}.
  \end{equation}
  If cluster $k$ is empty $(n_k = 0)$, then
  $\sigma_k^2 \mid \text{rest} \sim \mathsf{IG}\bigl(\nu_0/2,\nu_0\xi^2/2\bigr)$.

  \item $\bm{\beta} \mid \text{rest} \sim \mathsf{N}_p(\bm{m}_{\bm{\beta}},\bm{V}_{\bm{\beta}})$, with
  \begin{equation}
    \label{CHLRM_MCMC_Beta}
    \bm{m}_{\bm{\beta}}
    =
    \bm{V}_{\bm{\beta}}
    \left(
      \bm{\Lambda}_0^{-1} \bm{\mu}_0
      + \bm{\Sigma}^{-1}
        \sum_{k : n_k > 0} \bm{\beta}_k
    \right),
    \qquad
    \bm{V}_{\bm{\beta}}
    =
    \left(
      \bm{\Lambda}_0^{-1} + \kappa \bm{\Sigma}^{-1}
    \right)^{-1}.
  \end{equation}

  \item $\bm{\Sigma} \mid \text{rest} \sim \mathsf{IW}(n_{\Sigma},\bm{S}_{\Sigma}^{-1})$, with
  \begin{equation}
    \label{CHLRM_MCMC_Sigma}
    n_{\Sigma}
    =
    n_0 + \kappa,
    \qquad
    \bm{S}_{\Sigma}
    =
    \bm{S}_0
    + \sum_{k : n_k > 0}
      (\bm{\beta}_k - \bm{\beta})(\bm{\beta}_k - \bm{\beta})^{\top}.
  \end{equation}

   \item $\xi^2 \mid \text{rest} \sim \mathsf{G}(a_{\xi},b_{\xi})$, with
  \begin{equation}
    \label{CHLRM_MCMC_xi2}
    a_{\xi}
    =
    a_0 + \frac{\kappa \, \nu_0}{2},
    \qquad
    b_{\xi}
    =
    b_0 + \frac{\nu_0}{2}
    \sum_{k : n_k > 0} \sigma_k^{-2},
  \end{equation}
  where $\kappa = \#\{ k : n_k > 0\}$ is the number of nonempty clusters.
\end{itemize}}\hspace{-3pt}

\begin{algorithm}[!htb]
\footnotesize
\caption{MCMC for Clustering Hierarchical Linear Regression Model}
\label{alg_mcmc_chlrm}
\begin{algorithmic}
\State \textbf{Input} Initial values 
$\{\gamma_j^{(0)}\}_{j=1}^m,\ \bm{\omega}^{(0)},\ \{\bm{\beta}_k^{(0)}\}_{k=1}^K,\{(\sigma_k^2)^{(0)}\}_{k=1}^K,\ 
 \bm{\beta}^{(0)},\ \bm{\Sigma}^{(0)},\ (\xi^2)^{(0)}$, 
hyperparameters $\bm{\mu}_0$, $\bm{\Lambda}_0$, $n_0$, $\bm{S}_0$, $a_0$, $b_0$, $\nu_0$, 
data $\{\bm{y}_j,\bm{X}_j\}_{j=1}^m$
\For{$b = 1, \dots, B$}
  \For{$j = 1, \dots, m$}
    \State Sample $\gamma_j^{(b)} \sim 
    p\bigl(\gamma_j \mid \bm{\omega}^{(b-1)}, \{\bm{\beta}_k^{(b-1)}, (\sigma_k^2)^{(b-1)}\}, \bm{y}_j, \bm{X}_j\bigr)$
    with Eq.~\eqref{CHLRM_MCMC_gammaj}
  \EndFor
  \State Sample $\bm{\omega}^{(b)} \sim 
  p\bigl(\bm{\omega} \mid \{\gamma_j^{(b)}\}\bigr)$
  with Eq.~\eqref{CHLRM_MCMC_Omega}
  \For{$k = 1, \dots, K$}
    \State Sample $\bm{\beta}_k^{(b)} \sim 
    p\bigl(\bm{\beta}_k \mid \bm{\beta}^{(b-1)}, \bm{\Sigma}^{(b-1)}, (\sigma_k^2)^{(b-1)}, \{\bm{y}_j,\bm{X}_j,\gamma_j^{(b)}\}\bigr)$
    with Eq.~\eqref{CHLRM_MCMC_Betak}
    \State Sample $(\sigma_k^2)^{(b)} \sim 
    p\bigl(\sigma_k^2 \mid \bm{\beta}_k^{(b)}, (\xi^2)^{(b-1)}, \{\bm{y}_j,\bm{X}_j,\gamma_j^{(b)}\}\bigr)$
    with Eq.~\eqref{CHLRM_MCMC_sigma2k}
  \EndFor
  \State Sample $\bm{\beta}^{(b)} \sim 
  p\bigl(\bm{\beta} \mid \{\bm{\beta}_k^{(b)}\},\bm{\Sigma}^{(b-1)}\bigr)$
  with Eq.~\eqref{CHLRM_CAVI_Mu_Sigma_Beta}
  \State Sample $\bm{\Sigma}^{(b)} \sim 
  p\bigl(\bm{\Sigma} \mid \{\bm{\beta}_k^{(b)}\},\bm{\beta}^{(b)}\bigr)$
  with Eq.~\eqref{CHLRM_MCMC_Sigma}
  \State Sample $(\xi^2)^{(b)} \sim 
  p\bigl(\xi^2 \mid \{(\sigma_k^2)^{(b)}\}\bigr)$
  with Eq.~\eqref{CHLRM_MCMC_xi2}
\EndFor
\State \textbf{Output} Posterior draws 
$\big\{\gamma_j^{(b)},\bm{\omega}^{(b)},\bm{\beta}_k^{(b)},(\sigma_k^2)^{(b)},
\bm{\beta}^{(b)},\bm{\Sigma}^{(b)},(\xi^2)^{(b)}\big\}_{b=1}^B$
\end{algorithmic}
\end{algorithm}

\subsubsection{VI}

As before, we approximate the posterior distribution with a variational distribution under the
mean--field assumption \eqref{MFA} (see Algorithm \ref{alg_cavi_chlrm}). In this case yields
\begin{equation*}
\label{Q_CHLRM}
\begin{split}
q(\bm{\Theta})
&=
q(\bm{\omega})\,q(\bm{\beta})\,q(\bm{\Sigma})\,q(\xi^2)
\left(\prod_{k=1}^{K} q(\bm{\beta}_k)\,q(\sigma_k^2)\right)
\left(\prod_{j=1}^{m} q(\gamma_j)\right).
\end{split}
\end{equation*}
Using the CAVI result for exponential family models in Eq. \eqref{NMVFB_opt}, we obtain
closed-form updates for the variational parameters of each factor in the CLRM. These parameters naturally
decompose into local parameters, specific to each group $j$, and global parameters, associated
with latent variables shared across all groups, as shown below:
{\footnotesize
\begin{itemize}
  \item $q(\gamma_j)=\mathsf{Cat}\bigl(\rho_{j1},\dots,\rho_{jK}\bigr)$ for $j=1,\dots,m$, with
  \begin{equation}
    \label{CHLRM_CAVI_rhojk_gamma}
    \begin{split}
    \log \rho_{jk}
    &\propto
    -\frac{n_j}{2}\,\mathsf{E}_{\sigma_k^2}\bigl[\log\sigma_k^{2}\bigr]
    +\mathsf{E}_{\bm{\omega}}\bigl[\log\omega_k\bigr] \\
    &\quad\quad -\frac{1}{2}\,\mathsf{E}_{\sigma_k^2}\bigl[\sigma_k^{-2}\bigr]\,
      \mathsf{E}_{\bm{\beta}_k}
      \Bigl[
        (\bm{y}_{j}-\bm{X}_{j}\bm{\beta}_{k})^\top
        (\bm{y}_{j}-\bm{X}_{j}\bm{\beta}_{k})
      \Bigr]
    ,\qquad k=1,\dots,K.
     \end{split}
  \end{equation}
  
  \item $q(\bm{\omega})=\mathsf{Dir}\bigl(\alpha_{\bm{\omega}1},\dots,\alpha_{\bm{\omega}K}\bigr)$, with
  \begin{equation}
    \label{CHLRM_CAVI_Alpha_Omega}
    \alpha_{\bm{\omega}k}
    =
    \alpha_{0k}
    +
    \sum_{j=1}^{m}
    \mathsf{E}_{\gamma_j}\bigl[\,[\gamma_j = k]\,\bigr],
    \qquad k=1,\dots,K.
  \end{equation}
  where $[\cdot]$ denotes the Iverson bracket, so that $[\gamma_j = k] = 1$ if $\gamma_j = k$ and $[\gamma_j = k] = 0$ otherwise.

  \item $q(\bm{\beta}_k)=\mathsf{N}_p\bigl(\bm{\mu}_{\bm{\beta}_k},\bm{\Sigma}_{\bm{\beta}_k}\bigr)$ for $k=1,\dots,K$, with
  \begin{equation}
    \label{CHLRM_CAVI_Mu_Sigma_Betak}
    \begin{split}
    \bm{\mu}_{\bm{\beta}_k}
    &=
    \bm{\Sigma}_{\bm{\beta}_k}
    \left(
      \mathsf{E}_{\bm{\Sigma}}\bigl[\bm{\Sigma}^{-1}\bigr]\,
      \mathsf{E}_{\bm{\beta}}\bigl[\bm{\beta}\bigr]
      +
      \mathsf{E}_{\sigma_k^2}\bigl[\sigma_k^{-2}\bigr]\,
      \sum_{j=1}^{m}
        \mathsf{E}_{\gamma_j}\bigl[\,[\gamma_j = k]\,\bigr]\,
        \bm{X}_{j}^\top \bm{y}_{j}
    \right), \\
    \bm{\Sigma}_{\bm{\beta}_k}
    &=
    \left(
    \mathsf{E}_{\bm{\Sigma}}\bigl[\bm{\Sigma}^{-1}\bigr]
    +
    \mathsf{E}_{\sigma_k^2}\bigl[\sigma_k^{-2}\bigr]\,
    \sum_{j=1}^{m}
      \mathsf{E}_{\gamma_j}\bigl[\,[\gamma_j = k]\,\bigr]\,
      \bm{X}_{j}^\top \bm{X}_{j}
      \right)^{-1}.
    \end{split}
  \end{equation}

  \item $q(\sigma_k^2)=\mathsf{IG}\bigl(a_{\sigma_k^2},b_{\sigma_k^2}\bigr)$ for $k=1,\dots,K$, with
  \begin{equation}
    \label{CHLRM_CAVI_a_b_sigma2k}
    \begin{split}
    a_{\sigma_k^2}
    &=
    \frac{
      \nu_0
      +
      \sum_{j=1}^{m}
        \mathsf{E}_{\gamma_j}\bigl[\,[\gamma_j = k]\,\bigr]\,
        n_j
    }{2},\\
    b_{\sigma_k^2}
    &=
    \frac{
      \nu_0\,\mathsf{E}_{\xi^2}\bigl[\xi^2\bigr]
      +
      \sum_{j=1}^{m}
        \mathsf{E}_{\gamma_j}\bigl[\,[\gamma_j = k]\,\bigr]\,
        \mathsf{E}_{\bm{\beta}_k}
        \Bigl[
          (\bm{y}_{j} - \bm{X}_{j} \bm{\beta}_k)^\top
          (\bm{y}_{j} - \bm{X}_{j} \bm{\beta}_k)
        \Bigr]
    }{2}.
    \end{split}
  \end{equation}

  \item $q(\bm{\beta})=\mathsf{N}_p\bigl(\bm{\mu}_{\bm{\beta}},\bm{\Sigma}_{\bm{\beta}}\bigr)$, with
  \begin{equation}
    \label{CHLRM_CAVI_Mu_Sigma_Beta}
    \begin{split}
    \bm{\mu}_{\bm{\beta}}
    &=
    \bm{\Sigma}_{\bm{\beta}}
    \left(
      \bm{\Lambda}_0^{-1} \bm{\mu}_0
      +
      \mathsf{E}_{\bm{\Sigma}}\bigl[\bm{\Sigma}^{-1}\bigr]
      \sum_{k=1}^{K}
        \mathsf{E}_{\bm{\beta}_k}\bigl[\bm{\beta}_k\bigr]
    \right),\\
    \bm{\Sigma}_{\bm{\beta}}
    &=
    \left(
      \bm{\Lambda}_0^{-1}
      +
      \sum_{k=1}^{K}
        \mathsf{E}_{\bm{\Sigma}}\bigl[\bm{\Sigma}^{-1}\bigr]
    \right)^{-1}.
    \end{split}
  \end{equation}

  \item $q(\bm{\Sigma})=\mathsf{IW}\bigl(\nu_{\bm{\Sigma}},\bm{S}_{\bm{\Sigma}}^{-1}\bigr)$, with
  \begin{equation}
    \label{CHLRM_CAVI_nu_S_Sigma}
    \nu_{\bm{\Sigma}} = n_0 + K,
    \qquad
    \bm{S}_{\bm{\Sigma}}
    =
    \bm{S}_0
    +
    \sum_{k=1}^{K}
    \mathsf{E}_{\bm{\beta}_k,\bm{\beta}}
    \Bigl[
      \bigl(\bm{\beta}_k - \bm{\beta}\bigr)
      \bigl(\bm{\beta}_k - \bm{\beta}\bigr)^{\top}
    \Bigr].
  \end{equation}

  \item $q(\xi^2)=\mathsf{G}\bigl(a_{\xi^2},b_{\xi^2}\bigr)$, with
  \begin{equation}
    \label{CHLRM_CAVI_a_b_xi2}
    a_{\xi^2}
    =
    a_0 + \sum_{k=1}^{K} \frac{\nu_0}{2},
    \qquad
    b_{\xi^2}
    =
    b_0
    +
    \frac{\nu_0}{2}
    \sum_{k=1}^{K}
      \mathsf{E}_{\sigma_k^2}\bigl[\sigma_k^{-2}\bigr].
  \end{equation}
\end{itemize}}\hspace{-3pt}

The ELBO is
{\footnotesize
\begin{equation}
\label{CHLRM_CAVI_ELBO}
\begin{split}
\mathcal{L}\bigl(q(\bm{\Theta})\bigr)
&=
\mathsf{E}_{\bm{\Theta}}\bigl[\log p(\bm{y}\mid\bm{X},\bm{\gamma},\bm{\zeta},\bm{\sigma}^2)\bigr]
+
\sum_{j=1}^{m}
\Bigl(
\mathsf{E}_{\bm{\Theta}}\bigl[\log p(\gamma_j\mid\bm{\omega})\bigr]
-
\mathsf{E}_{\gamma_j}\bigl[\log q(\gamma_j)\bigr]
\Bigr)
\\
&\quad\quad
+
\Bigl(
\mathsf{E}_{\bm{\Theta}}\bigl[\log p(\bm{\omega})\bigr]
-
\mathsf{E}_{\bm{\omega}}\bigl[\log q(\bm{\omega})\bigr]
\Bigr)
+
\sum_{k=1}^{K}
\Bigl(
\mathsf{E}_{\bm{\Theta}}\bigl[\log p(\bm{\beta}_k\mid\bm{\beta},\bm{\Sigma})\bigr]
-
\mathsf{E}_{\bm{\beta}_k}\bigl[\log q(\bm{\beta}_k)\bigr]
\Bigr)
\\
&\quad\quad\quad
+
\Bigl(
\mathsf{E}_{\bm{\Theta}}\bigl[\log p(\bm{\beta})\bigr]
-
\mathsf{E}_{\bm{\beta}}\bigl[\log q(\bm{\beta})\bigr]
\Bigr)
+
\Bigl(
\mathsf{E}_{\bm{\Theta}}\bigl[\log p(\bm{\Sigma})\bigr]
-
\mathsf{E}_{\bm{\Sigma}}\bigl[\log q(\bm{\Sigma})\bigr]
\Bigr)
\\
&\quad\quad\quad\quad
+
\sum_{k=1}^{K}
\Bigl(
\mathsf{E}_{\bm{\Theta}}\bigl[\log p(\sigma_k^2 \mid \xi^2)\bigr]
-
\mathsf{E}_{\sigma_k^2}\bigl[\log q(\sigma_k^2)\bigr]
\Bigr)
+
\Bigl(
\mathsf{E}_{\bm{\Theta}}\bigl[\log p(\xi^2)\bigr]
-
\mathsf{E}_{\xi^2}\bigl[\log q(\xi^2)\bigr]
\Bigr).
\end{split}
\end{equation}
}\hspace{-3pt}
with
{\footnotesize
\begin{align*}
\mathsf{E}_{\bm{\Theta}}\bigl[\log p(\bm{y}\mid\bm{X},\bm{\gamma},\bm{\zeta},\bm{\sigma}^2)\bigr]
&=
\sum_{j=1}^{m}\sum_{k=1}^{K}
\rho_{jk}
\Bigg(
-\frac{n_j}{2}\log(2\pi)
-\frac{n_j}{2}\bigl(\log b_{\sigma_k^2}-\psi(a_{\sigma_k^2})\bigr) \\
&\quad\quad
-\frac{1}{2}\left(\frac{a_{\sigma_k^2}}{b_{\sigma_k^2}}\right)
\Big[
(\bm{y}_j-\mathbf{X}_j\bm{\mu}_{\bm{\beta}_k})^\top(\bm{y}_j-\mathbf{X}_j\bm{\mu}_{\bm{\beta}_k})
+\operatorname{tr}\bigl(\mathbf{X}_j^\top\mathbf{X}_j\,\bm{\Sigma}_{\bm{\beta}_k}\bigr)
\Big]
\Bigg),\\[6pt]
\mathsf{E}_{\bm{\Theta}}\bigl[\log p(\gamma_j\mid\bm{\omega})\bigr]
&=
\sum_{k=1}^{K}\rho_{jk}
\Bigg[
\psi\Bigl(\alpha_{0k}+\sum_{j=1}^{m}\rho_{jk}\Bigr)
-
\psi\Bigl(\sum_{k=1}^{K}\bigl[\alpha_{0k}+\sum_{j=1}^{m}\rho_{jk}\bigr]\Bigr)
\Bigg],\\[6pt]
\mathsf{E}_{\gamma_j}\bigl[\log q(\gamma_j)\bigr]
&=
\sum_{k=1}^{K}\rho_{jk}\log\rho_{jk},\\[6pt]
\mathsf{E}_{\bm{\Theta}}\bigl[\log p(\bm{\omega})\bigr]
&=
\log\Gamma\Bigl( \sum_{k=1}^{K} \alpha_{0k} \Bigr)
-\sum_{k=1}^{K} \log\Gamma\bigl(\alpha_{0k}\bigr)
+\sum_{k=1}^{K}\bigl(\alpha_{0k}-1\bigr)
\Big[
\psi\bigl(\alpha_{0k}\bigr)
-\psi\Bigl(\sum_{k=1}^{K}\alpha_{0k}\Bigr)
\Big],\\[6pt]
\mathsf{E}_{\bm{\omega}}\bigl[\log q(\bm{\omega})\bigr]
&=
\log\Gamma\Bigl( \sum_{k=1}^{K} \bigl[\alpha_{0k}+\sum_{j=1}^{m}\rho_{jk} \bigr]\Bigr)
-\sum_{k=1}^{K} \log\Gamma\Bigl(\alpha_{0k}+\sum_{j=1}^{m}\rho_{jk}\Bigr)\\
&\quad\quad
+\sum_{k=1}^{K}\Bigl(\alpha_{0k}+\sum_{j=1}^{m}\rho_{jk} -1\Bigr)
\Big[
\psi\Bigl(\alpha_{0k}+\sum_{j=1}^{m}\rho_{jk}\Bigr)
-\psi\Bigl(\sum_{k=1}^{K}\bigl[\alpha_{0k}+\sum_{j=1}^{m}\rho_{jk} \bigr]\Bigr)
\Big],
\end{align*}
}\hspace{-3pt}

{\footnotesize
\begin{align*}
\mathsf{E}_{\bm{\Theta}}\bigl[\log p(\bm{\beta}_k\mid\bm{\beta},\bm{\Sigma})\bigr]
&=
-\frac{p}{2}\log(2\pi)
-\frac{1}{2}\Bigg(
-\sum_{i=1}^{p}\psi\Bigl(\frac{\nu_{\bm{\Sigma}}+1-i}{2}\Bigr)
-p\log 2+\log\bigl|\bm{S}_{\bm{\Sigma}}^{-1}\bigr|
\Bigg) \\
&\quad\quad
-\frac{1}{2}\operatorname{tr}\Bigl(\bm{S}_{\bm{\Sigma}}^{-1}\big[
(\bm{\mu}_{\bm{\beta}_k}-\bm{\mu}_{\bm{\beta}})
(\bm{\mu}_{\bm{\beta}_k}-\bm{\mu}_{\bm{\beta}})^\top
+\bm{\Sigma}_{\bm{\beta}_k}+\bm{\Sigma}_{\bm{\beta}}
\big]\Bigr),\\[6pt]
\mathsf{E}_{\bm{\beta}_k}\bigl[\log q(\bm{\beta}_k)\bigr]
&=
-\frac{p}{2}\log(2\pi e)
-\frac{1}{2}\log\bigl|\bm{\Sigma}_{\bm{\beta}_k}\bigr|,\\[6pt]
\mathsf{E}_{\bm{\Theta}}\bigl[\log p(\bm{\beta})\bigr]
&=
-\frac{p}{2}\log(2\pi)
-\frac{1}{2}\log\bigl|\bm{\Lambda}_0\bigr|
-\frac{1}{2}\Big[
(\bm{\mu}_{\bm{\beta}}-\bm{\mu}_0)^\top\bm{\Lambda}_0^{-1}(\bm{\mu}_{\bm{\beta}}-\bm{\mu}_0)
+\operatorname{tr}\bigl(\bm{\Lambda}_0^{-1}\bm{\Sigma}_{\bm{\beta}}\bigr)
\Big],\\[6pt]
\mathsf{E}_{\bm{\beta}}\bigl[\log q(\bm{\beta})\bigr]
&=
-\frac{p}{2}\log(2\pi e)
-\frac{1}{2}\log\bigl|\bm{\Sigma}_{\bm{\beta}}\bigr|,\\[6pt]
\mathsf{E}_{\bm{\Theta}}\bigl[\log p(\bm{\Sigma})\bigr]
&=
\frac{n_0}{2}\log\bigl|\bm{S}_0\bigr|
-\frac{n_0 p}{2}\log 2
-\log\Gamma_p\Bigl(\frac{n_0}{2}\Bigr) \\
&\quad\quad
-\frac{n_0+p+1}{2}\Bigg(
-\sum_{i=1}^{p}\psi\Bigl(\frac{\nu_{\bm{\Sigma}}+1-i}{2}\Bigr)
-p\log 2+\log\bigl|\bm{S}_{\bm{\Sigma}}^{-1}\bigr|
\Bigg)
-\frac{1}{2}\operatorname{tr}\bigl(\bm{S}_0\,\bm{S}_{\bm{\Sigma}}^{-1}\bigr),\\[6pt]
\mathsf{E}_{\bm{\Sigma}}\bigl[\log q(\bm{\Sigma})\bigr]
&=
\frac{\nu_{\bm{\Sigma}}}{2}\log\bigl|\mathbf{S}_{\bm{\Sigma}}\bigr|
- \frac{\nu_{\bm{\Sigma}} p}{2}\log 2
- \log\Gamma_p\Bigl(\frac{\nu_{\bm{\Sigma}}}{2}\Bigr) \\
&\quad\quad
- \frac{\nu_{\bm{\Sigma}}+p+1}{2}\Big(
-\sum_{i=1}^{p}\psi\Bigl(\frac{\nu_{\bm{\Sigma}}+1-i}{2}\Bigr)
-p\log 2+\log\bigl|\mathbf{S}_{\bm{\Sigma}}^{-1}\bigr|
\Big)
- \frac{\nu_{\bm{\Sigma}}}{2}\operatorname{tr}\bigl(\mathbf{S}_{\bm{\Sigma}}^2\bigr),\\[6pt]
\mathsf{E}_{\bm{\Theta}}\bigl[\log p(\sigma_k^2 \mid \xi^2)\bigr]
&=
\frac{\nu_0}{2}\log\Bigl(\frac{\nu_0}{2}\Bigr)
+\frac{\nu_0}{2}\bigl(\psi(a_{\xi^2})-\log b_{\xi^2}\bigr)
-\log\Gamma\Bigl(\frac{\nu_0}{2}\Bigr)\\
&\quad\quad
-\Bigl(\frac{\nu_0}{2}+1\Bigr)\bigl(\log b_{\sigma_k^2}-\psi(a_{\sigma_k^2})\bigr)
-\frac{\nu_0\sigma_0^2}{2}
\Bigl(\frac{a_{\xi^2}}{b_{\xi^2}}\Bigr)
\Bigl(\frac{a_{\sigma_k^2}}{b_{\sigma_k^2}}\Bigr),\\[6pt]
\mathsf{E}_{\sigma_k^2}\bigl[\log q(\sigma_k^2)\bigr]
&=
-\log b_{\sigma_k^2}
-\log\Gamma(a_{\sigma_k^2})
+(a_{\sigma_k^2}+1)\psi(a_{\sigma_k^2})
-a_{\sigma_k^2},\\[6pt]
\mathsf{E}_{\bm{\Theta}}\bigl[\log p(\xi^2)\bigr]
&=
a_0\log b_0
-\log\Gamma(a_0)
+(a_0-1)\bigl(\psi(a_{\xi^2})-\log b_{\xi^2}\bigr)
-b_0\Bigl(\frac{a_{\xi^2}}{b_{\xi^2}}\Bigr),\\[6pt]
\mathsf{E}_{\xi^2}\bigl[\log q(\xi^2)\bigr]
&=
\log b_{\xi^2}
-\log\Gamma(a_{\xi^2})
+(a_{\xi^2}-1)\psi(a_{\xi^2})
-a_{\xi^2}.
\end{align*}}\hspace{-3pt}
where $\Gamma(\cdot)$, $\Gamma_p(\cdot)$, and $\psi(\cdot)$ denote the gamma, multivariate gamma and digamma functions, respectively.

\begin{algorithm}[!htb]
\footnotesize
\caption{VI for Clustering Hierarchical Linear Regression Model}\label{alg_cavi_chlrm}
\label{CAVI_CHLRM_alg}
\begin{algorithmic}
\State \textbf{Input} Initial values 
$\{\bm{\mu}_{\bm{\beta}_k}^{(0)}, \bm{\Sigma}_{\bm{\beta}_k}^{(0)}\}_{k=1}^K,
 \bm{\mu}_{\bm{\beta}}^{(0)}, \bm{\Sigma}_{\bm{\beta}}^{(0)},
 \nu_{\bm{\Sigma}}^{(0)}, \bm{S}_{\bm{\Sigma}}^{(0)},
 a_{\xi^2}^{(0)}, b_{\xi^2}^{(0)},
 \{a_{\sigma_k^2}^{(0)}, b_{\sigma_k^2}^{(0)}\}_{k=1}^K,
 \{\rho_{jk}^{(0)}\}_{j=1}^m,
 \bm{\alpha}^{(0)}$, 
hyperparameters $\bm{\mu}_0$, $\bm{\Lambda}_0$, $n_0$, $\bm{S}_0$, $a_0$, $b_0$, $\nu_0$, 
and data $\{\bm{y}_j,\bm{X}_j\}_{j=1}^m$
\While{ELBO has not converged}
  \For{$j = 1,\dots,m$}
    \State \textbf{Local updates}
    \State Update $q(\gamma_j)$ setting 
    $\{\rho_{jk}\}_{k=1}^K$ with 
    Eq.~\eqref{CHLRM_CAVI_rhojk_gamma}
  \EndFor
  \State \textbf{Global updates}
  \State Update $q(\bm{\omega})$ setting 
  $\bm{\alpha}_{\bm{\omega}}$ with 
  Eq.~\eqref{CHLRM_CAVI_Alpha_Omega}
  \For{$k = 1,\dots,K$}
    \State Update $q(\bm{\beta}_k)$ setting 
    $\bm{\mu}_{\bm{\beta}_k}$ and $\bm{\Sigma}_{\bm{\beta}_k}$ with 
    Eqs.~\eqref{CHLRM_CAVI_Mu_Sigma_Betak}
    \State Update $q(\sigma_k^2)$ setting 
    $a_{\sigma_k^2}$ and $b_{\sigma_k^2}$ with 
    Eqs.~\eqref{CHLRM_CAVI_a_b_sigma2k}
  \EndFor
  \State Update $q(\bm{\beta})$ setting 
  $\bm{\mu}_{\bm{\beta}}$ and $\bm{\Sigma}_{\bm{\beta}}$ with 
  Eqs.~\eqref{CHLRM_CAVI_Mu_Sigma_Beta}
  \State Update $q(\bm{\Sigma})$ setting 
  $\nu_{\bm{\Sigma}}$ and $\bm{S}_{\bm{\Sigma}}$ with 
  Eqs.~\eqref{CHLRM_CAVI_nu_S_Sigma}
  \State Update $q(\xi^2)$ setting 
  $a_{\xi^2}$ and $b_{\xi^2}$ with 
  Eqs.~\eqref{CHLRM_CAVI_a_b_xi2}
  \State Compute ELBO $\mathcal{L}\bigl(q(\bm{\Theta})\bigr)$ with 
  Eq.~\eqref{CHLRM_CAVI_ELBO}
\EndWhile
\State \textbf{Output} Optimal variational parameters 
$\{\bm{\mu}_{\bm{\beta}_k},\bm{\Sigma}_{\bm{\beta}_k}\}_{k=1}^K,
 \bm{\mu}_{\bm{\beta}},\bm{\Sigma}_{\bm{\beta}},
 \nu_{\bm{\Sigma}},\bm{S}_{\bm{\Sigma}},
 a_{\sigma_k^2}, b_{\sigma_k^2},
 a_{\xi^2}, b_{\xi^2},
 \{\rho_{jk}\}_{j=1}^m,\bm{\alpha}$
\end{algorithmic}
\end{algorithm}

\subsubsection{SVI}\label{eq_global_alg_svi}

Since the CHLRM satisfies the assumptions of the stochastic variational inference framework (see Section~\ref{SVI}), the required updates follow from applying the Robbins--Monro scheme in Eq.~\eqref{Robbins-Monro} to the VI updates in Eqs.~\eqref{CHLRM_CAVI_rhojk_gamma}--\eqref{CHLRM_CAVI_a_b_xi2}. In this setting, the local variational factors, associated with the group allocation variables and the cluster specific regression and variance parameters, are updated within each iteration by performing a single CAVI step on a minibatch of groups, which produces unbiased noisy estimates of the corresponding full data updates. The global variational factors, which control the mixture weights, the population level regression coefficients, the covariance matrix and the common variance parameter, are then refined by stochastic approximation, combining past values with the intermediate noisy updates through the Robbins--Monro learning rate schedule. The resulting stochastic variational algorithm for the CHLRM is summarized in Algorithm~\ref{SVI_CHLRM_alg}, in which intermediate global parameters are computed as:
{\footnotesize
\begin{align*}
  \widehat{\bm{\mu}}_{\bm{\beta}_k}
  &=
  \widehat{\bm{\Sigma}}_{\bm{\beta}_k}
  \left(
    \bm{\Lambda}_0^{-1} \bm{\mu}_0 
    +
    \frac{m}{|\mathcal{S}|}
    \sum_{j \in \mathcal{S}} \sum_{k=1}^K
      \mathsf{E}_{\bm{\Sigma}}\bigl[\bm{\Sigma}^{-1}\bigr]\,
      \mathsf{E}_{\bm{\beta}_k}\bigl[\bm{\beta}_k\bigr]
  \right),
  \\
  \widehat{\bm{\Sigma}}_{\bm{\beta}_k}
  &=
  \left(\mathsf{E}_{\bm{\Sigma}}\bigl[\bm{\Sigma}^{-1}\bigr]
  +
  \mathsf{E}_{\sigma_k^2}\bigl[\sigma_k^{-2}\bigr]\,
  \frac{m}{|\mathcal{S}|}
  \sum_{j \in \mathcal{S}}
    \mathsf{E}_{\gamma_j}\bigl[\,[\gamma_j = k]\,\bigr]\,
    \bm{X}_{j}^\top \bm{X}_{j}\right)^{-1},
  \\
  \widehat{a}_{\sigma_k^2}
  &=
  \frac{\nu_0}{2}
  +
  \frac{1}{2}\,
  \frac{m}{|\mathcal{S}|}
  \sum_{j \in \mathcal{S}}
    \mathsf{E}_{\gamma_j}\bigl[\,[\gamma_j = k]\,\bigr]\,
    n_j,
  \\
  \widehat{b}_{\sigma_k^2}
  &=
  \frac{\nu_0}{2}\,
  \mathsf{E}_{\xi^2}\bigl[\xi^2\bigr]
  +
  \frac{1}{2}\,
  \frac{m}{|\mathcal{S}|}
  \sum_{j \in \mathcal{S}}
    \mathsf{E}_{\gamma_j}\bigl[\,[\gamma_j = k]\,\bigr]\,
    \mathsf{E}_{\bm{\beta}_k}
    \Bigl[
      (\bm{y}_{j} - \bm{X}_{j} \bm{\beta}_k)^\top
      (\bm{y}_{j} - \bm{X}_{j} \bm{\beta}_k)
    \Bigr],
  \\
  \widehat{\bm{\mu}}_{\bm{\beta}}
  &=
  \widehat{\bm{\Sigma}}_{\bm{\beta}}
  \left(
    \bm{\Lambda}_0^{-1} \bm{\mu}_0
    +
    \sum_{k=1}^K
      \mathsf{E}_{\bm{\Sigma}}\bigl[\bm{\Sigma}^{-1}\bigr]\,
      \mathsf{E}_{\bm{\beta}_k}\bigl[\bm{\beta}_k\bigr],
  \right)\\
  \widehat{\bm{\Sigma}}_{\bm{\beta}}
  &=
  \left(\bm{\Lambda}_0^{-1}
  +
  \sum_{k=1}^K
    \mathsf{E}_{\bm{\Sigma}}\bigl[\bm{\Sigma}^{-1}\bigr]\right)^{-1},
  \\
  \widehat{\bm{S}}_{\bm{\Sigma}}
  &=
  \bm{S}_0
  +
  \sum_{k=1}^K
    \mathsf{E}_{\bm{\beta}_k,\bm{\beta}}
    \Bigl[
      (\bm{\beta}_k - \bm{\beta})
      (\bm{\beta}_k - \bm{\beta})^\top
    \Bigr],
  \\
  \widehat{b}_{\xi^2}
  &=
  b_0
  +
  \frac{\nu_0}{2}
  \sum_{k=1}^K
    \mathsf{E}_{\sigma_k^2}\bigl[\sigma_k^{-2}\bigr],
  \\
  \widehat{\bm{\alpha}}_{\bm{\omega}}
  &=
  \bm{\alpha}_0
  +
  \frac{m}{|\mathcal{S}|}
  \sum_{j \in \mathcal{S}}
    \bigl(
      \mathsf{E}_{\gamma_j}\bigl[\,[\gamma_j = 1]\,\bigr]
      \dots,
      \mathsf{E}_{\gamma_j}\bigl[\,[\gamma_j = K]\,\bigr]
    \bigr)^\top.
  \end{align*}
}

\begin{algorithm}[!htb]
\footnotesize
\caption{SVI for Clustering Hierarchical Linear Regression Model}
\label{SVI_CHLRM_alg}
\begin{algorithmic}
\State \textbf{Input} Initial values 
$\{\bm{\mu}_{\bm{\beta}_k}^{(0)},\bm{\Sigma}_{\bm{\beta}_k}^{(0)}\}_{k=1}^K,
 \bm{\mu}_{\bm{\beta}}^{(0)},\bm{\Sigma}_{\bm{\beta}}^{(0)},
 \nu_{\bm{\Sigma}},\bm{S}_{\bm{\Sigma}}^{(0)},
 a_{\xi^2}, b_{\xi^2}^{(0)},
 \{a_{\sigma_k^2}^{(0)}, b_{\sigma_k^2}^{(0)}\}_{k=1}^K,
 \{\rho_{jk}^{(0)}\}_{j=1}^m,
 \bm{\alpha}_{\bm{\omega}}^{(0)}$, 
hyperparameters $\bm{\mu}_0,\bm{\Lambda}_0,n_0,\bm{S}_0,a_0,b_0,\nu_0,\bm{\alpha}_0$, 
learning–rate parameters $\tau,\chi$, minibatch size $|\mathcal{S}|$, number of iterations $T$
\For{$t = 1,\dots,T$}
  \State \textbf{Local updates}
  \State Sample a minibatch of groups $\mathcal{S} \subset \{1,\dots,m\}$
  \For{$j \in \mathcal{S}$}
    \State Update $q(\gamma_j)$ setting $\rho_{jk}^{(t)}$ with 
    Eq.~\eqref{CHLRM_CAVI_rhojk_gamma}
  \EndFor
  \State Compute step size $\rho_t = (t+\tau)^{-\chi}$
  \State \textbf{Noisy global parameters}
  \State Compute intermediate global parameters given in Sec. \ref{eq_global_alg_svi}
  \State \textbf{Global updates}
  \For{$k = 1,\dots,K$}
    \State Update $q(\bm{\beta}_k)$ setting
    \begin{align*}
      \bm{\mu}_{\bm{\beta}_k}^{(t)}
      &\leftarrow
      (1-\rho_t)\,\bm{\mu}_{\bm{\beta}_k}^{(t-1)}
      + \rho_t\,\widehat{\bm{\mu}}_{\bm{\beta}_k}\\
      \bigl(\bm{\Sigma}_{\bm{\beta}_k}^{-1}\bigr)^{(t)}
      &\leftarrow
      (1-\rho_t)\bigl(\bm{\Sigma}_{\bm{\beta}_k}^{-1}\bigr)^{(t-1)}
      + \rho_t\,\widehat{\bm{\Sigma}}_{\bm{\beta}_k}^{-1}
    \end{align*}
    \State Update $q(\sigma_k^2)$ setting
    \begin{align*}
      a_{\sigma_k^2}^{(t)}
      &\leftarrow
      (1-\rho_t)\,a_{\sigma_k^2}^{(t-1)}
      + \rho_t\,\widehat{a}_{\sigma_k^2}\\
      b_{\sigma_k^2}^{(t)}
      &\leftarrow
      (1-\rho_t)\,b_{\sigma_k^2}^{(t-1)}
      + \rho_t\,\widehat{b}_{\sigma_k^2}
    \end{align*}
  \EndFor
  \State Update $q(\bm{\omega})$ setting
  \[
    \bm{\alpha}_{\bm{\omega}}^{(t)}
    \leftarrow
    (1-\rho_t)\,\bm{\alpha}_{\bm{\omega}}^{(t-1)}
    + \rho_t\,\widehat{\bm{\alpha}}_{\bm{\omega}}
  \]
  \State Update $q(\bm{\beta})$ setting
  \begin{align*}
    \bm{\mu}_{\bm{\beta}}^{(t)}
    &\leftarrow
    (1-\rho_t)\,\bm{\mu}_{\bm{\beta}}^{(t-1)}
    + \rho_t\,\widehat{\bm{\mu}}_{\bm{\beta}}\\
    \bigl(\bm{\Sigma}_{\bm{\beta}}^{-1}\bigr)^{(t)}
    &\leftarrow
    (1-\rho_t)\bigl(\bm{\Sigma}_{\bm{\beta}}^{-1}\bigr)^{(t-1)}
    + \rho_t\,\widehat{\bm{\Sigma}}_{\bm{\beta}}^{-1}
  \end{align*}
  \State Update $q(\bm{\Sigma})$ setting
  \[
    \bm{S}_{\bm{\Sigma}}^{(t)}
    \leftarrow
    (1-\rho_t)\,\bm{S}_{\bm{\Sigma}}^{(t-1)}
    + \rho_t\,\widehat{\bm{S}}_{\bm{\Sigma}}
  \]
  \State Update $q(\xi^2)$ setting
  \[
    b_{\xi^2}^{(t)}
    \leftarrow
    (1-\rho_t)\,b_{\xi^2}^{(t-1)}
    + \rho_t\,\widehat{b}_{\xi^2}
  \]
\EndFor
\State \textbf{Output} Approximate optimal variational parameters 
$\{\bm{\mu}_{\bm{\beta}_k}$, $\bm{\Sigma}_{\bm{\beta}_k}\}_{k=1}^K$,
$\bm{\mu}_{\bm{\beta}}$, $\bm{\Sigma}_{\bm{\beta}}$,
$\bm{S}_{\bm{\Sigma}}$,
$\{a_{\sigma_k^2}$, $b_{\sigma_k^2}\}_{k=1}^K$,
$a_{\xi^2}$, $b_{\xi^2}$,
$\{\rho_{jk}\}_{j=1}^m$,
$\bm{\alpha}_{\bm{\omega}}$
\end{algorithmic}
\end{algorithm}

\begin{figure}[!htb]
    \centering
    \includegraphics[scale=0.31]{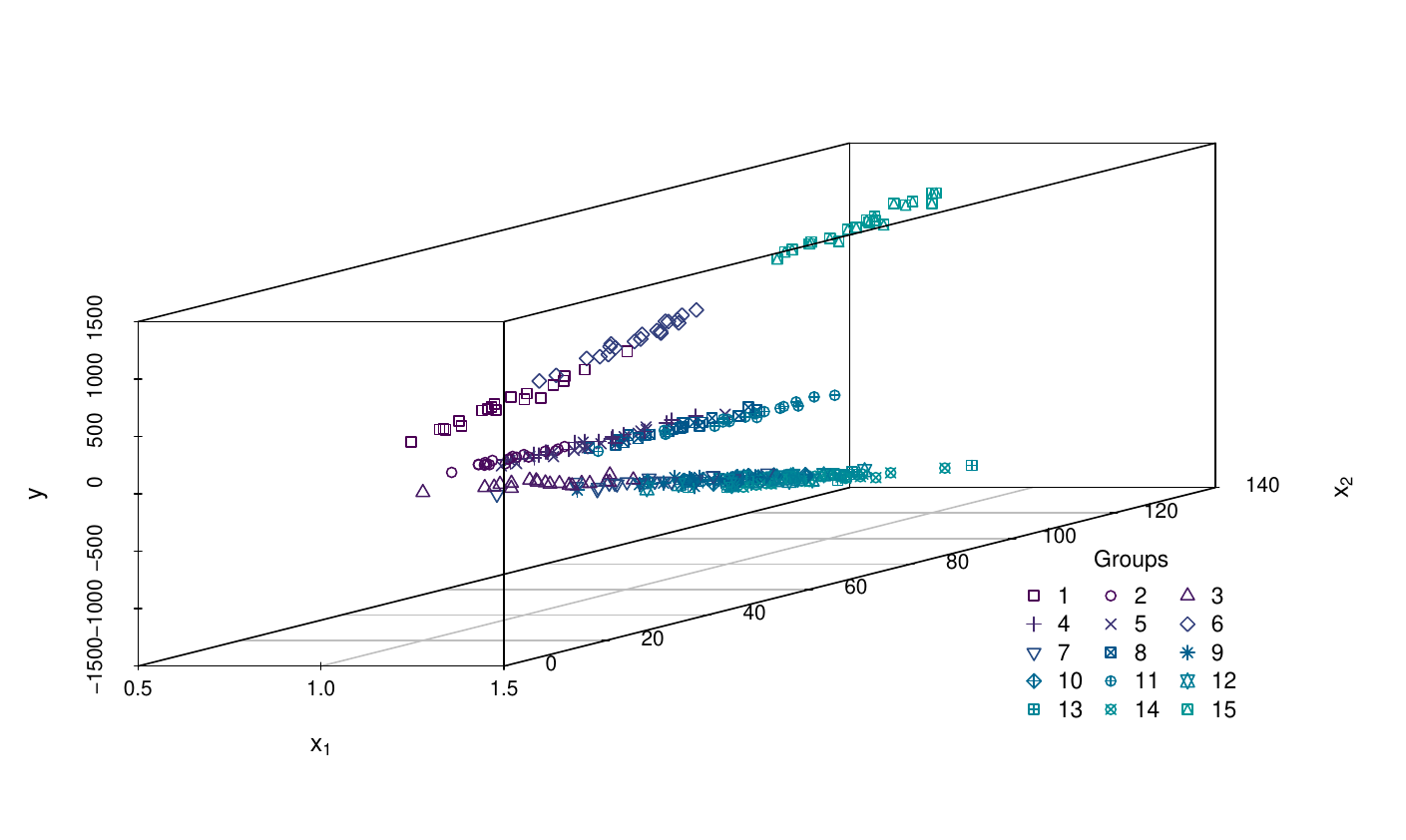}
    \includegraphics[scale=0.31]{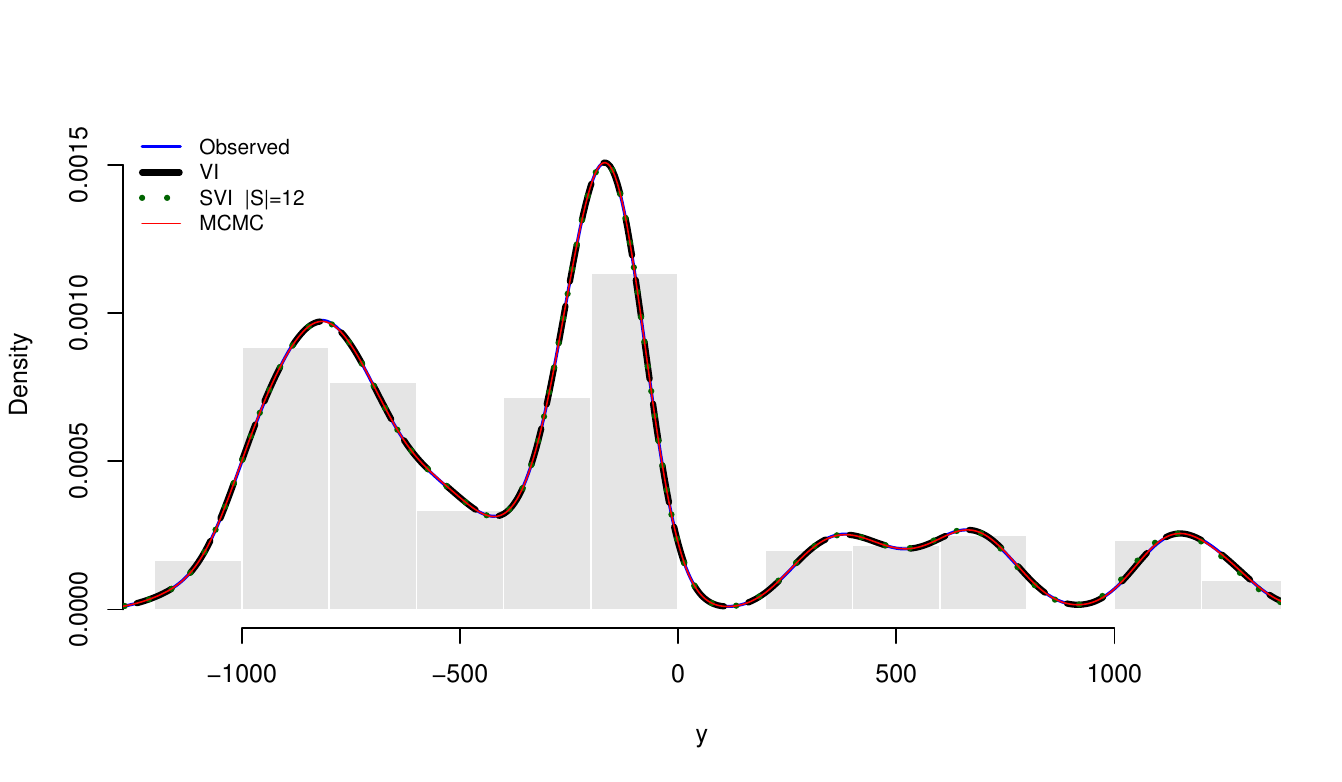}
    \caption{Left: Scatterplot of simulated CHLRM data, response $y$ versus covariates $x_1$ and $x_2$, with groups distinguished by color. Right: Observed response $y$ and posterior predictive densities under MCMC, VI, and SVI, simulated data.}
    \label{fig:chlrm_scatter}
\end{figure}

\subsubsection{Illustration: Simulation}

We assess the performance of MCMC, VI, and SVI under a controlled simulation scenario generated from a CHLRM with \(K = 3\) latent regression components, \(p = 3\) covariates (including an intercept), \(m = 15\) groups, and \(n_j = 20\) observations per group, with mixing weights \(\bm{\omega} = (0.4,\,0.3,\,0.3)^\top\). The component parameters are chosen to induce clearly heterogeneous linear effects across clusters, $\bm{\beta}_1=\left(-5,\ 8,\ 3\right)^\top$, $\bm{\beta}_2=\left(10,\ -1,\ -2\right)^\top$, and $\bm{\beta}_3=\left(35,\ -8,\ -2\right)^\top$, together with within–component error variances $\sigma_1^2=16$, $\sigma_2^2=9$, $\sigma_3^2=4$, so that both the signal level and noise level differ across clusters. Left panel of Figure~\ref{fig:chlrm_scatter} displays the simulated data in covariate–response space, where groups are color coded and the three mixture components are visibly separated, while still exhibiting realistic within–group variability. This geometric heterogeneity facilitates identifiability of the cluster specific parameters \((\bm{\beta}_k,\sigma_k^2)\) and provides a convenient benchmark to study parameter recovery, clustering accuracy, and predictive performance of MCMC, VI, and SVI under a CHLRM data generating mechanism.

\begin{table}[!htb]
\centering
\footnotesize
\caption{Goodness of fit and runtime for the CHLRM with 1000 posterior samples based, simulated data.}
\label{tab:fit_chlrm}
\begin{tabular}{@{} l l c c c c c @{}}
\toprule
Method & Settings & WAIC & DIC & MSE & $\text{R}^2$ & Time (s)\\
\midrule
MCMC &
\begin{tabular}{@{}l@{}}
samples: 50000\\
burn-in: 10000\\
thin: 10
\end{tabular}
& 1453.311 & 1453.671 & 8.023 & 1 & 466.055\\
\midrule
VI &
\begin{tabular}{@{}l@{}}
iterations: 20\\
$\epsilon=1\times10^{-7}\%$
\end{tabular}
& 1453.323 & 1453.570 & 8.022 & 1 & 0.198\\
\midrule
SVI &
\begin{tabular}{@{}l@{}}
iterations: 15\\
$\epsilon=1\times10^{-7}\%$\\
$|S| = 12,\ \chi = 0.7,\ \tau = 25.8$
\end{tabular}
& 1983.810 & 2028.781 & 11.344 & 1 & 0.02\\
\bottomrule
\end{tabular}
\end{table}

\begin{figure}[!htb]
    \centering
    \includegraphics[scale=0.31]{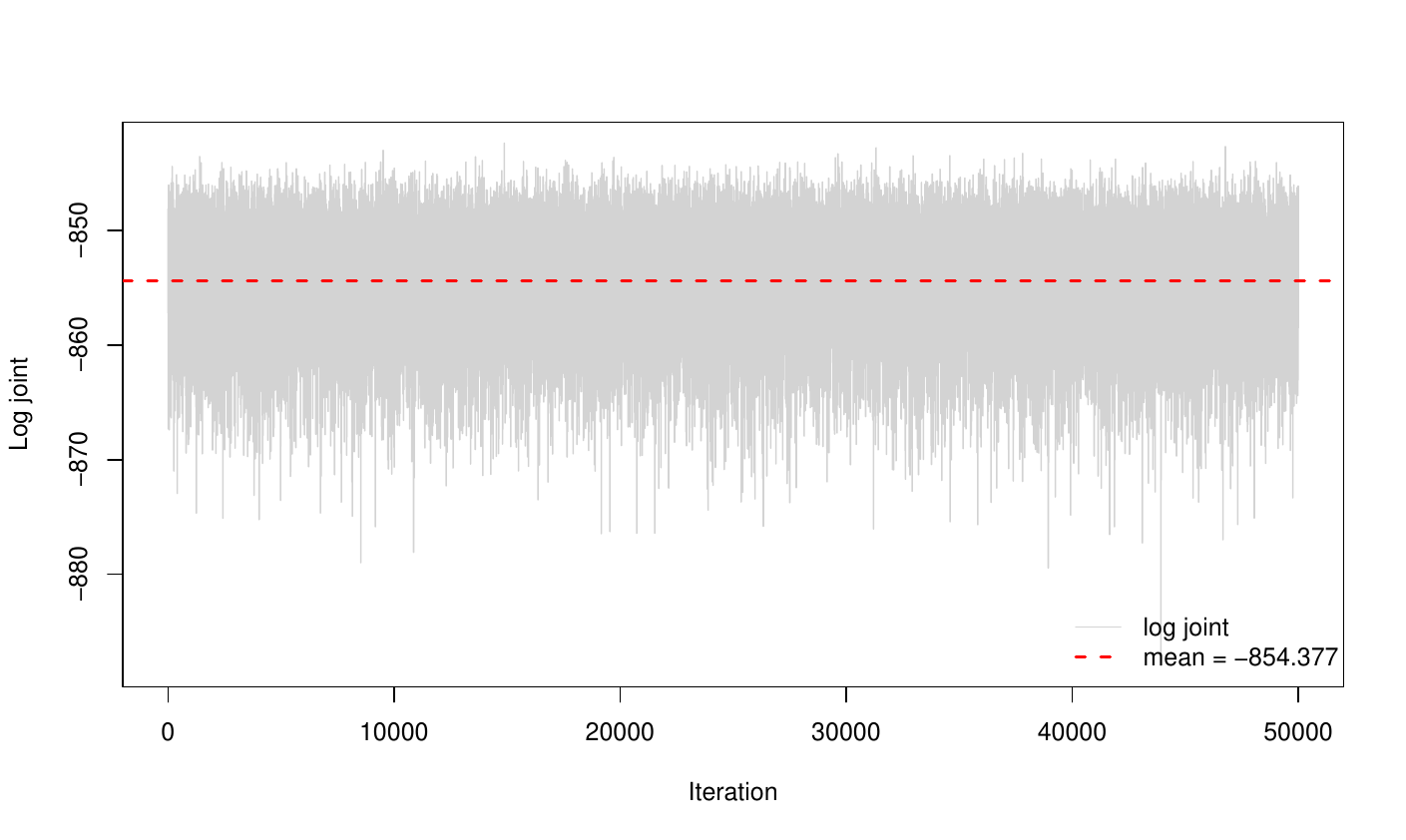}
    \includegraphics[scale=0.31]{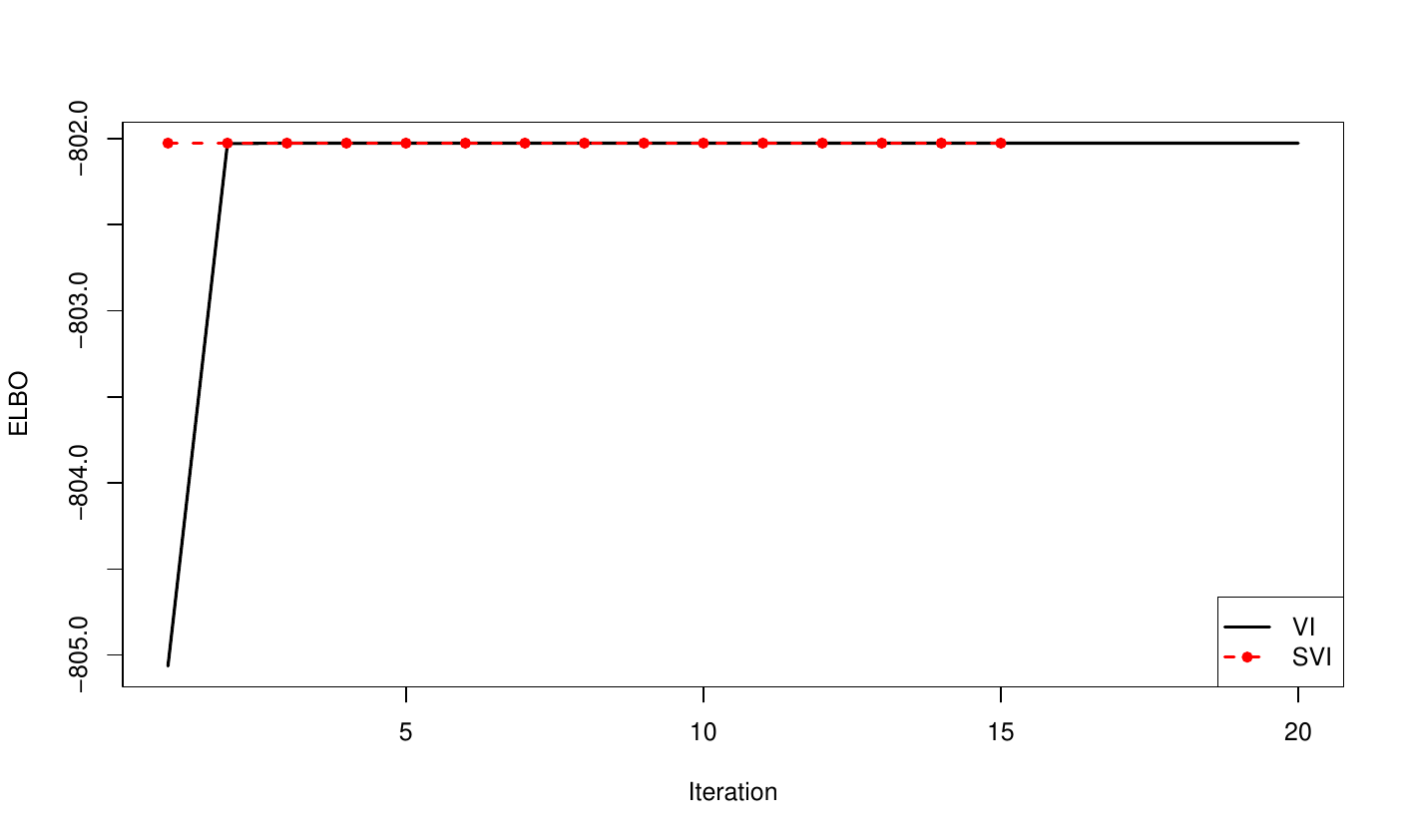}
    \caption{Left: Trace of the log–joint distribution under MCMC estimation, simulated data. Right: ELBO across iterations for VI and SVI with $|\mathcal{S}| = 12$, $\epsilon = 1\times 10^{-7}\%$, $\chi = 0.7$ and $\tau = 71.2$, simulated data.}
    \label{fig:chlrm_logjoint_chain}
\end{figure}

Figures~\ref{fig:chlrm_logjoint_chain} assess convergence. The MCMC log–joint chain stabilizes after burn–in with a stationary mean close to $-854$, and shows good mixing. The ELBO under VI increases smoothly, whereas SVI converges more erratically toward a similar bound. Together with the runtimes reported in Table~\ref{tab:fit_chlrm}, these results highlight a clear trade–off in which MCMC delivers faithful posterior samples at high computational cost (around 466 seconds), while VI and SVI achieve comparable predictive performance in only fractions of a second.

Figures of 95\% credible intervals for the regression coefficients and variance components confirm accurate recovery of the true slopes and error variances. Posterior means align closely with the generating values across methods, although credible interval widths differ: MCMC yields wider and better–calibrated intervals, whereas VI and SVI underestimate uncertainty, as expected under factorized approximations. This discrepancy is most pronounced for Cluster~1, where higher variability amplifies uncertainty. Nevertheless, the accuracy metrics in Table~\ref{tab:fit_chlrm} confirm that all three methods deliver adequate predictive performance.

Taken together, Tables~\ref{tab:fit_chlrm} and~\ref{tab:ppp} show that the three inference strategies attain essentially identical predictive accuracy. WAIC and DIC are virtually indistinguishable across MCMC, VI, and SVI, and both MSE and $\text{R}^2$ indicate that the fitted models explain almost all variability in the simulated data. Posterior predictive $p$-values (ppp's) are all close to $0.5$, indicating that replicated datasets are statistically consistent with the observed outcomes. Overall, the stability of the information criteria, the negligible predictive error, and the well–balanced ppp's values demonstrate that, despite relying on stronger approximating assumptions and delivering substantial computational savings, VI and SVI do not compromise predictive adequacy relative to the MCMC benchmark.

\begin{table}[!h]
\centering
\footnotesize
\caption{\label{tab:ppp}Posterior predictive $p$-values for CHLRM (1000 replications), simulated data.}
\begin{tabular}{@{} lrrrrrr @{}}
\toprule
Method & min & max & IQR & mean & median & sd \\
\midrule
MCMC & 0.75 & 0.79 & 0.37 & 0.48 & 0.65 & 0.48\\
VI   & 0.73 & 0.78 & 0.35 & 0.48 & 0.65 & 0.50\\
SVI  & 0.73 & 0.79 & 0.23 & 0.50 & 0.61 & 0.53\\
\bottomrule
\end{tabular}
\end{table}

\begin{figure}[!htb]
    \centering
    \includegraphics[scale=0.32]{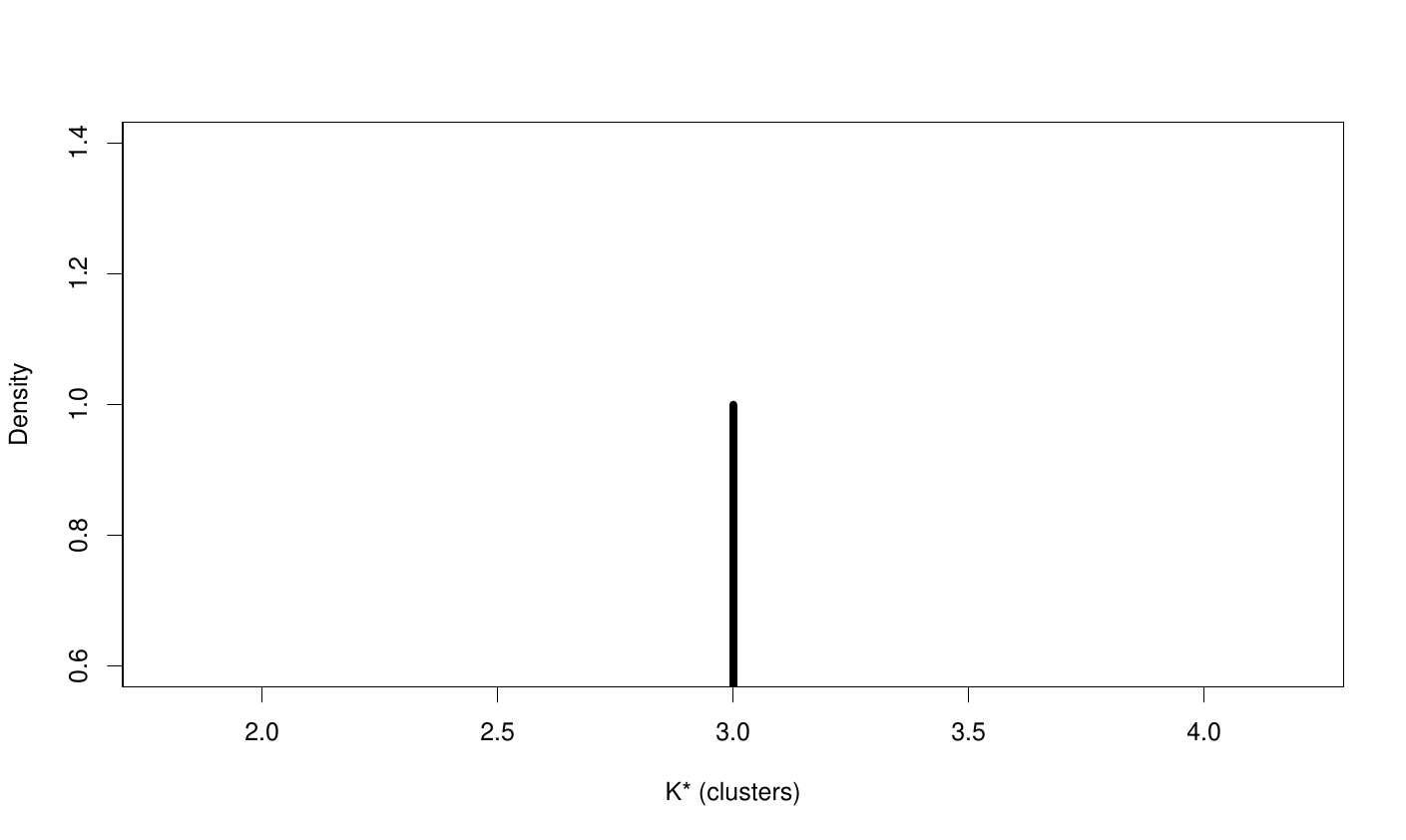}
    \includegraphics[scale=0.32]{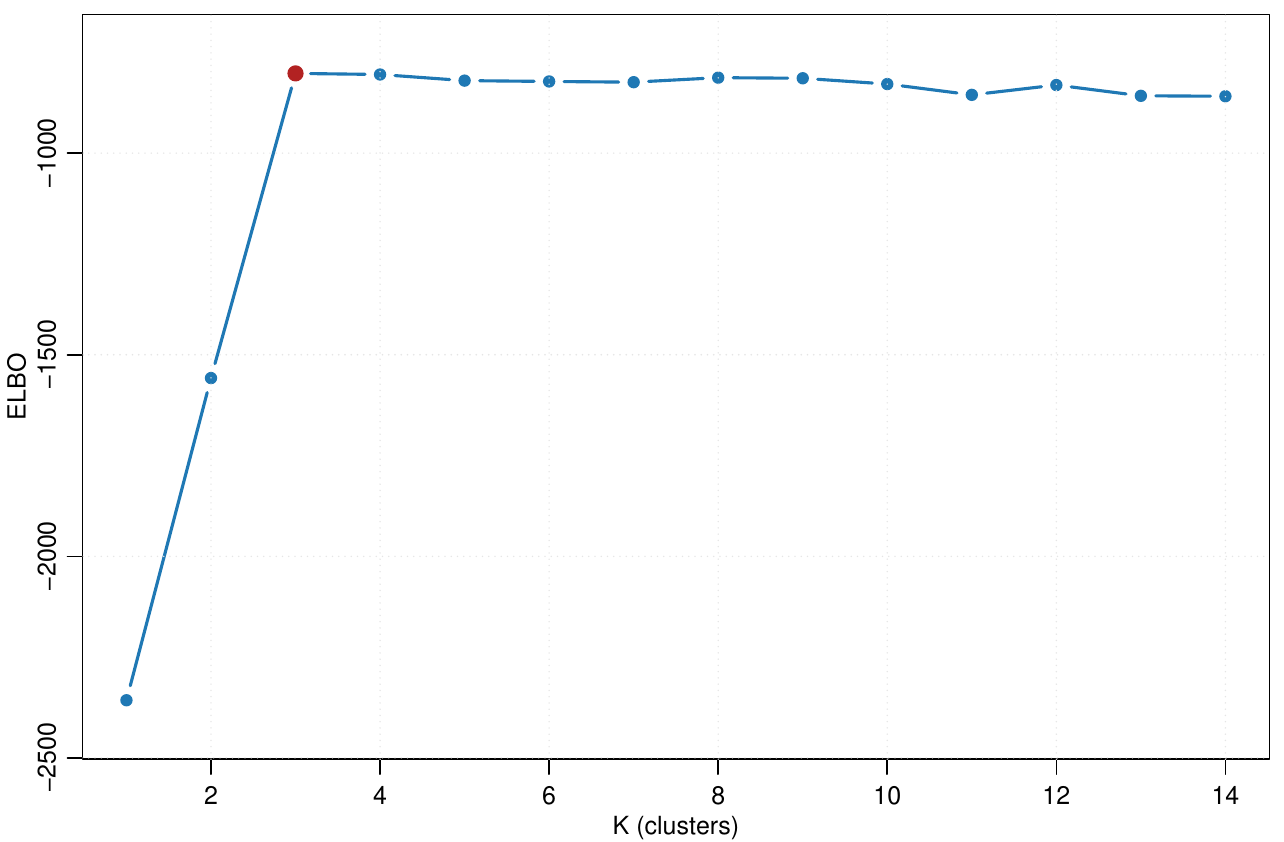}
    \caption{Right: Posterior density of the number of nonempty clusters under MCMC with $K = 14$, simulated data. Left: ELBO values as a function of $K$, simulated data.}
    \label{fig:chlrm_mcmc_k}
\end{figure}

To determine the partition size, two complementary analyses are carried out. First, an MCMC run with $K = 14$ is used to estimate the posterior distribution of the number of nonempty clusters, which concentrates its mass at $K = 3$ (left panel of Figure~\ref{fig:chlrm_mcmc_k}). Second, VI is fitted for $K = 1,\dots,14$ under the settings of Table~\ref{tab:fit_chlrm}, and the ELBO attains its maximum at $K = 3$ and deteriorates for larger values (right panel of Figure~\ref{fig:chlrm_mcmc_k}). Both diagnostics consistently recover the true number of clusters.

\subsubsection{Illustration: Farms Data}

\begin{figure}[!htb]
    \centering
    \includegraphics[scale=0.35]{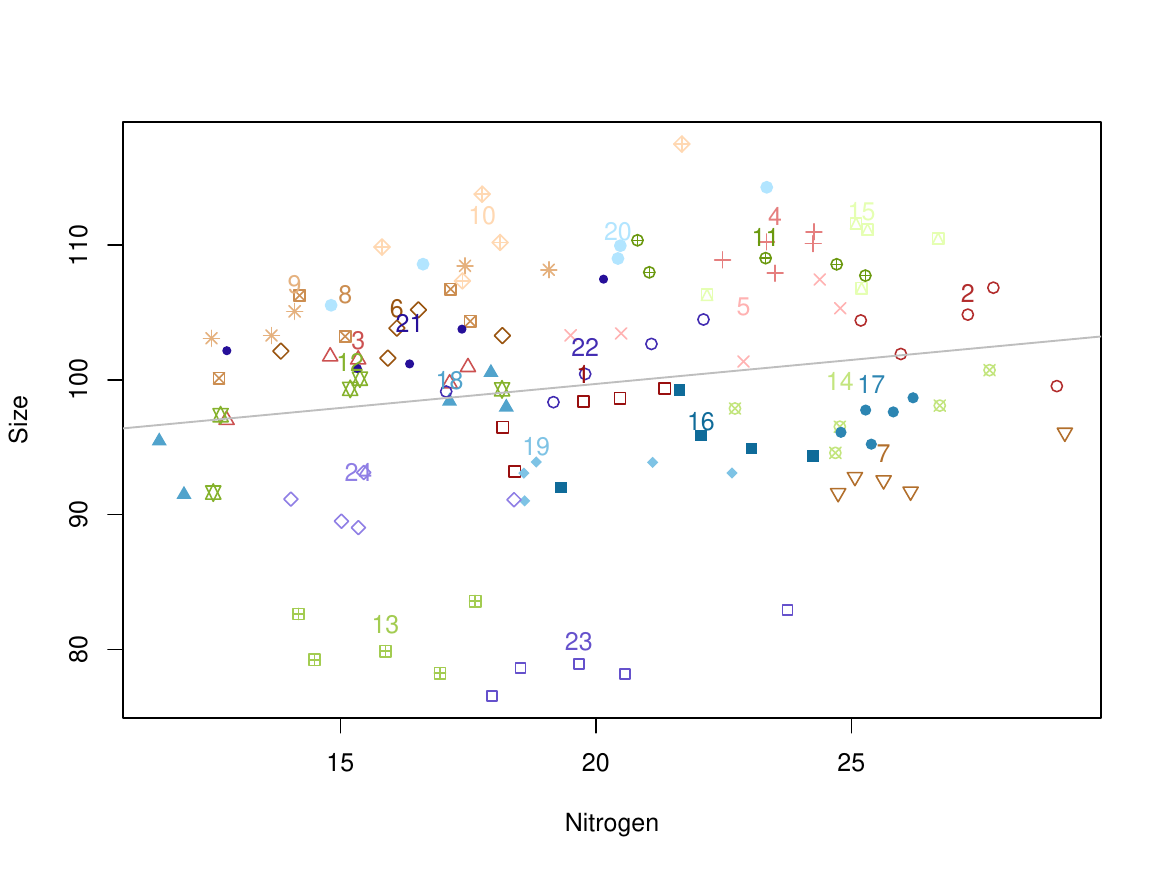}
    \includegraphics[scale=0.35]{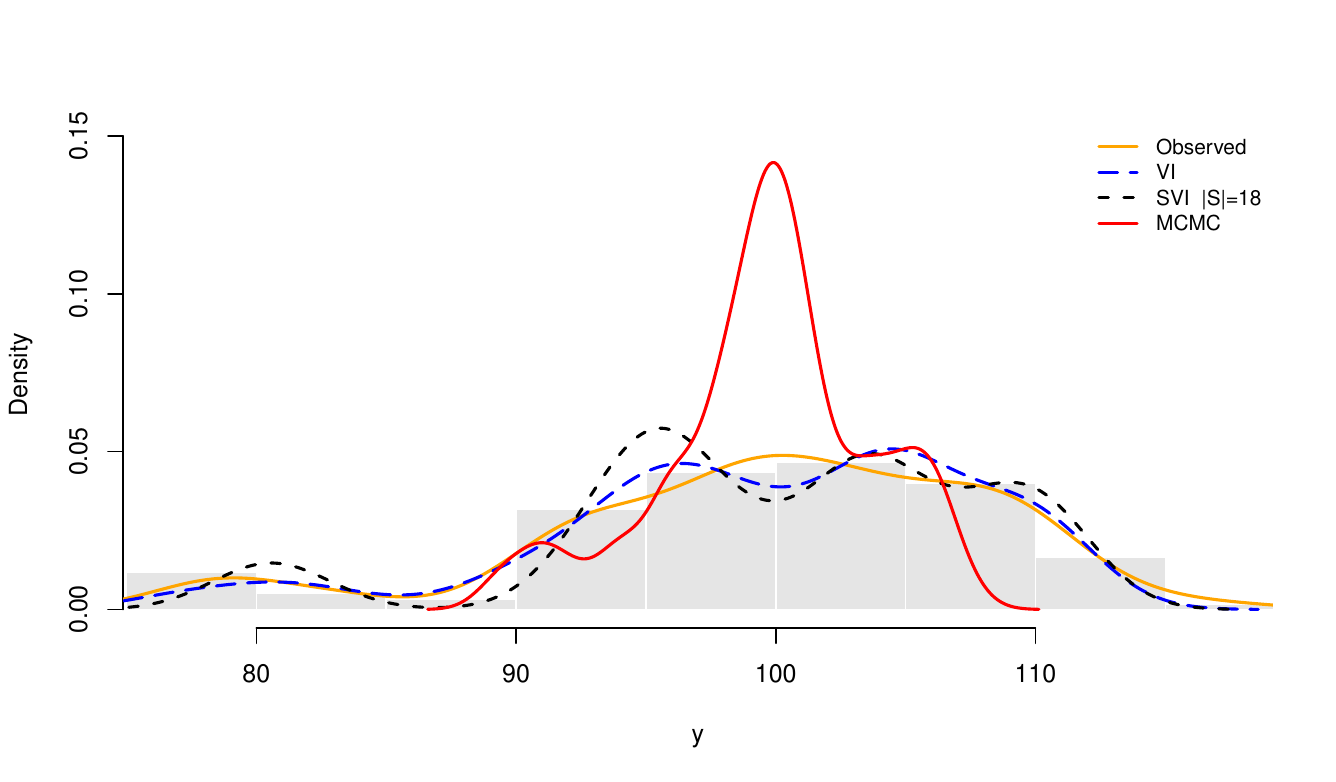}
    \caption{Left: Scatter plot of plant size versus soil nitrogen concentration for 24 farms, with each farm represented by a different symbol and the global ordinary least squares (OLS) regression line providing a baseline fit. Right: Observed response $y$ and posterior predictive densities under MCMC, VI, and SVI with $K = 5$ clusters, farms dataset.}\label{fig:hlrm_ols_farms}
\end{figure}

We illustrate the proposed algorithms using an agricultural dataset on soil nitrogen and plant size originally presented in \cite{Crawley2012}. The data consist of $m = 24$ farms, each contributing $n = 5$ paired observations of soil nitrogen concentration $(N_{ij})$ and plant size $(S_{ij})$, for $i = 1,\dots,5$ and $j = 1,\dots,24$. Nitrogen is a key macro-nutrient for plant development, with higher levels generally associated with increased growth and yield, although excessively high concentrations may be harmful. This structure naturally motivates a hierarchical regression model in which the effect of nitrogen on plant size varies across farms while borrowing strength through population-level priors. We use this example to compare MCMC, VI, and SVI in terms of estimation accuracy, computational efficiency, and predictive performance, emphasizing the trade-offs between exact and approximate Bayesian inference in hierarchical regression. Figure~\ref{fig:hlrm_ols_farms} displays the relationship between soil nitrogen and plant size for the 24 farms, with each farm represented by a distinct color and symbol and a global OLS regression line providing a baseline fit that highlights farm-level heterogeneity and motivates a hierarchical specification.

\begin{figure}[!htb]
    \centering
    \includegraphics[scale=0.32]{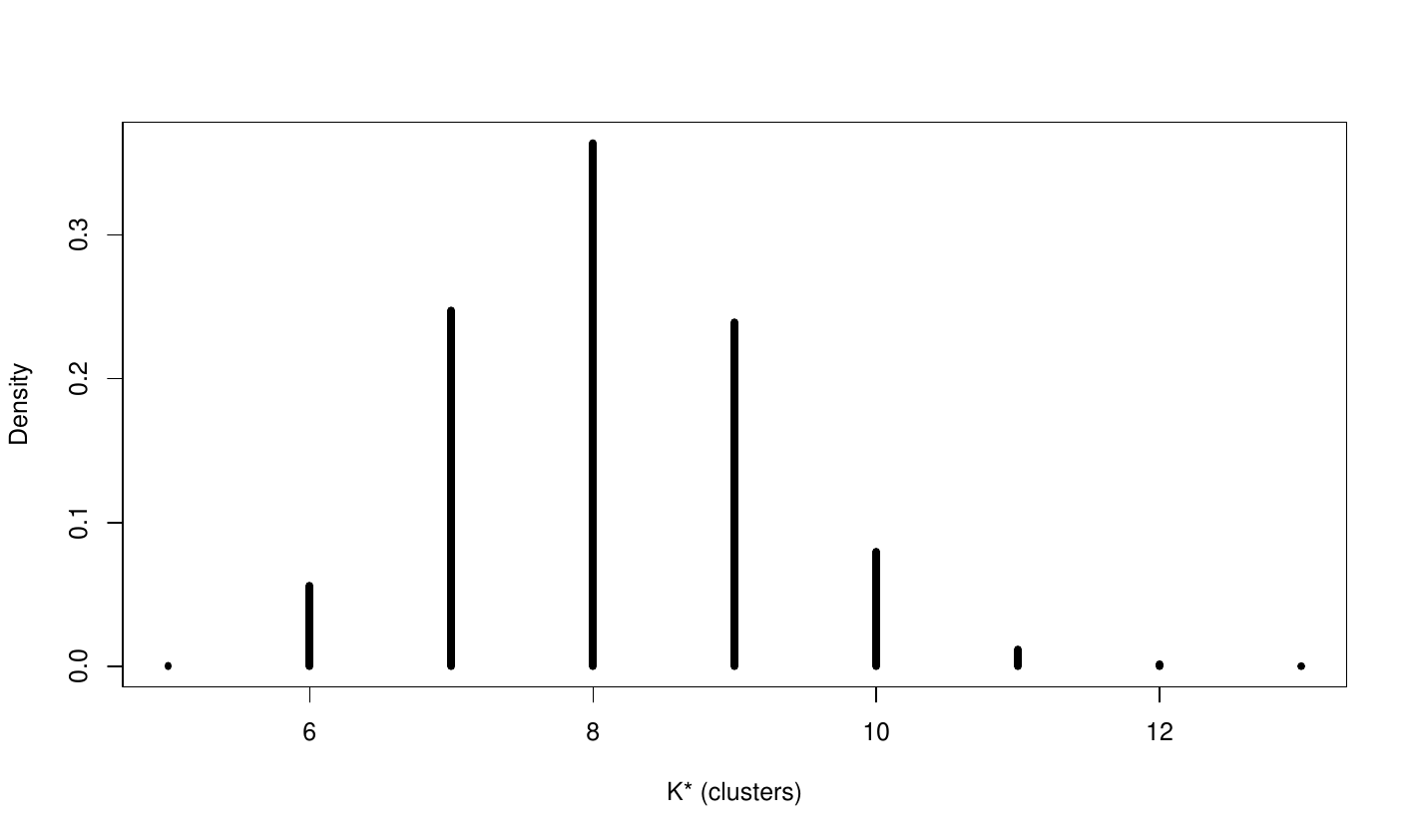}
    \includegraphics[scale=0.35]{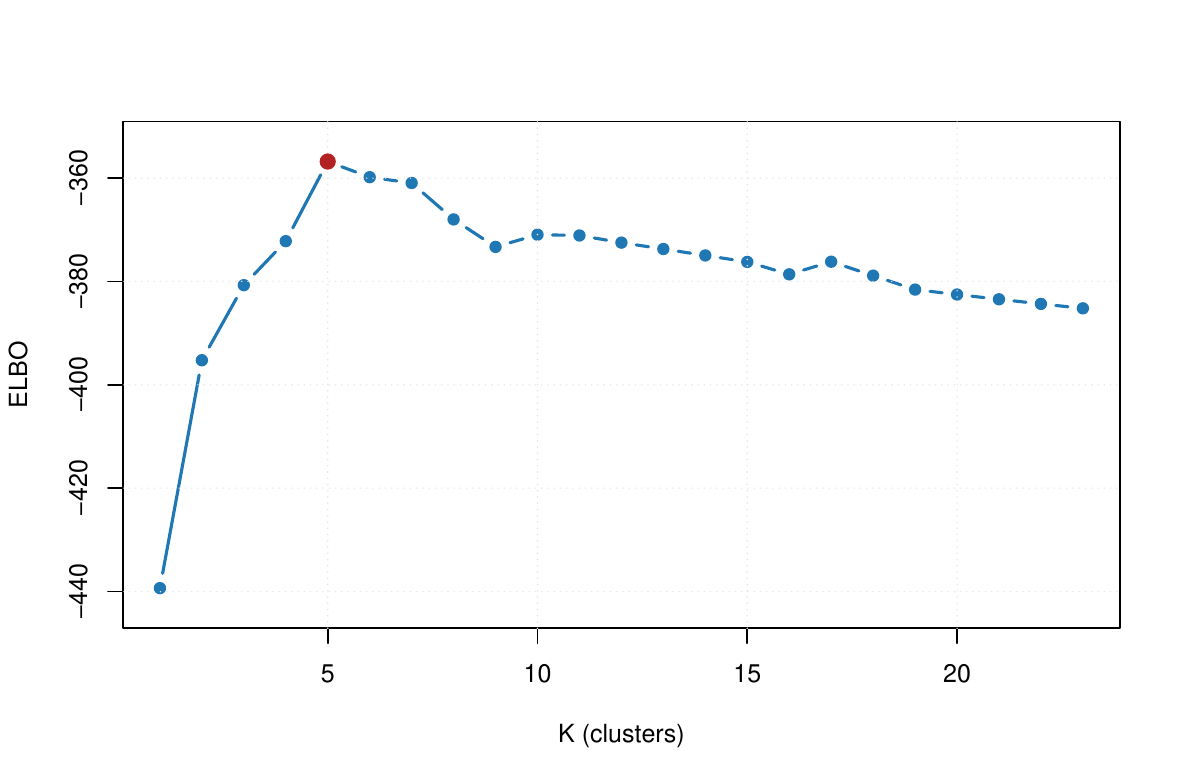}
    \caption{Left: Posterior density of the number of nonempty clusters under MCMC with $K = 23$ (50{,}000 samples, 10{,}000 burn-in iterations, thinning every 10 iterations), farms dataset. Right: ELBO values versus $K$ under SVI ($|\mathcal{S}| = 18$, $\epsilon = 1\times 10^{-7}\%$, $\chi = 0.7$, $\tau = 71.2$), farms dataset.}
    \label{fig:chlrm_mcmc_k_farms}
\end{figure}

We apply MCMC, VI, and SVI to the CHLRM under a range of values of $K$ to assess their performance when cluster separation is not visually evident as in the simulated data, with the goal of obtaining a partition that explains farm clustering in plant size adjusted by soil nitrogen concentration. The left panel of Figure~\ref{fig:chlrm_mcmc_k_farms} displays the posterior density of the number of nonempty clusters, with most mass around $K \in \{7,8,9\}$, which reflects the tendency of MCMC to favor richer cluster structures under weak empirical separation. The posterior distribution remains diffuse, indicating substantial uncertainty in the partition, and stabilization of the number of clusters requires long chains and careful thinning, which together with the considerable computational cost is evident in the runtime comparison of Table~\ref{tab:fit_chlrm_farms}, where MCMC is several orders of magnitude slower than VI and SVI. The right panel of Figure~\ref{fig:chlrm_mcmc_k_farms} shows the evolution of the ELBO as a function of $K$, with a maximum at $K = 5$, favoring a more parsimonious partition than the one suggested by MCMC and illustrating the smooth deterministic convergence of VI and SVI with low sensitivity to sampling noise. Maximizing the ELBO does not guarantee exact recovery of the true number of clusters, but it balances model complexity and predictive adequacy and provides computational efficiency at the cost of underestimating posterior variance and uncertainty in cluster allocation.

The posterior predictive densities in Figure~\ref{fig:hlrm_ols_farms} show that all three methods recover the global shape of the observed $y$, but with systematic differences. VI and SVI yield sharper, more concentrated predictive distributions that align more closely with the empirical data, consistent with their lower MSE and higher $\text{R}^2$ in Table~\ref{tab:fit_chlrm_farms}, whereas MCMC retains wider tails and greater dispersion, providing a fuller representation of posterior uncertainty at the expense of predictive precision. Table~\ref{tab:fit_chlrm_farms} quantifies this trade-off, since MCMC attains the smallest WAIC and DIC but requires over 3000 seconds, while VI and SVI reduce runtime to fractions of a second or a few seconds, achieve better MSE and $\mathsf{R}^2$, and display slightly larger WAIC and DIC, reflecting the bias introduced by variational approximations. Overall, VI and SVI offer substantial computational gains and strong predictive performance, at the cost of some posterior fidelity when cluster separation is weak.

\begin{table}[!htb]
\centering
\footnotesize
\caption{Goodness of fit and runtime for the CHLRM with 1000 posterior samples based, farms data.}
\label{tab:fit_chlrm_farms}
\begin{tabular}{@{} l l c c c c c @{}}
\toprule
Method & Settings & WAIC & DIC & MSE & $\text{R}^2$ & Time (s)\\
\midrule
MCMC &
\begin{tabular}{@{}l@{}}
samples: 120000\\
burn–in: 30000\\
thin: 20
\end{tabular}
& 570.278 & 583.640 & 30.711 & 0.579 & 3039.362\\
\midrule
VI &
\begin{tabular}{@{}l@{}}
iterations: 11\\
$\epsilon=1\times10^{-7}\%$
\end{tabular}
& 574.070 & 579.002 & 6.330 & 0.913 & 0.020\\
\midrule
SVI &
\begin{tabular}{@{}l@{}}
iterations: 2496\\
$\epsilon=1\times10^{-1}\%$\\
$|\mathcal{S}| = 18,\ \chi = 0.7,\ \tau = 71.2$
\end{tabular}
& 615.305 & 624.503 & 5.948 & 0.918 & 3.848\\
\bottomrule
\end{tabular}
\end{table}

Posterior predictive $p$–values in Table~\ref{tab:ppp_farms} confirm the adequacy of all three methods, with means and medians close to 0.5, indicating well–calibrated predictive distributions. SVI shows slightly higher central values, suggesting a somewhat optimistic fit to the observed statistics \citep{HOFF}. MCMC yields more balanced $p$–values across discrepancy measures, albeit with greater variability, whereas VI produces stable but slightly conservative predictive assessments.

\begin{table}[!htb]
\centering
\footnotesize
\caption{Posterior predictive $p$-values for CHLRM (1000 replications), farms dataset.}\label{tab:ppp_farms}
\begin{tabular}{@{} lcccccc @{}}
\toprule
Method & Min & Max & IQR & Mean & Median & SD \\
\midrule
MCMC & 0.34 & 0.33 & 0.23 & 0.50 & 0.46 & 0.56\\
VI   & 0.12 & 0.31 & 0.33 & 0.50 & 0.52 & 0.56\\
SVI  & 0.13 & 0.19 & 0.30 & 0.56 & 0.58 & 0.64\\
\bottomrule
\end{tabular}
\end{table}

\begin{figure}[!htb]
	\centering
	\setlength{\tabcolsep}{0pt}
	\begin{tabular}{ccc}
		& Initial & Final \\
		\begin{sideways} \hspace{1.1cm} MCMC \end{sideways}          &
		\includegraphics[scale = 0.28]{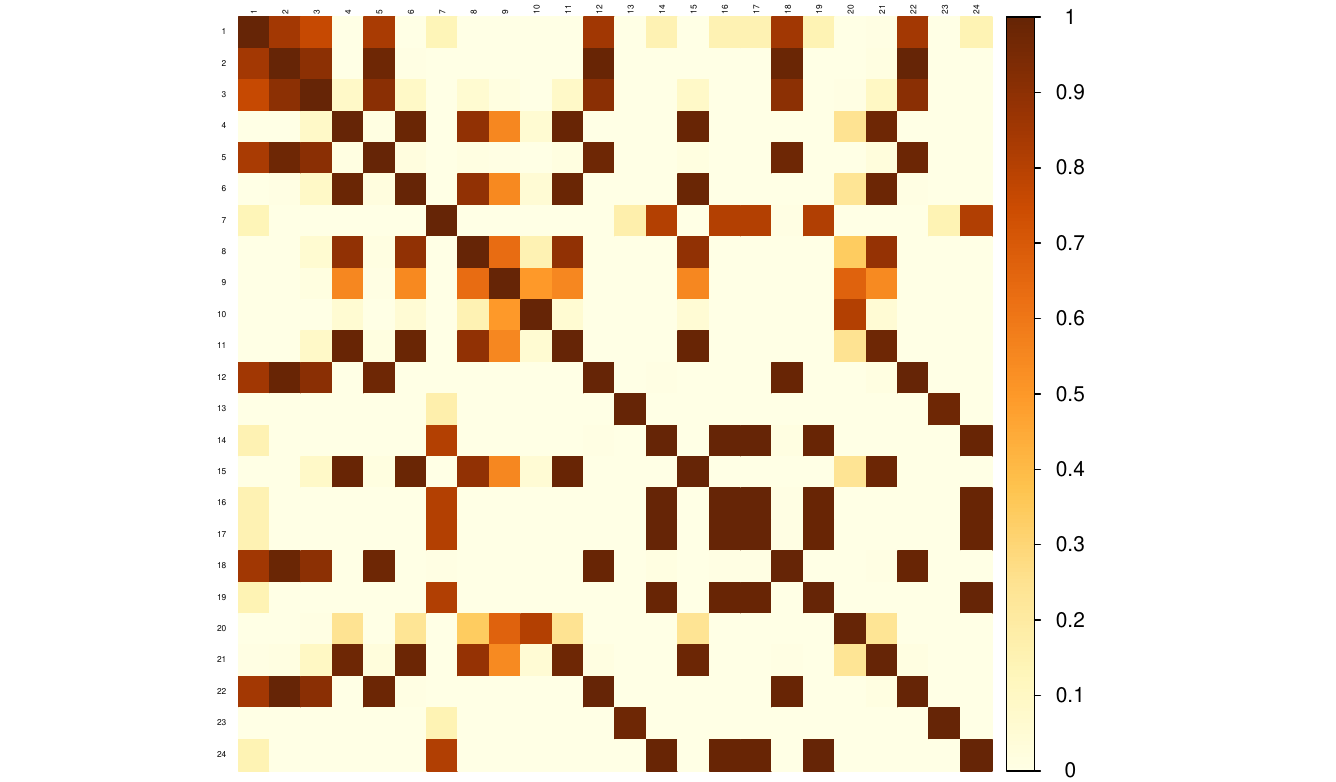}  &
		\includegraphics[scale = 0.28]{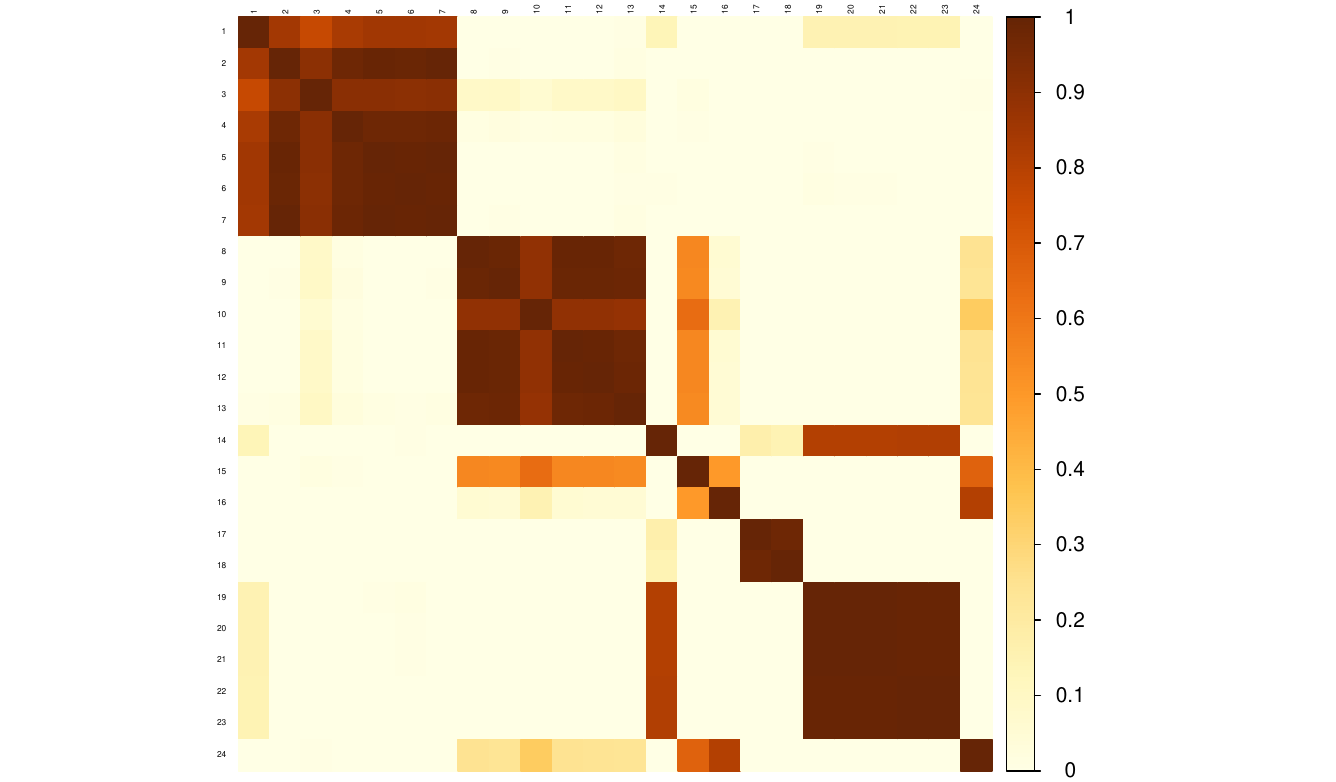}   \\
		\begin{sideways} \hspace{1.5cm} VI \end{sideways}            &
		\includegraphics[scale = 0.28]{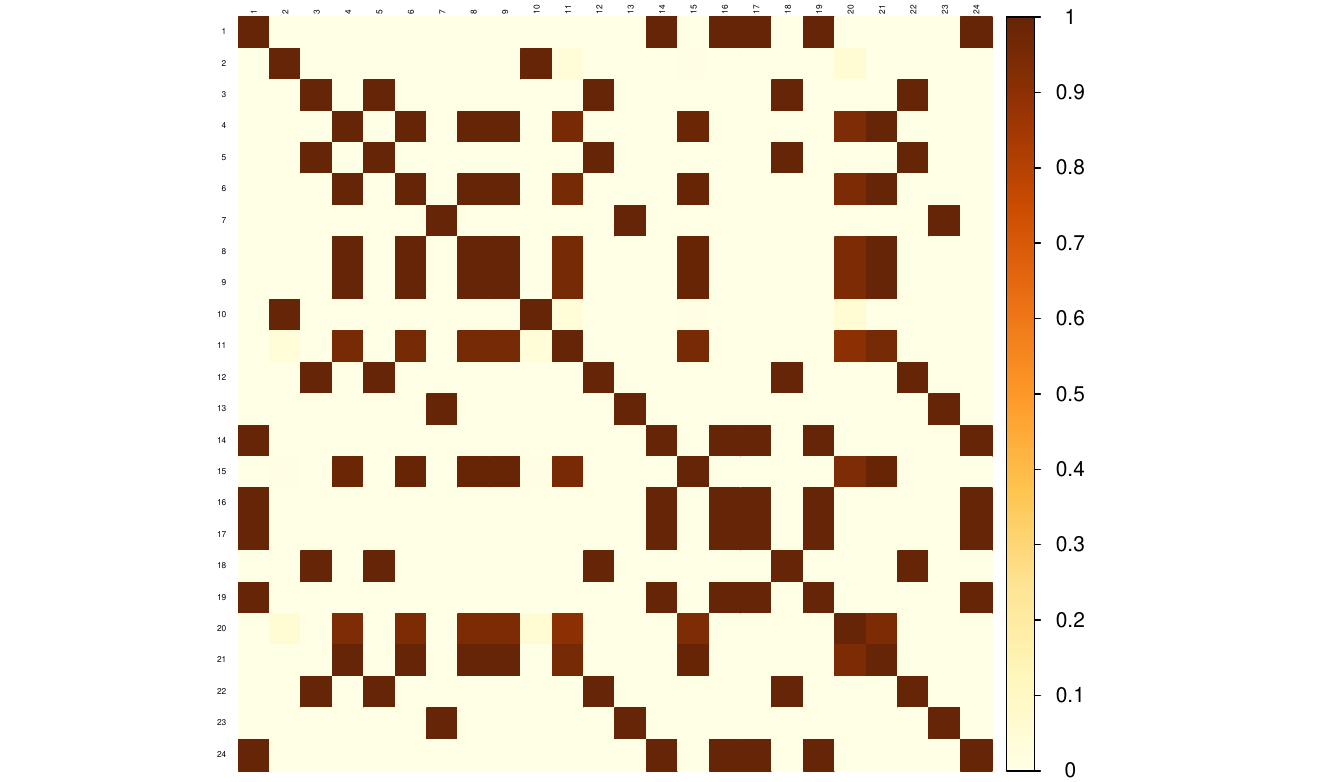}    &
		\includegraphics[scale = 0.28]{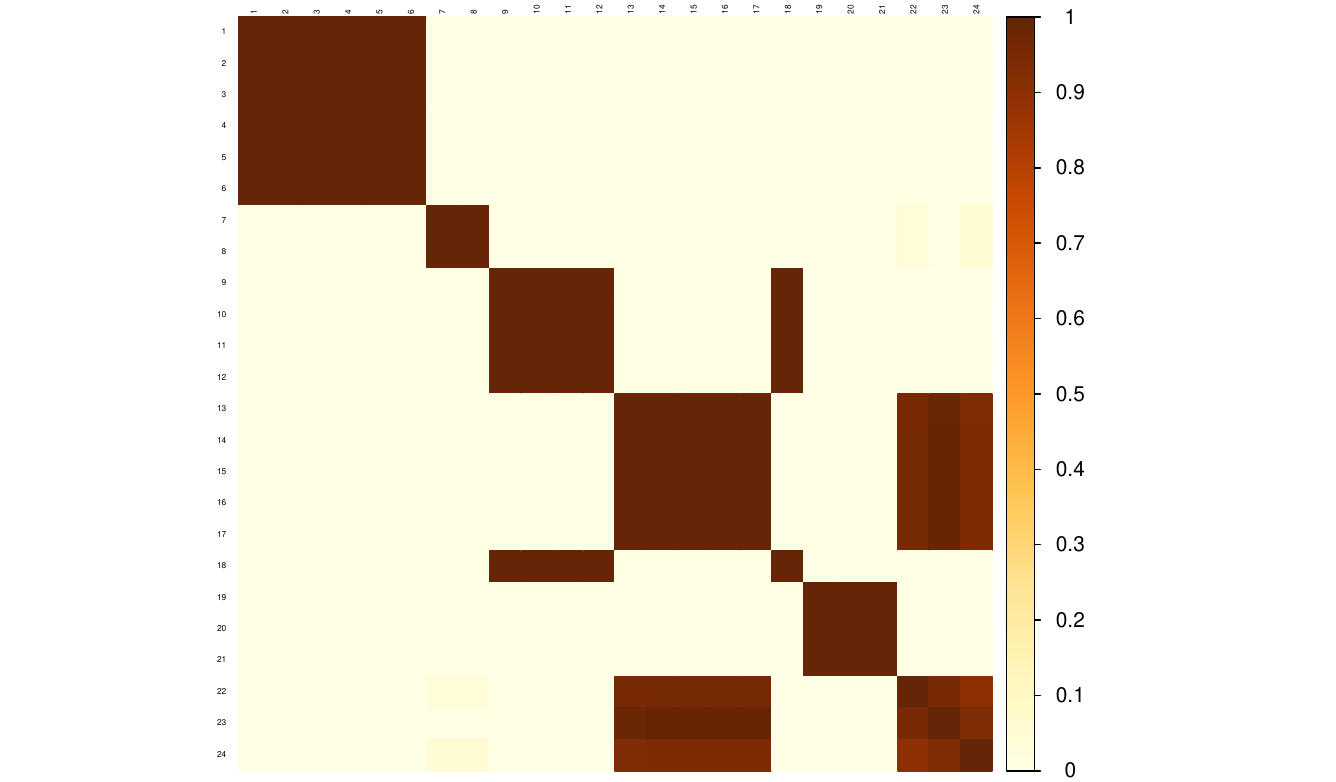}    \\
        \begin{sideways} \hspace{1.5cm} SVI \end{sideways}           &
		\includegraphics[scale = 0.28]{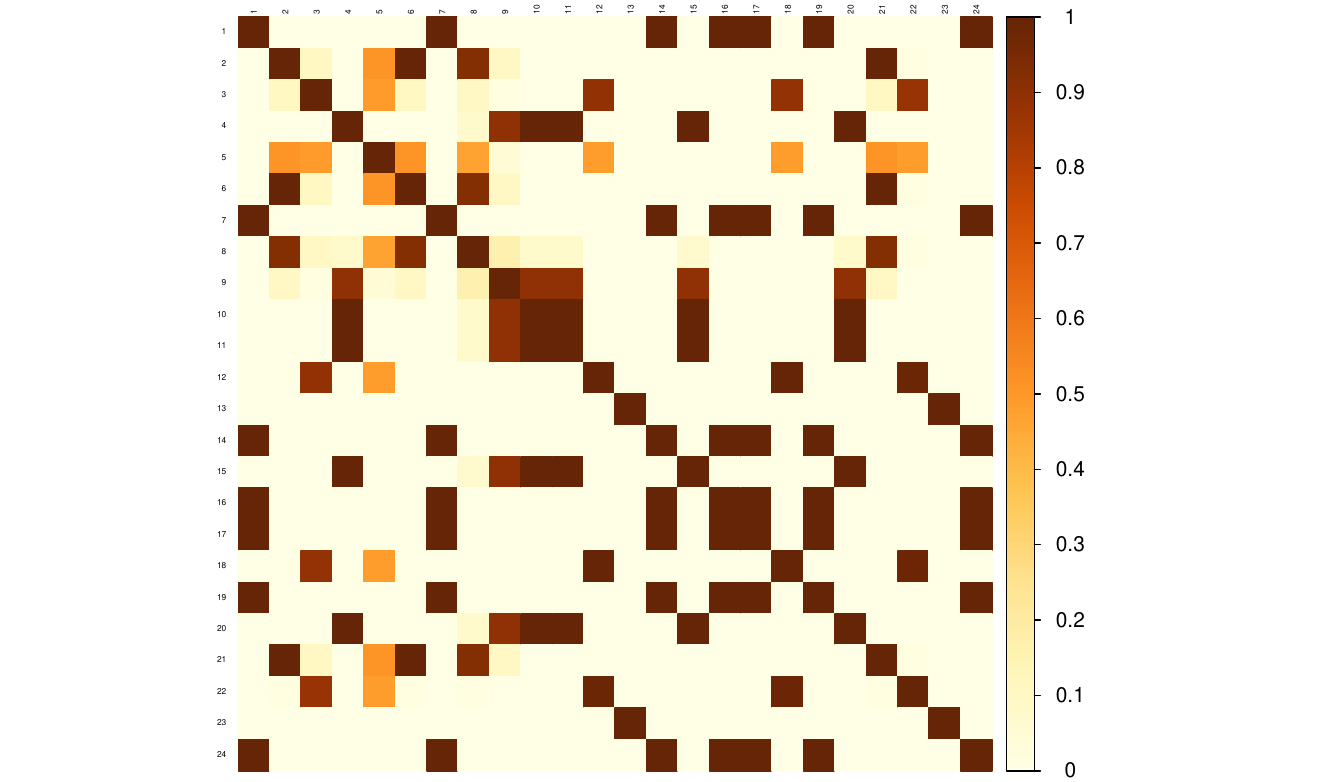}   &
		\includegraphics[scale = 0.28]{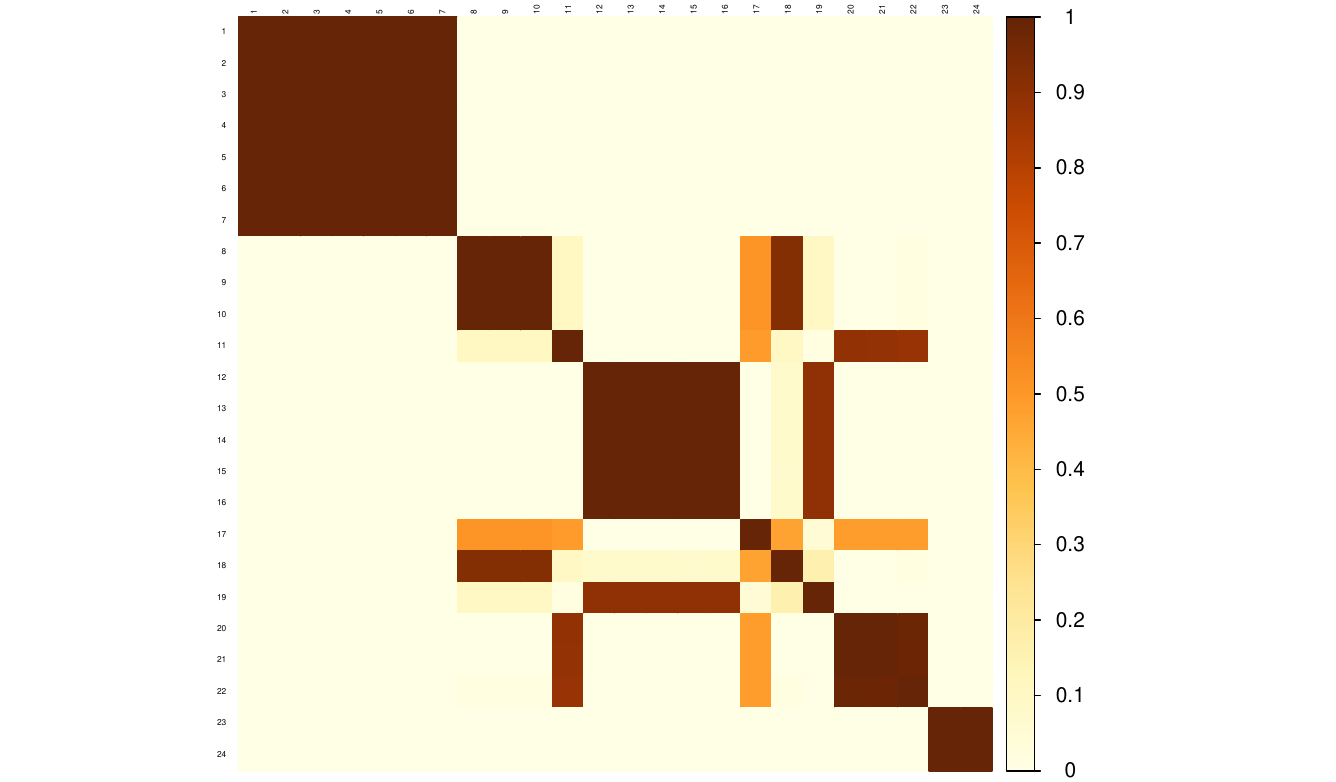}   \\
	\end{tabular}
	\caption{Co–clustering probabilities between groups, farms data.}
	\label{fig_adjacency}
\end{figure}

The adjacency matrices in Figure~\ref{fig_adjacency} display posterior co-assignment probabilities $\Pr(\gamma_j = \gamma_k \mid \bm{X},\bm{y})$ for $1\leq j\leq k \leq m$, evolving from diffuse patterns to a clear block-diagonal structure that recovers the true $K = 3$ partition. All three methods converge to similar clustering solution. MCMC gradually sharpens the block structure and provides the most reliable uncertainty quantification, VI attains rigid partitions more quickly but underrepresents uncertainty, and SVI offers clear cluster delineation with shorter runtime at the cost of additional stochastic variability from mini-batch updates.

In summary, the farms dataset underscores the central trade–offs among MCMC, VI, and SVI for clustered hierarchical regression. MCMC delivers the most faithful posterior inference, as indicated by lower information criteria and richer adjacency structures, but at a prohibitive computational cost. VI provides fast, stable convergence and sharp predictive fits, though it systematically underestimates posterior variance. SVI occupies an intermediate position, since it scales efficiently, attains predictive accuracy comparable to VI, and recovers cluster structure more robustly than VI, while still exhibiting stochastic noise and some bias in uncertainty quantification. Overall, MCMC remains the benchmark for posterior fidelity, whereas VI and SVI constitute practical alternatives when runtime and scalability are paramount, especially in settings with weak cluster separation.

\section{Discussion}

This paper studies Bayesian inference for regression models of increasing complexity, from the Linear Regression Model (LRM) to the Clustered Hierarchical Linear Regression Model (CHLRM). We compare Markov chain Monte Carlo (MCMC) with Variational Inference (VI) and Stochastic Variational Inference (SVI) in order to evaluate the trade off between computational scalability and statistical fidelity, and to clarify when optimization based approximations can replace simulation based inference.

The analysis of the LRM shows that, under simple conjugate structures, VI and SVI produce posterior summaries and predictive distributions that are almost indistinguishable from those obtained with MCMC. Criteria such as WAIC, DIC, $\text{R}^2$ and MSE indicate negligible loss in statistical accuracy, while computation times decrease by one or more orders of magnitude. This agrees with the literature \citep{Ormerod2010-bc, Blei2016VariationalIA, zhang2018advancesvariationalinference}, which reports that in low dimensional or conjugate models variational methods provide an efficient and reliable surrogate for MCMC.

Our results also confirm a well known limitation of mean–field variational methods \citep{Blei2006-kd, Wainwright2007-ul, wang22g}. By factorizing the posterior into independent components, VI and SVI reduce posterior dependencies and tend to underestimate variances, particularly for variance components and random effects. The discussion in \cite{Blei2016VariationalIA} links this behavior to the Kullback–Leibler divergence used in the variational objective, which penalizes placing mass in regions of low posterior density more strongly than it rewards covering regions of high posterior support. In practice, this produces approximations that reproduce central tendencies well but attenuate dispersion, a pattern that appears throughout our LRM and CHLRM experiments.

The clustered extension yields a more balanced comparison. As reported in Table~\ref{tab:fit_chlrm_farms}, MCMC attains the best WAIC and DIC, while VI and SVI deliver values of similar magnitude and substantially lower runtime. All three methods recover essentially the same clustering structure, as illustrated by the co-clustering matrices in Figure~\ref{fig_adjacency}. Thus, although VI and SVI distort posterior uncertainty, they still identify global features such as cluster memberships with high accuracy, in line with \cite{Blei2006-kd, HOFFMAN, beal2003variational}. This robustness is relevant in applications where clustering performance is more important than exact calibration of posterior intervals.

A notable empirical finding is the similarity between VI and SVI in all scenarios considered. Both methods optimize the same Evidence Lower Bound within the same mean–field family, with SVI introducing stochastic natural gradients and mini–batching \cite{HOFFMAN, VAE}. In the sample sizes studied here, gradient noise is moderate and both procedures converge to almost identical solutions. This explains why SVI does not show a clear advantage over VI in our experiments, even though it remains crucial for very large or streaming datasets.

Overall, the results illustrate the fundamental trade off of Bayesian computation. MCMC preserves dependence structures and provides well calibrated credible intervals, but at a high computational cost. VI and SVI recast inference as optimization over restricted families and achieve drastic gains in speed and scalability, at the expense of underestimating posterior uncertainty and producing less reliable information criteria, echoing the theoretical analyses in \cite{Wainwright2007-ul, Blei2016VariationalIA, zhang2018advancesvariationalinference}. Our empirical study makes these contrasts explicit in linear and clustered regression settings.

From a methodological point of view, the strengths and weaknesses of each algorithm are now clear. MCMC is statistically robust and widely applicable, but slow. VI and SVI are fast and scalable, but variance underestimation and distorted posteriors are inherent to the mean–field framework. The paper contributes pedagogically by presenting full derivations, simulation studies and reproducible implementations that highlight these contrasts. The construction of VI and SVI algorithms required careful work on coordinate updates and natural gradients, together with practical decisions on initialization and convergence monitoring, underscoring the value of strong mathematical foundations in scalable Bayesian computation.

Future research could extend these results in several directions. Richer variational families, such as structured VI \cite{wang22g} or deep generative approximations based on variational autoencoders \cite{VAE}, could be adapted to the CHLRM to mitigate variance underestimation while retaining scalability. Hybrid schemes in which VI or SVI provide fast initializations and short MCMC runs refine the posterior \cite{Blei2016VariationalIA, zhang2018advancesvariationalinference} merit systematic evaluation, particularly in high dimensional settings. Extensions that combine CHLRM structure with nonparametric clustering, for instance Dirichlet process mixtures or Pitman–Yor priors, would increase flexibility in modeling latent heterogeneity and connect this framework with the broader literature on Bayesian mixture modeling \cite{Blei2006-kd, beal2003variational}. Finally, applying the proposed algorithms to large scale educational or biomedical datasets would test robustness under severe imbalance and model misspecification, and would help to establish practical guidelines for choosing between MCMC, VI and SVI in applied Bayesian regression.

\section*{Statements and declarations}

The authors declare that they have no known competing financial interests or personal relationships that could have appeared to influence the work reported in this article.

All R and C++ code required to reproduce our results is publicly available at \url{https://github.com/ccparra/VariationalBayes-HLRM/}. The repository includes a detailed README with step by step instructions, and the scripts are well documented. All datasets used in the applications and cross validation exercises are also included in the repository.

During the preparation of this work the authors used ChatGPT-5-Thinking in order to improve language and readability. After using this tool, the authors reviewed and edited the content as needed and take full responsibility for the content of the publication.

\bibliography{references.bib}

@article{watanabe2010asymptotic,
  title={Asymptotic equivalence of Bayes cross validation and widely applicable information criterion in singular learning theory.},
  author={Watanabe, Sumio and Opper, Manfred},
  journal={Journal of machine learning research},
  volume={11},
  number={12},
  year={2010}
}

@article{spiegelhalter2014deviance,
  title={The deviance information criterion: 12 years on},
  author={Spiegelhalter, David J and Best, Nicola G and Carlin, Bradley P and Linde, Angelika},
  journal={Journal of the Royal Statistical Society Series B: Statistical Methodology},
  volume={76},
  number={3},
  pages={485--493},
  year={2014},
  publisher={Oxford University Press}
}

@article{spiegelhalter2002bayesian,
  title={Bayesian measures of model complexity and fit},
  author={Spiegelhalter, David J and Best, Nicola G and Carlin, Bradley P and Van Der Linde, Angelika},
  journal={Journal of the royal statistical society: Series b (statistical methodology)},
  volume={64},
  number={4},
  pages={583--639},
  year={2002},
  publisher={Wiley Online Library}
}

@article{neal2011mcmc,
  title={MCMC using Hamiltonian dynamics},
  author={Neal, Radford M and others},
  journal={Handbook of markov chain monte carlo},
  volume={2},
  number={11},
  pages={2},
  year={2011},
  publisher={Chapman and Hall/CRC}
}

@article{gelfand2000gibbs,
  title={Gibbs sampling},
  author={Gelfand, Alan E},
  journal={Journal of the American statistical Association},
  volume={95},
  number={452},
  pages={1300--1304},
  year={2000},
  publisher={Taylor \& Francis}
}

@article{chib1995understanding,
  title={Understanding the metropolis-hastings algorithm},
  author={Chib, Siddhartha and Greenberg, Edward},
  journal={The american statistician},
  volume={49},
  number={4},
  pages={327--335},
  year={1995},
  publisher={Taylor \& Francis}
}

@book{rizzo2019statistical,
  title={Statistical computing with R},
  author={Rizzo, Maria L},
  year={2019},
  publisher={Chapman and Hall/CRC}
}

@article{watanabe2013widely,
  title={A widely applicable Bayesian information criterion},
  author={Watanabe, Sumio},
  journal={The Journal of Machine Learning Research},
  volume={14},
  number={1},
  pages={867--897},
  year={2013},
  publisher={JMLR. org}
}

@book{gelfand2000calculus,
  title={Calculus of variations},
  author={Gelfand, Izrail Moiseevitch and Silverman, Richard A and others},
  year={2000},
  publisher={Courier Corporation}
}

@book{bernardo1994bayesian,
  title={Bayesian theory},
  author={Bernardo, Jos{\'e} M and Smith, Adrian FM and Berliner, Mark},
  volume={586},
  year={1994},
  publisher={Wiley Online Library}
}

@book{albert2009bayesian,
  title={Bayesian computation with R},
  author={Albert, Jim and others},
  volume={2},
  year={2009},
  publisher={Springer}
}

@ARTICLE{SOSA,
 author={Sosa, Juan. and Aristizabal, Jeimmy.},
 title={Some Developments in Bayesian Hierarchical
Linear Regression Modeling},
 journal= {Revista Colombiana de Estadística - Applied Statistics
},
 volume={45},
 year={2022},
 pages={231-255}
 }

@ARTICLE{Blei2006-kd,
  title     = "Variational inference for Dirichlet process mixtures",
  author    = "Blei, David M and Jordan, Michael I",
  journal   = "Bayesian Anal.",
  publisher = "Institute of Mathematical Statistics",
  volume    =  1,
  number    =  1,
  pages     = "121--143",
  month     =  mar,
  year      =  2006
}

@ARTICLE{Diaconis1979-lb,
  title     = "Conjugate priors for exponential families",
  author    = "Diaconis, Persi and Ylvisaker, Donald",
  journal   = "Ann. Stat.",
  publisher = "Institute of Mathematical Statistics",
  volume    =  7,
  number    =  2,
  pages     = "269--281",
  month     =  mar,
  year      =  1979
}

@BOOK{Christensen2010-ps,
  title     = "Bayesian ideas and data analysis",
  author    = "Christensen, Ronald and Johnson, Wesley O and Branscum, Adam J
               and Hanson, Timothy E",
  publisher = "CRC Press",
  series    = "Chapman \& Hall/CRC Texts in Statistical Science",
  month     =  jul,
  year      =  2010,
  address   = "Boca Raton, FL"
}

@inproceedings{NIPS2000_77369e37,
 author = {Ghahramani, Zoubin and Beal, Matthew},
 booktitle = {Advances in Neural Information Processing Systems},
 editor = {T. Leen and T. Dietterich and V. Tresp},
 pages = {},
 publisher = {MIT Press},
 title = {Propagation Algorithms for Variational Bayesian Learning},
 url = {https://proceedings.neurips.cc/paper_files/paper/2000/file/77369e37b2aa1404f416275183ab055f-Paper.pdf},
 volume = {13},
 year = {2000}
}

@ARTICLE{Wainwright2007-ul,
  title     = "Graphical models, exponential families, and variational
               inference",
  author    = "Wainwright, Martin J and Jordan, Michael I",
  
  journal   = "Found. Trends\textregistered{} Mach. Learn.",
  publisher = "Now Publishers",
  volume    =  1,
  number    = "1-2",
  pages     = "1--305",
  year      =  2007,
  language  = "en"
}

@InProceedings{wang22g,
  title = 	 { Structured variational inference in Bayesian state-space models },
  author =       {Wang, Honggang and Bhattacharya, Anirban and Pati, Debdeep and Yang, Yun},
  booktitle = 	 {Proceedings of The 25th International Conference on Artificial Intelligence and Statistics},
  pages = 	 {8884--8905},
  year = 	 {2022},
  editor = 	 {Camps-Valls, Gustau and Ruiz, Francisco J. R. and Valera, Isabel},
  volume = 	 {151},
  series = 	 {Proceedings of Machine Learning Research},
  month = 	 {28--30 Mar},
  publisher =    {PMLR},
  url = 	 {https://proceedings.mlr.press/v151/wang22g.html} 
}

@BOOK{Crawley2012,
  title     = "The {R} Book",
  author    = "Crawley, Michael J",
  publisher = "Wiley-Blackwell",
  edition   =  2,
  month     =  dec,
  year      =  2012,
  address   = "Hoboken, NJ",
  language  = "en"
}

@inbook{Zellner1986,
author = {Zellner, A.},
title= {On assessing prior distributions and Bayesian regression analysis using g-prior distributions},
year = {1986},
booktitle = {Bayesian Inference and Decision Techniques: Essays in Honor of Bruno de Finetti},
address = {Amsterdam},
publisher = {NorthHolland},
pages = {233--243},
}

@ARTICLE{KassWasserman1995,
  title     = "A reference Bayesian test for nested hypotheses and its
               relationship to the Schwarz criterion",
  author    = "Kass, Robert E and Wasserman, Larry",
  journal   = "J. Am. Stat. Assoc.",
  publisher = "Informa UK Limited",
  volume    =  90,
  number    =  431,
  pages     = "928--934",
  month     =  sep,
  year      =  1995,
  language  = "en"
}

@misc{VAE,
      title={Auto-Encoding Variational Bayes}, 
      author={Diederik P Kingma and Max Welling},
      year={2022},
      eprint={1312.6114},
      archivePrefix={arXiv},
      primaryClass={stat.ML},
      url={https://arxiv.org/abs/1312.6114}, 
}

@ARTICLE{Ormerod2010-bc,
  title     = "Explaining variational approximations",
  author    = "Ormerod, J T and Wand, M P",
  journal   = "The American Statistician",
  publisher = "Informa UK Limited",
  volume    =  64,
  number    =  2,
  pages     = "140--153",
  month     =  may,
  year      =  2010,
  language  = "en"
}

@misc{zhang2018advancesvariationalinference,
      title={Advances in Variational Inference}, 
      author={Cheng Zhang and Judith Butepage and Hedvig Kjellstrom and Stephan Mandt},
      year={2018},
      eprint={1711.05597},
      archivePrefix={arXiv},
      primaryClass={cs.LG},
      url={https://arxiv.org/abs/1711.05597}, 
}

@BOOK{GAMERMAN,
  title     = "Markov chain Monte Carlo",
  author    = "Gamerman, Dani and Lopes, Hedibert F",
  publisher = "Chapman \& Hall/CRC",
  series    = "Chapman \& Hall/CRC Texts in Statistical Science",
  edition   =  2,
  month     =  may,
  year      =  2006,
  address   = "Philadelphia, PA"
}

@book{beal2003variational,
  title={Variational algorithms for approximate Bayesian inference},
  author={Beal, Matthew James},
  year={2003},
  publisher={University of London, University College London (United Kingdom)}
}

@article{Blei2016VariationalIA,
  title={Variational Inference: A Review for Statisticians},
  author={David M. Blei and Alp Kucukelbir and Jon D. McAuliffe},
  journal={Journal of the American Statistical Association},
  year={2016},
  volume={112},
  pages={859 - 877},
  url={https://api.semanticscholar.org/CorpusID:3554631}
}

@misc{REGUEIRO,
      title={Stochastic Gradient Variational Bayes in the Stochastic Blockmodel}, 
      author={Pedro Regueiro and Abel Rodríguez and Juan Sosa},
      year={2024},
      eprint={2410.02649},
      archivePrefix={arXiv},
      primaryClass={stat.ME},
      url={https://arxiv.org/abs/2410.02649}, 
}

@ARTICLE{AMARI,
  author={Amari, Shun-ichi},
  journal={Neural Computation}, 
  title={Natural Gradient Works Efficiently in Learning}, 
  year={1998},
  volume={10},
  number={2},
  pages={251-276},
  doi={10.1162/089976698300017746}}

@BOOK{HOFF,
 author={Hoff, Peter D.},
 title={A First Course in Bayesian Statistical Methods},
 publisher={Springer},
 address={Seattle WA},
 year={2009},
 }

@BOOK{GELMAN,
 author={Gelman, Andrew. and Hill, Jennifer.},
 title={Data Analysis Using Regression and
Multilevel/Hierarchical Models},
 publisher={Cambridge},
 address={New York},
 year={2007},
 }

@BOOK{GELMAN14,
 author={Gelman, Andrew. and Carlin, John. and Stern, Hal. and Dunson, David. and Vehtari, Aki. and Rubin, Donald.},
 title={Bayesian Data Analysis},
 publisher={CRC Press},
 address={Florida},
 year={2014},
 }

@ARTICLE{KL,
 author={ Kullback, S. and Leibler, R. A.},
 title={Information and Sufficiency},
 journal= {The Annals of Mathematical Statistics},
 volume={22},
 year={1951},
 pages={79-86}
 }

@TECHREPORT{RANGAN,
  author      = {Ranganath, Rajesh and Gerrish, Sean and Blei, David},
  title       = {Black Box Variational Inference},
  institution = {Princeton University},
  year        = {2014}
}

@ARTICLE{HOFFMAN,
 author={Matt Hoffman and David M. Blei and Chong Wang and John Paisley},
 title={Stochastic Variational Inference},
 journal= {Journal of Machine Learning Research},
 volume={14 No. 4},
 year={2013},
 pages={1303-1347}
 }

@book{BISHOP,
  author = {Bishop, Christopher M.},
  edition = {1},
  publisher = {Springer},
  title = {Pattern Recognition and Machine Learning (Information Science and Statistics)},
  year ={2007}
}

@article{Robbins&Monro:1951,
  author = {Robbins, H. and Monro, S.},
  journal = {Annals of Mathematical Statistics},
  pages = {400-407},
  title = {A stochastic approximation method},
  volume ={ 22},
  year = {1951}
}

@MASTERSTHESIS{ZETER,
 author={Zetterström, Victor},
 title={Variational Bayes as a Computer Intensive Method for Bayesian Regression},
 school={Uppsala University},
 year={2021},
 }

@BOOK{ML_MURPHY,
author = {Murphy, Kevin P.},
title = {Machine Learning: A Probabilistic Perspective},
year = {2012},
publisher = {The MIT Press}
}
\bibliographystyle{apalike}

\appendix

\section{Notation}\label{Appendix_A}

A probability density or mass function $p(\bm{x} \mid \bm{\theta})$ belongs to the natural exponential family, denoted $\bm{x}\sim \mathsf{NEF}(\bm{\eta})$, if it can be written as
\begin{equation}
\label{NEF}
p(\bm{x} \mid \bm{\eta}) 
= h(\bm{x}) \exp\left\{ \bm{\eta}^\top \bm{t}(\bm{x}) - a(\bm{\eta}) \right\},
\end{equation}
where
\begin{itemize}
\item $\bm{\eta} = g(\bm{\theta})$, with $g(\cdot)$ a mapping from the parameter vector $\bm{\theta}$ to the natural parameter vector $\bm{\eta}$,
\item $h(\bm{x})$ is the base measure,
\item $\bm{t}(\bm{x})$ is the sufficient statistic,
\item $a(\bm{\eta})$ is the log partition function that enforces normalization and satisfies
\[
\int h(\bm{x}) \exp\left\{ \bm{\eta}^\top \bm{t}(\bm{x}) - a(\bm{\eta}) \right\} \, \textsf{d}\bm{x} = 1
\quad\text{i.e.}\quad
a(\bm{\eta}) 
= \log\left( \int h(\bm{x}) \exp\left\{ \bm{\eta}^\top \bm{t}(\bm{x}) \right\} \, \textsf{d}\bm{x} \right).
\]
\end{itemize}
The log partition function satisfies
$\nabla_{\bm{\eta}} a(\bm{\eta}) = \mathsf{E}_{\bm{\eta}}\bigl[\bm{t}(\bm{x})\bigr]$ 
and
$\nabla_{\bm{\eta}}^2 a(\bm{\eta}) = \mathsf{Var}_{\bm{\eta}}\bigl[\bm{t}(\bm{x})\bigr]$.

\subsection{Distributions}

\textbf{Multivariate normal}

A $d \times 1$ random vector $\bm{X} = (X_1, \dots, X_d)^\top$ has a multivariate Normal distribution with mean $\bm{\mu}$ and covariance $\bm{\Sigma}$, denoted $\bm{X} \mid \bm{\mu}, \bm{\Sigma} \sim \mathsf{N}_d(\bm{\mu}, \bm{\Sigma})$, if
\[
p(\bm{x} \mid \bm{\mu}, \bm{\Sigma})
= (2\pi)^{-d/2} |\bm{\Sigma}|^{-1/2}
  \exp\left\{ -\frac{1}{2} (\bm{x} - \bm{\mu})^\top \bm{\Sigma}^{-1} (\bm{x} - \bm{\mu}) \right\}.
\]
\begin{itemize}
\item Sufficient statistics:
\[
\bm{t}(\bm{x}) = \bigl(\bm{x},\, \bm{x}\bm{x}^\top\bigr)^\top.
\]
\item Natural parameter:
\[
\bm{\eta}
=
\bigl(\bm{\Sigma}^{-1}\bm{\mu},\ -\tfrac{1}{2}\bm{\Sigma}^{-1}\bigr)^\top.
\]
\item Inverse mapping for $\bm{\eta} = (\bm{\eta}_1,\bm{\eta}_2)^\top$:
\[
\bm{\Sigma} = -\tfrac{1}{2}\bm{\eta}_2^{-1},
\qquad
\bm{\mu} = -\tfrac{1}{2}\bm{\eta}_2^{-1}\bm{\eta}_1.
\]
\item Key expectations:
\[
\mathsf{E}[\bm{X}] = \bm{\mu},
\qquad
\mathsf{E}[\bm{X}\bm{X}^\top] = \bm{\Sigma} + \bm{\mu}\bm{\mu}^\top,
\qquad
\mathsf{E}[\log p(\bm{X} \mid \bm{\mu}, \bm{\Sigma})]
=
-\frac{d}{2}\log(2\pi e) - \frac{1}{2}\log|\bm{\Sigma}|.
\]
\end{itemize}

\textbf{Gamma}

A random variable $X$ has a Gamma distribution with parameters $\alpha,\beta>0$, denoted $X \mid \alpha,\beta \sim \mathsf{G}(\alpha,\beta)$, if
\[
p(x \mid \alpha,\beta)
=
\frac{\beta^\alpha}{\Gamma(\alpha)} x^{\alpha - 1} e^{-\beta x},
\qquad x>0.
\]
\begin{itemize}
\item Sufficient statistics:
\[
\bm{t}(x) = \bigl(\log x,\ x\bigr)^\top.
\]
\item Natural parameter:
\[
\bm{\eta} = (\alpha - 1,\ -\beta)^\top.
\]
\item Inverse mapping for $\bm{\eta} = (\eta_1,\eta_2)^\top$:
\[
\alpha = \eta_1 + 1,
\qquad
\beta = -\eta_2.
\]
\item Key expectations:
\[
\mathsf{E}[X] = \frac{\alpha}{\beta},
\qquad
\mathsf{E}[\log X] = \psi(\alpha) - \log\beta,
\]
\[
\mathsf{E}[\log p(X \mid \alpha,\beta)]
=
\log\beta - \log\Gamma(\alpha) + (\alpha - 1)\psi(\alpha) - \alpha.
\]
\end{itemize}

\textbf{Inverse Gamma}

A random variable $X$ has an Inverse Gamma distribution with parameters $\alpha,\beta>0$, denoted $X \mid \alpha,\beta \sim \mathsf{IG}(\alpha,\beta)$, if
\[
p(x \mid \alpha,\beta)
=
\frac{\beta^\alpha}{\Gamma(\alpha)} x^{-(\alpha + 1)}
\exp\left\{ -\frac{\beta}{x} \right\},
\qquad x>0.
\]
\begin{itemize}
\item Sufficient statistics:
\[
\bm{t}(x) = \left(\log x,\ \frac{1}{x}\right)^\top.
\]
\item Natural parameter:
\[
\bm{\eta} = (-\alpha - 1,\ -\beta)^\top.
\]
\item Inverse mapping for $\bm{\eta} = (\eta_1,\eta_2)^\top$:
\[
\alpha = -\eta_1 - 1,
\qquad
\beta = -\eta_2.
\]
\item Key expectations:
\[
\mathsf{E}[X] = \frac{\beta}{\alpha - 1}\quad(\alpha>1),
\qquad
\mathsf{E}\Bigl[\frac{1}{X}\Bigr] = \frac{\alpha}{\beta},
\qquad
\mathsf{E}[\log X] = \log\beta - \psi(\alpha),
\]
\[
\mathsf{E}[\log p(X \mid \alpha,\beta)]
=
-\log\beta - \log\Gamma(\alpha)
+ (\alpha + 1)\psi(\alpha) - \alpha.
\]
\end{itemize}

\textbf{Inverse Wishart}

A $d \times d$ random matrix $\bm{W}$ has an Inverse Wishart distribution with parameters $\nu$ and $\bm{S}^{-1}$, denoted $\bm{W} \mid \nu,\bm{S}^{-1} \sim \mathsf{IW}(\nu,\bm{S}^{-1})$, if
\[
p(\bm{W} \mid \nu,\bm{S}^{-1})
=
\frac{|\bm{S}|^{\nu/2}}
     {2^{d\nu/2}\,\Gamma_d(\nu/2)}
|\bm{W}|^{-(\nu + d + 1)/2}
\exp\left\{ -\frac{1}{2}\operatorname{tr}(\bm{S}\bm{W}^{-1}) \right\},
\quad \nu>0,\ \bm{S}>0.
\]
\begin{itemize}
\item Sufficient statistics:
\[
\bm{t}(\bm{W}) = \bigl(\bm{W}^{-1},\ \log|\bm{W}|\bigr)^\top.
\]
\item Natural parameter:
\[
\bm{\eta}
=
\left(-\frac{1}{2}\bm{S},\ -\frac{\nu + d + 1}{2}\right).
\]
\item Inverse mapping for $\bm{\eta} = (\bm{\eta}_1,\eta_2)$:
\[
\bm{S} = -2\bm{\eta}_1,
\qquad
\nu = -2\eta_2 - d - 1.
\]
\item Key expectations:
\[
\mathsf{E}[\bm{W}^{-1}] = \nu\bm{S},
\qquad
\mathsf{E}[\bm{W}] = \frac{\bm{S}^{-1}}{\nu - d - 1},
\]
\[
\mathsf{E}[\log|\bm{W}|]
=
-\sum_{i=1}^{d}\psi\!\left(\frac{\nu + 1 - i}{2}\right)
- d\log 2 + \log|\bm{S}^{-1}|,
\]
\[
\begin{split}
\mathsf{E}[\log p(\bm{W} \mid \nu,\bm{S}^{-1})]
&=
\frac{\nu}{2}\log|\bm{S}|
- \frac{\nu d}{2}\log 2
- \log\Gamma_d\!\left(\frac{\nu}{2}\right) \\
&\quad
- \frac{\nu + d + 1}{2}
\Biggl[
-\sum_{i=1}^{d}\psi\!\left(\frac{\nu + 1 - i}{2}\right)\\
&\quad- d\log 2 + \log|\bm{S}^{-1}|
\Biggr]
- \frac{\nu}{2}\operatorname{tr}(\bm{S}^2).
\end{split}
\]
\end{itemize}

\textbf{Dirichlet}

A $K \times 1$ random vector $\bm{X} = (X_1,\dots,X_K)^\top$ has a Dirichlet distribution with parameter $\bm{\alpha} = (\alpha_1,\dots,\alpha_K)^\top$, $\alpha_k>0$, denoted $\bm{X} \mid \bm{\alpha} \sim \mathsf{Dir}(\bm{\alpha})$, if
\[
p(\bm{x} \mid \bm{\alpha})
=
\frac{\Gamma\bigl(\sum_{k=1}^{K}\alpha_k\bigr)}
     {\prod_{k=1}^{K}\Gamma(\alpha_k)}
\prod_{k=1}^{K} x_k^{\alpha_k - 1},
\quad
\sum_{k=1}^{K} x_k = 1.
\]
\begin{itemize}
\item Sufficient statistics:
\[
\bm{t}(\bm{x}) = (\log x_1,\dots,\log x_K)^\top.
\]
\item Natural parameter:
\[
\bm{\eta} = (\alpha_1 - 1,\dots,\alpha_K - 1)^\top.
\]
\item Inverse mapping for $\bm{\eta} = (\eta_1,\dots,\eta_K)^\top$:
\[
\alpha_k = \eta_k + 1,
\qquad k=1,\dots,K.
\]
\item Key expectations:
\[
\mathsf{E}[X_k] = \frac{\alpha_k}{\sum_{\ell=1}^{K}\alpha_\ell},
\qquad
\mathsf{E}[\log X_k] = \psi(\alpha_k) - \psi\!\left(\sum_{\ell=1}^{K}\alpha_\ell\right),
\]
\[
\mathsf{E}[\log p(\bm{X} \mid \bm{\alpha})]
=
\log\Gamma\!\left(\sum_{k=1}^{K}\alpha_k\right)
- \sum_{k=1}^{K}\log\Gamma(\alpha_k)
+ \sum_{k=1}^{K}(\alpha_k - 1)
\Bigl[\psi(\alpha_k) - \psi\!\left(\sum_{\ell=1}^{K}\alpha_\ell\right)\Bigr].
\]
\end{itemize}

\textbf{Categorical}

A random variable $X \in \{1,\dots,K\}$ has a categorical distribution with parameter $\bm{p} = (p_1,\dots,p_K)^\top$, $\sum_{k=1}^{K}p_k=1$, denoted $X \mid \bm{p} \sim \mathsf{Cat}(\bm{p})$, with
\[
p(x \mid \bm{p}) = \prod_{k=1}^{K} p_k^{[x = k]}.
\]
\begin{itemize}
\item Sufficient statistics:
\[
\bm{t}(x) = \bigl([x=1],\dots,[x=K]\bigr)^\top.
\]
\item Natural parameter:
\[
\bm{\eta} = (\log p_1,\dots,\log p_K)^\top.
\]
\item Inverse mapping for $\bm{\eta} = (\eta_1,\dots,\eta_K)^\top$:
\[
p_k = \frac{\exp(\eta_k)}{\sum_{\ell=1}^{K}\exp(\eta_\ell)},
\qquad k=1,\dots,K.
\]
\item Key expectations:
\[
\mathsf{E}[\, [X = k]\,] = p_k,
\qquad
\mathsf{E}[\log p(X \mid \bm{p})]
= \sum_{k=1}^{K} p_k \log p_k.
\]
\end{itemize}

\textbf{Multinomial}

A $K \times 1$ random vector $\bm{X} = (X_1,\dots,X_K)^\top$ with $\sum_{k=1}^{K}X_k = n$ has a multinomial distribution with parameter $\bm{p} = (p_1,\dots,p_K)^\top$, $\sum_{k=1}^{K}p_k=1$, denoted $\bm{X} \mid n,\bm{p} \sim \mathsf{Mult}(n,\bm{p})$, with
\[
p(\bm{x} \mid n,\bm{p})
=
\frac{n!}{x_1!\cdots x_K!}
\prod_{k=1}^{K} p_k^{x_k},
\quad
\sum_{k=1}^{K} x_k = n.
\]
\begin{itemize}
\item Sufficient statistics:
\[
\bm{t}(\bm{x}) = (x_1,\dots,x_K)^\top.
\]
\item Natural parameter:
\[
\bm{\eta} = (\log p_1,\dots,\log p_K)^\top.
\]
\item Inverse mapping for $\bm{\eta} = (\eta_1,\dots,\eta_K)^\top$:
\[
p_k = \frac{\exp(\eta_k)}{\sum_{\ell=1}^{K}\exp(\eta_\ell)},
\qquad k=1,\dots,K.
\]
\item Key expectations:
\[
\mathsf{E}[X_k] = n p_k,
\]
\[
\mathsf{E}[\log p(\bm{X} \mid n,\bm{p})]
=
\log(n!)
- \sum_{k=1}^{K} \mathsf{E}[\log(X_k!)]
+ n \sum_{k=1}^{K} p_k \log p_k.
\]
\end{itemize}

Here $\Gamma(\cdot)$ and $\Gamma_d(\cdot)$ denote the univariate and multivariate Gamma functions, respectively,
\[
\Gamma(z) = \int_{0}^{\infty} t^{z-1} e^{-t}\,dt,
\qquad
\Gamma_d(\alpha) = \pi^{d(d-1)/4} \prod_{j=1}^{d} \Gamma\bigl(\alpha - (j-1)/2\bigr),
\]
and $\psi(\cdot)$ is the digamma function,
\[
\psi(z) = \frac{\textsf{d}}{\textsf{d}z}\log\Gamma(z).
\]
The Iverson bracket $[A]$ equals $1$ if condition $A$ holds and $0$ otherwise.

\end{document}